\documentclass[11pt]{article}
\usepackage[letterpaper,margin=1in]{geometry}
\usepackage[T1]{fontenc}
\usepackage{amsmath,amsthm,mathtools}
\usepackage{newtxtext,newtxmath}
\usepackage{tikz}
\usepackage{microtype,booktabs,array,enumitem,graphicx,placeins}

\usetikzlibrary{arrows.meta}
\usepackage{authblk}
\usepackage{algorithm,algpseudocode}
\algrenewcommand{\algorithmicrequire}{\textbf{Input:}}
\algrenewcommand{\algorithmicensure}{\textbf{Output:}}
\usepackage[round,authoryear]{natbib}
\usepackage{xcolor}
\usepackage[colorlinks=true,linkcolor=blue!45!black,citecolor=blue!45!black,urlcolor=blue!45!black]{hyperref}
\newtheorem{theorem}{Theorem}[section]
\newtheorem{proposition}[theorem]{Proposition}
\newtheorem{lemma}[theorem]{Lemma}
\newtheorem{corollary}[theorem]{Corollary}
\theoremstyle{definition}
\newtheorem{assumption}[theorem]{Assumption}
\newtheorem{example}[theorem]{Example}
\newcommand{\M}{\mathcal M}
\newcommand{\E}{\mathbb E}
\newcommand{\R}{\mathbb R}
\newcommand{\law}{\mathcal L}

\newcommand{\Pcal}{\mathcal P}
\newcommand{\norm}[1]{\left\lVert #1\right\rVert}
\newcommand{\Wp}[2]{W_2\!\left(#1\,;#2\right)}
\DeclareMathOperator*{\argmin}{arg\,min}
\DeclareMathOperator{\diam}{diam}

\DeclareMathOperator{\supp}{supp}

\DeclareMathOperator{\Lip}{Lip}

\usepackage{xcolor}
\title{Fr\'echetFlow: Flow Matching for Non-Euclidean Data}
\author[1]{Yidong Zhou}
\author[2]{Muqing Cui}
\author[2]{Hans-Georg M\"uller\thanks{Corresponding author: Hans-Georg M\"uller, One Shields Avenue, Davis, CA 95616, USA. Email: \href{mailto:hgmueller@ucdavis.edu}{hgmueller@ucdavis.edu}.}}
\affil[1]{School of Statistics, University of Minnesota}
\affil[2]{Department of Statistics and Data Science, University of California, Davis}
\date{}
\begin{document}
\maketitle

\begin{abstract}
Scientific observations such as probability distributions, trees and covariance matrices call for generative models that respect their non-Euclidean geometry. While flow matching has been extended to manifolds by utilizing geodesics and velocities, such manifold structures  are  not available in a general metric space. We introduce \emph{Fr\'echetFlow}, a generative modeling framework for metric space-valued data that requires neither linear operations nor a smooth manifold structure. A key feature of the proposed approach is that we  utilize well-established Euclidean flows to learn the joint distribution of distances to a set of landmarks, whereupon new  objects are reconstructed from generated landmark representations. The reconstruction step utilizes  weighted Fr\'echet means of reference objects, drawing on Fr\'echet regression. We establish Wasserstein error bounds that connect accuracy of the generated object distributions to the placement of landmarks, flow approximation and reconstruction error and prove consistency under finite second moments. This leads to convergence rates that depend on the metric complexity of the target objects,  tail moments and fitting accuracy. Experiments show competitive accuracy with established flow matching methods in Euclidean simulations and lower Wasserstein errors than specifically tailored methods for several scientific datasets. Across simulation settings, we find that generation error decreases as sample size increases.
\end{abstract}


\section{Introduction}\label{sec:introduction}
\subsection{General framework and overview} 
Increasingly, data encountered  in  scientific studies are of high complexity  and may encompass   probability distributions \citep{petersen2022modeling}, covariance matrices \citep{dryden2009covariance}, phylogenetic trees \citep{billera2001geometry} or  locations on a curved manifold \citep{bhattacharya2003manifolds}. Such data are referred to as  random objects, which are random elements located in a metric space \citep{marron2014ooda,muller2016random,dubey2024profiles}. We study the problem of estimating  the probability distribution of such random objects from a given sample, aiming to generate new observations that follow the same law. 

Generative adversarial networks and diffusion-based models have been well established and provide flexible methods for learning distributions in Euclidean spaces \citep{goodfellow2014generative,ho2020denoising,song2021score}. Flow matching has emerged as an alternative method that trains a time-dependent vector field by regression along conditional probability paths \citep{lipman2023flow,tong2024cfm}, with related implementations through rectified flows and stochastic interpolants \citep{liu2023rectified,albergo2023interpolants}. Riemannian and Wasserstein flow matching extend this approach using paths and velocities adapted to manifolds and probability distributions \citep{chen2024riemannian,haviv2025wasserstein}. While an underlying metric enables comparisons between random objects situated in a metric space through their distances, there are generally no  vector operations or tangent spaces for defining the intrinsic dynamics on which previous formulations of flow matching rely. The challenge for extending flow matching to the case of general random objects  is to retain the flexibility of flow matching without requiring a linear or smooth manifold structure on the data space.  %

Here we introduce Fr\'echetFlow, a generative modeling framework for metric space-valued data. The key idea is to represent each random object by its distances to a common set of landmarks in the object space. The resulting landmark representation then yields a numerical vector that is different from but nevertheless related to dissimilarity learning \citep{pekalska2001dissimilarity}. Subsequently, a Euclidean flow learns the joint distribution of these vectors and then generates new vectors from the fitted distribution.  The generated vectors specify the distances that a new object should reproduce. We then construct the corresponding newly generated object as a weighted Fr\'echet mean of reference objects, the metric counterpart of an ordinary weighted average \citep{frechet1948elements,sturm2003barycenters} that is related to Fr\'echet regression \citep{petersen2019frechet,zhou2026e2m}. We employ a neural network to learn the weights by minimizing a normalized combination of the squared supremum norm distance between generated and reconstructed landmark representations and the mean squared reconstruction error on observed objects. During fitting and generation, each weighted Fr\'echet mean is obtained by minimizing a weighted sum of squared distances. Both representation and reconstruction are therefore defined through the metric and proceed without relying on linear operations or a smooth manifold structure in  the object space. It is important to note that the actual flow matching takes place in a Euclidean vector space defined by the distances from objects to the landmarks and therefore one can utilize the most promising established algorithms for this step.

Our main contributions are as follows:
\begin{itemize}[leftmargin=*,itemsep=4pt,topsep=4pt]
\item We propose Fr\'echetFlow, a general generative modeling framework for random objects, i.e.  metric space-valued data,  that does not require any linear or smooth manifold structure. It is based on representing each object by its distances to $k$ selected landmarks, yielding a vector in $\mathbb{R}^k$, utilizing existing flow matching for these vectors, and learning a reconstruction map from Euclidean vector space to the object space. %
\item We establish Wasserstein bounds that show how sampling, landmark selection, flow fitting and object reconstruction affect the accuracy of the generated object distribution. %
Utilizing empirical medoid landmarks, we prove consistency on Polish spaces under  finite second moments as the number of landmarks increases and fitting errors vanish. Under additional covering and moment conditions, we provide convergence rates that quantify the effects of support complexity, landmark number and fitting accuracy.
\item We show that Fr\'echetFlow matches the flow matching baseline on eight Gaussians and two moons, with slightly lower mean Wasserstein errors at larger sample sizes. It also achieves lower mean Wasserstein error than the corresponding baselines for other data such as  earthquakes, volcanoes, MERFISH cell neighborhoods and brain networks. 
\item For non-Euclidean data such as  metric graphs, phylogenetic trees and data in cubical configuration spaces, we  show through simulations that the proposed Fr\'echetFlow captures branch, topology and cell probabilities along  with variation in positions and branch lengths, with  errors   decreasing with increasing sample size. 
\item We develop an algorithm to learn the weights used for the Fr\'echet reconstruction of the objects that is a key step for  Fr\'echetFlow, by training a neural network to match the reconstructed object's landmark representation with the generated Euclidean distance vector, also utilizing  an additional loss that quantifies the accuracy of the reconstruction of observed objects from their own landmark representations. %
\end{itemize}

\subsection{Related work}\label{sec:related-work}
Over the past several years, statistical methods for random objects in the area of metric statistics have been developed for  estimation, prediction and inference, combining  geometric features with the underlying probability measure of the metric measure space where the data are located \citep{Song2026Non}. Early work established large sample theory for intrinsic and extrinsic means on smooth Riemannian manifolds \citep{bhattacharya2003manifolds} and an empirical regression approach based on distance matrices  \citep{faraway2014distance}. Subsequently, Fr\'echet regression introduced  global and local models for metric space-valued responses in a regression setting \citep{petersen2019frechet,chen2022uniform}, while distance profiles support distributional exploration and hypothesis testing \citep{dubey2024profiles}. 

More recently, flexible prediction approaches have been introduced through deep Fr\'echet regression, which learns Euclidean representations and maps predictions back by local Fr\'echet regression \citep{iao2025deep} and Fr\'echet geodesic boosting, which constructs ensembles using geodesics as residuals \citep{zhou2025boosting}. End-to-end metric regression \citep{zhou2026e2m} and related neural methods \citep{kim2025dfnn} use weighted Fr\'echet means for prediction and learn the weights by minimizing mean squared distance between predicted and observed responses; the latter methods have been shown to have superior performance over previous Fr\'echet regression approaches in the reported experiments.  

The proposed Fr\'echetFlow generates new objects from a learned distribution of landmark representations. The reconstruction map is trained to match each generated vector with the reconstructed object's landmark representation and to reconstruct observed objects from their own landmark representations. %

Geometric generative models adapt continuous normalizing flows and score-based diffusions to manifolds \citep{mathieu2020rcnf,debortoli2022riemannian}. Generative models have also been developed for stratified supports \citep{martinez2026deep}. Riemannian flow matching learns tangent velocities along geodesic or premetric paths \citep{chen2024riemannian}. Related developments include formulations using Lie group operations or data-dependent Riemannian metrics obtained through manifold learning \citep{kapusniak2024metric,sherry2025lie} and associated theoretical analysis \citep{kumar2026flow}. Flow matching has also been extended to probability measures through measure-dependent transport dynamics under the Wasserstein metric \citep{haviv2025wasserstein,piening2026generalized,haviv2026riemannian} and to functional data through function space structure or models of functional dependence \citep{kerrigan2024functional,tan2026smooth}. For samples of phylogenetic trees, PhylaFlow \citep{ektefaie2026phylaflow} combines continuous branch length dynamics with discrete topology changes in Billera--Holmes--Vogtmann space \citep{billera2001geometry}.  Autoencoder-based generators and latent flow matching have been proposed for high-dimensional vector data. These approaches learn distributions in a representation space and reconstruct the high-dimensional data from the generated vectors \citep{tolstikhin2018wae,dao2023latent}. Related approaches are  DiffeoCFM, which  uses global diffeomorphisms and pullback metrics for matrix valued observations \citep{collas2025diffeocfm} and  Cartan flow matching, which uses Lie algebra and quotient structure of Riemannian symmetric spaces \citep{ruscelli2026cartan}.
Locations on a fixed metric graph have been modeled using ambient embeddings, neural entropic optimal transport and projection back to the graph \citep{micheli2026metric_graph}. 

Fr\'echetFlow applies a common flow objective and reconstruction criterion across these types of random objects, where the underlying metric structure enters through distances and weighted Fr\'echet means. Since it is based on underlying latent Euclidean dynamics, it takes advantage of established Euclidean flow models and ordinary differential equation (ODE) solvers, while not requiring linear operations or a smooth manifold structure for the metric space where the random objects reside.  In graph, whole-tree and cubical configuration spaces that we use to illustrate the approach, this reconstruction also determines branch, topology or cell membership without requiring a  separate model for discrete transitions.
The landmark representations we propose here  do not require a smooth invertible coordinate map; the weighted means with learned weights yield valid random objects that are situated in  the original
metric space. The Wasserstein bounds connect the generated object law with the error in the fitted flow, the attained reconstruction residual and distance to the nearest landmark, %
explicitly accounting for the numerical reconstruction error. %

\section{Generating random objects}\label{sec:method}
\subsection{Preliminaries and landmark representation}
Let $(\M,d,P)$ be a metric measure space \citep{heinonen2001analysis}, where $P$ is a Borel probability measure on $(\M,d)$. We are given  a sample consisting  of random objects  $Y_1,\ldots,Y_n$ distributed according to the  unknown law $P$. Our goal is to construct a generator such that the distribution of the generated objects  approximates the unknown data distribution $P$. Write $P_n=n^{-1}\sum_{i=1}^n\delta_{Y_i}$ for the empirical distribution of the observed sample, where $\delta_y$ denotes the unit point mass at $y\in\M$. 

For a metric space $\mathcal X$, let $\Pcal_2(\mathcal X)$ denote the Borel probability measures with finite second moments. For two probability distributions $\mu$ and $\nu$ on the same metric space, both with finite second moments, $\Wp{\mu,\nu}{\rho}$ denotes their 2-Wasserstein distance with respect to a given metric $\rho$, which could be the metric $d$ or some other metric; we  write $\Wp{\mu,\nu}{d}$ on $\M$ and $\Wp{\mu,\nu}{\norm{\cdot}_2}$ or $\Wp{\mu,\nu}{\norm{\cdot}_\infty}$ for Euclidean or supremum norm distance on Euclidean spaces. The metric $\rho$  is not emphasized when it is clear from the context. For a measurable map $\Psi:\mathcal M_1\to\mathcal M_2$ between metric spaces $\M_1$ and $\M_2$ and a Borel probability measure $P$ on $\mathcal M_1$, the pushforward $\Psi_\#P$ is the probability distribution on $\mathcal M_2$ of $\Psi(Y)$ for $Y\sim P$. Either space may be a non-Euclidean object space $\M$ or %
$\R^p$.

\begin{figure}[!t]
\centering
\includegraphics[width=\textwidth]{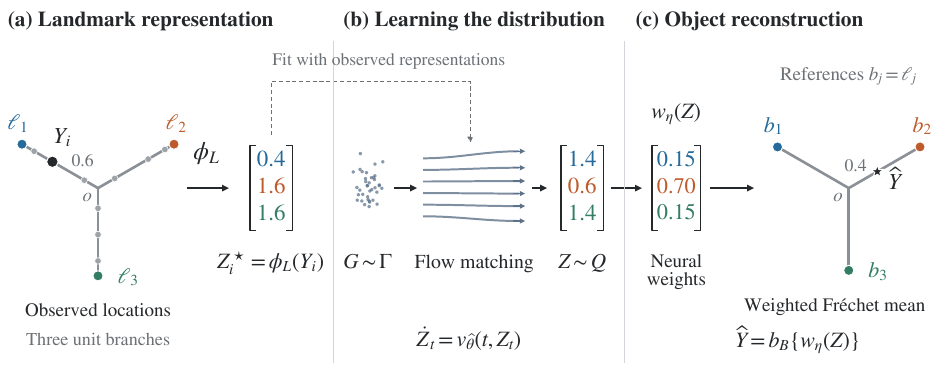}
\caption{Fr\'echetFlow for locations on a tree with three unit branches meeting at $o$ and shortest-path distance. (a) Each observed point gives a vector $Z_i^\star=\phi_L(Y_i)$ of distances to the three leaves. Colors link leaves and vector entries. (b) The dashed arrow supplies observed representations for flow fitting; Gaussian draws produce new representations $Z\sim Q$. (c) Neural weights $w_\eta(Z)$ yield a weighted Fr\'echet mean, using the same leaves as references. The displayed distances and weights give a generated point 0.4 units along the right branch; its distances to the three leaves match the sampled representation $Z$.}
\label{fig:frechetflow-overview}
\end{figure}

\begin{assumption}%
\label{ass:object-law}
The metric space $(\M,d)$ is complete and separable (Polish) with $P\in\Pcal_2(\M)$, i.e. $\E_Pd(Y,y_0)^2<\infty$ for some $y_0\in\M$.
\end{assumption}
The Polish space assumption implies the existence of  regular conditional distributions that are needed for the  coupling arguments in the proofs; finite second moments ensure finiteness of the Wasserstein distance $W_2$  \citep[Chapters~1 and~6]{villani2009optimal}.

A key feature of Fr\'echetFlow is that objects are represented by their distances to a common collection of landmarks; see Figure~\ref{fig:frechetflow-overview}(a). Let $L=(\ell_1,\ldots,\ell_k)$ be a finite nonempty collection of $k$ landmarks, which are elements of $\M$. Both the landmarks and their number may depend on the sample size $n$; we suppress this dependence in our notation.  The landmark representation of a random element $Y\in\M$ and the population distribution of these landmark representations are  given by
\begin{equation}\label{eq:coordinates}\phi_L(Y)=\{d(Y,\ell_1),\ldots,d(Y,\ell_k)\}^{\mathsf T},\qquad \mu_L=(\phi_L)_\#P,\end{equation}
where each entry of the vector $\phi_L(Y)$ is indexed by a specified landmark and $\mu_L$ describes the joint variation of these distances. 

The motivation for applying this device is to obtain a tractable vector representation of random objects that records their distances to the selected landmarks,  with a forward map $\phi_L$ that encodes the distance vector for any given random object, and a more challenging reverse map from landmark representations to objects. This device makes it possible to tap into vector flow methodology when learning flows for random objects. By the reverse triangle inequality,  $\norm{\phi_L(x)-\phi_L(y)}_\infty\le d(x,y)$, so that nearby objects also have nearby landmark representations.
Distances to reference objects are also used in dissimilarity learning and distance profiles \citep{pekalska2001dissimilarity,dubey2024profiles,tan2026dpe}. Gaussian and exponential transformations turn distances into similarities, connecting such representations to kernel features \citep{muandet2017kernel}.

For selecting landmarks from the observed sample, we use farthest-point selection (FPS) \citep{gonzalez1985clustering}. Starting with a randomly chosen observation, FPS successively adds the observation farthest from its nearest selected landmark until $k$ landmarks have been selected. Uniform random subsampling provides an alternative.

\subsection{Learning the distribution of landmark representations}
The observed vectors $Z_i^\star=\phi_L(Y_i)$ provide the data for learning the landmark representation law $\mu_L$ in $\R^k$, as defined in \eqref{eq:coordinates}. Flow matching learns this law by transporting draws from a given source distribution through Euclidean space \citep{lipman2023flow,liu2023rectified,albergo2023interpolants}; see Figure~\ref{fig:frechetflow-overview}(b). We use a $k$-dimensional Gaussian source distribution $$\Gamma \equiv N(0,I_k),$$ and a neural network $v_\theta(t,x)\in\R^k$,  with parameters $\theta$ to describe the velocity at position $x$ and time $t\in[0,1]$. Time runs from the Gaussian source law $\Gamma$ at $t=0$ toward the target representation law $\mu_L$ at $t=1$.

To learn the velocity field, flow matching uses straight paths for which the velocities are known. At the population level, we draw $G\sim\Gamma$, $Y\sim P$ and $T\sim\operatorname{Unif}[0,1]$ independently and set $Z^\star=\phi_L(Y)$. The path from the Gaussian draw $G$ to the target vector $Z^\star$ and the associated velocity are
\begin{equation}\label{eq:fm-training-path}X_t=(1-t)G+tZ^\star,\qquad U=Z^\star-G,\qquad 0\le t\le1.\end{equation}
At time $T$, the regression input is $(T,X_T)$ and the response is the path velocity $U=dX_t/dt$. Both are computed directly from the sampled endpoints using \eqref{eq:fm-training-path}. We fit the network by predicting this velocity given  time and position, aiming to minimize the  population loss $\E\norm{v_\theta(T,X_T)-U}_2^2$. The expectation is taken  over the independent draws $Y$, $G$ and $T$. At a given time and position, different paths are associated with different velocities. The population minimizer $u_t(x)=\E(U\mid X_t=x)$ averages these velocities and describes the local movement of probability mass along the interpolation. %

Given data,  we obtain estimates of the landmark representation law through the empirical distribution of the sample $Z_1^\star,\ldots,Z_n^\star$. Each regression sample is a uniform draw of these vectors that is paired with independent Gaussian and time draws. For the training of the model we employ  batches of the resulting inputs and responses to fit $v_\theta$ by stochastic gradient updates. Since the training paths and velocities are explicitly available,  this step does not require numerical integration. A possible option is to replace $Z^\star$ in \eqref{eq:fm-training-path} by $Z^\star+\sigma H/\sqrt{k}$, where $\sigma\ge0$ and $H\sim N(0,I_k)$ is independent of the other draws \citep{lipman2023flow}. We use $\sigma=0$ in our experiments, so that the paths end at  observed landmark representations, and this also avoids the need to select an additional tuning parameter (see Section~\ref{app:flow} of the Supplement). %

Given fitted parameters $\widehat\theta$, we then draw a new $G\sim\Gamma$ and solve
\begin{equation}\label{eq:fm-generation}\dot Z_t=v_{\widehat\theta}(t,Z_t),\qquad Z_0=G,\qquad 0\le t\le1.\end{equation}
Numerical integration produces an endpoint $\widetilde Z_1$; independent Gaussian starting points lead to  new landmark representations. If there are known constraints for the data, we consider a closed convex set $\mathcal Z\subseteq\R^k$ containing the true landmark representations and set
\begin{equation}\label{eq:clipped-code}Z=\Pi_{\mathcal Z}(\widetilde Z_1),\qquad Q=\law(Z),\end{equation}
where $\Pi_{\mathcal Z}$ denotes Euclidean projection onto the constrained space $\mathcal Z$.  Pertinent scenarios  are the choices $\mathcal Z=[0,\infty)^k$, which enforces nonnegative distances,  $\mathcal Z=\R^k$, which  leaves the endpoint unchanged and $\mathcal Z=[0,R]^k$ when all true landmark distances are known to be bounded by $R>0$. If flow fitting utilizes  an invertible affine scaling $S$ to standardize the landmark distances \citep[Section~4.2]{papamakarios2017masked}, we apply $S^{-1}$ before projection, returning to the original distance units.

\subsection{Generating objects from landmark representations}
A reconstruction map $D_\eta$ is a key ingredient of Fr\'echetFlow. It maps a generated landmark representation $Z\sim Q$ to an object in $\M$. For a realization $z$, the reconstruction target is to match the returned object's landmark representation to $z$, using the same representation as the flow fitting. We construct these objects as weighted Fr\'echet means in analogy to Fr\'echet regression \citep{frechet1948elements,sturm2003barycenters,petersen2019frechet,zhou2026e2m}.
Denote by $B=(b_1,\ldots,b_r)$ a finite collection of objects in $\M$, which we refer to as the reference set and the elements $b_j$ as references in the following, and by $\Delta^{r-1}=\{w\in[0,1]^r:\sum_jw_j=1\}$ the simplex. The landmarks $L$ determine the representation, while the references $B$ and the set of allowed weights determine the range of reconstructed objects. Sets $L$ and $B$ may be selected from observations or constructed using the data and known geometry and may overlap or coincide.

The permitted weight vectors form a set $\Omega_B\subseteq\Delta^{r-1}$, which may be the full simplex. Specific choices of $\Omega_B$ are described in Section~\ref{app:geometry-examples} of the Supplement. Write $b_B(w)$ for a selected weighted Fr\'echet mean:
\begin{equation}\label{eq:barycenter}b_B(w)\in\argmin_{x\in\M}\sum_{j=1}^r w_jd(x,b_j)^2,\qquad w\in\Omega_B.\end{equation}
Given the reference set $B$ and the set of allowable weights $\Omega_B$, the reconstruction target for a landmark vector $z$ is %
\begin{equation}\label{eq:reconstruction-target}\min_{w\in\Omega_B}\norm{\phi_L\{b_B(w)\}-z}_\infty^2,\end{equation}
 i.e., the weights $w$ for which the landmark vector of the reconstructed object matches the generated landmark vector as closely as possible in the sup norm. %
 To approximate this target, we adopt a neural weight map $w_\eta$ with parameters $\eta$ to implement  the reconstruction map $D_\eta(z)=b_B\{w_\eta(z)\}$, as illustrated in Figure~\ref{fig:frechetflow-overview}(c). A closely related approach in \citet{zhou2026e2m} that produced excellent results for Fr\'echet regression with vector predictors and random object responses  %
motivates this approach.

The softmax transformation exponentiates the network's output scores and normalizes them to sum to one, providing weights located in the simplex. The generated object and its conditional law are then 
\begin{equation}\label{eq:decoder}\widehat Y=D_\eta(Z),\qquad Z\sim Q,\qquad P_{\widehat\theta,\eta}=(D_\eta)_\#Q.\end{equation}
The  proposed reconstruction method attains a  reconstruction residual that in view of \eqref{eq:reconstruction-target} can be quantified as 
\begin{equation}\label{eq:decoder-residual}r_\eta(z)=\norm{\phi_L\{D_\eta(z)\}-z}_\infty,\end{equation}
which corresponds to the largest discrepancy between corresponding elements of the generated and reconstructed landmark vectors. %
In Section~\ref{sec:theory-transfer} we use these residuals to assess the generated object law. %

After fitting the flow on landmark representations, we draw $K$ independent vectors from the resulting fitted law $Q$ to train the reconstruction parameters $\eta$; the flow parameters $\widehat\theta$ remain unchanged. The population and empirical representation losses are
\begin{equation}\label{eq:decoder-risk}\mathcal J_Q(D_\eta)=\E_Qr_\eta(Z)^2,\qquad \widehat{\mathcal J}_K(D_\eta)=K^{-1}\sum_{i=1}^K r_\eta(Z_i)^2,\quad Z_i\overset{\rm iid}{\sim}Q.\end{equation}
To guide estimation of the weight map, we also use all observations $Y_i$ and their corresponding landmark representation $Z_i^\star$ and then quantify the reconstruction error %
\begin{equation}\label{eq:observed-decoder-risk}\widehat{\mathcal R}_n(D_\eta)=n^{-1}\sum_{i=1}^n d\{D_\eta(Z_i^\star),Y_i\}^2.\end{equation}

To adjust for potential differences in scale, %
we further implement normalization by average pairwise squared distances, which are proportional to metric variance \citep{mull:20:10}
as follows:
Draw $J$ pairs of distinct training indices $(i_j,i_j^{\prime})$ uniformly, using the same pairs to compute
\begin{equation}\label{eq:decoder-scales}s_Y^2=J^{-1}\sum_{j=1}^Jd(Y_{i_j},Y_{i_j^{\prime}})^2,\qquad s_Z^2=J^{-1}\sum_{j=1}^J\norm{Z_{i_j}^\star-Z_{i_j^{\prime}}^\star}_\infty^2.\end{equation}
For positive scales, initialize the weight map using the loss for reconstructing observed objects and then use the normalized loss function %
\begin{equation}\label{eq:normalized-decoder-loss}\widehat{\mathcal L}_{K,n}(\eta)=\frac{\widehat{\mathcal J}_K(D_\eta)}{s_Z^2}+\frac{\widehat{\mathcal R}_n(D_\eta)}{s_Y^2}.\end{equation}
For a generated landmark vector, there is no paired observed object, so the first term assesses reconstruction by matching landmark distances. The observed pairs provide the corresponding objects, allowing the second term to guide reconstruction directly through distances in the object space. The scales are computed once from training pairs and held fixed during fitting, so the criterion is invariant to changes %
of distance units. Gradients are computed through the relevant distance and mean calculations to estimate the network parameters. Once fitted, the reconstruction map generates new objects from independent flow outputs. Algorithm~\ref{alg:distance-fm} summarizes the neural fitting procedure; details on implementations and numerical settings can be found in Section~\ref{app:experiment-implementation} of the Supplement.

\begin{algorithm}[!ht]
\caption{Fitting and sampling for Fr\'echetFlow}
\label{alg:distance-fm}
\small
\begin{algorithmic}[1]
\Require Observations $Y_1,\ldots,Y_n$; metric $d$; landmarks $L$; references $B$ and allowed weights $\Omega_B$; networks $v_\theta,w_\eta$; Gaussian source $\Gamma$; fitting settings, including scaling $S$ and constraint set $\mathcal Z$; reconstruction sample size $K$; output size $m$.
\Ensure Generated objects $\widehat Y_1,\ldots,\widehat Y_m$.
\State Compute $Z_i^\star=\phi_L(Y_i)$ and standardized representations $\bar Z_i^\star=S(Z_i^\star)$.
\Statex \textit{Fit the flow on landmark representations.}
\For{each flow update}
\State Draw independent batch indices $I$, Gaussian vectors $G\sim\Gamma$ and times $T$ from the chosen distribution.
\State Set $X_T=(1-T)G+T\bar Z_I^\star$, $U=\bar Z_I^\star-G$ and update $\theta$ to reduce the batch mean of $\norm{v_\theta(T,X_T)-U}_2^2$.
\EndFor
\State Fix the selected flow parameters $\widehat\theta$ for reconstruction fitting and sampling.
\Statex \textit{Fit the neural reconstruction map.}
\State Compute $s_Y^2,s_Z^2$ using \eqref{eq:decoder-scales} and initialize $w_\eta$ using \eqref{eq:observed-decoder-risk}.
\State Generate $K$ vectors from the fitted ODE, undo $S$ and project onto $\mathcal Z$.
\For{each joint reconstruction update}
\State Draw batches of generated vectors and observed objects with their landmark representations.
\State Evaluate $D_\eta=b_B\circ w_\eta$ and update $\eta$ to reduce the normalized loss \eqref{eq:normalized-decoder-loss}.
\EndFor
\State Set $\widehat\eta$ to the parameters at the prescribed final update.
\Statex \textit{Generate new objects.}
\For{$j=1,\ldots,m$}
\State Draw $G_j\sim\Gamma$, integrate the fitted ODE to obtain $\widetilde Z_{1,j}$ and set $Z_j=\Pi_{\mathcal Z}\{S^{-1}(\widetilde Z_{1,j})\}$.
\State Set $\widehat Y_j=D_{\widehat\eta}(Z_j)$.
\EndFor
\State \Return $\widehat Y_1,\ldots,\widehat Y_m$.
\end{algorithmic}
\end{algorithm}

\section{Theoretical properties}\label{sec:theory}

\subsection{From representation error to object error}\label{sec:theory-transfer}

 Let $Q\in\Pcal_2(\R^k)$ be the generated representation law in \eqref{eq:clipped-code} and write $D:\R^k\to\M$ for a generic measurable reconstruction map; a pertinent scenario is the neural reconstruction map $D_\eta$ defined in \eqref{eq:decoder}. We assess a fitted generator through $\Wp{D_\#Q,P}{d}$, the Wasserstein distance between the generated output law and the target law. Recall that $\mu_L$ in \eqref{eq:coordinates} is the law of the landmark representation under $P$. Define
\begin{equation}\label{eq:coordinate-errors}\delta:=\Wp{Q,\mu_L}{\norm{\cdot}_\infty},\qquad e_Q(D):=\left\{\E_Q\norm{\phi_L\{D(Z)\}-Z}_\infty^2\right\}^{1/2}.\end{equation}
The first quantity measures error in the generated representation law; the second is the root mean squared reconstruction residual corresponding to the loss in \eqref{eq:decoder-risk}. Under Assumption~\ref{ass:object-law}, $\mu_L$ also belongs to $\Pcal_2(\R^k)$, so $\delta$ is finite.

For the placement of the landmarks $L$ we aim at minimizing the root mean squared distance of a random object to the nearest landmark:
\begin{equation}\label{eq:landmark-coverage}q_2(P,L):=\{\E_P\min_jd(Y,\ell_j)^2\}^{1/2}.\end{equation}
For any $x,y\in\M$ and landmark $\ell$, the triangle inequality yields $d(x,y)\le|d(x,\ell)-d(y,\ell)|+2d(y,\ell)$. Choosing $\ell$ nearest to $y$ bounds the distance between objects $x,y$ by $2\min_jd(y,\ell_j)$ plus the largest discrepancy between their distances to the landmarks. When a landmark $\ell$ is close to $y$, small discrepancies in their distances to the landmarks therefore imply that $x$ and $y$ are close. Coupling generated and true vectors of landmark distances and taking root mean squares leads to the following bound on the Wasserstein distance between the generated and target object laws.

\begin{theorem}\label{thm:transfer}
Under Assumption~\ref{ass:object-law}, if the set of landmarks $L$ is finite and nonempty, then for any $Q\in\Pcal_2(\R^k)$ and measurable reconstruction map $D:\R^k\to\M$ with finite residual $e_Q(D)$ in \eqref{eq:coordinate-errors},
\begin{equation}\label{eq:distance-transfer}\Wp{D_\#Q,P}{d}\le e_Q(D)+\delta+2q_2(P,L).\end{equation}
\end{theorem}
The three terms correspond to reconstruction error, error in the generated representation law and the distance to the nearest landmark. The bound applies to the object actually returned by a numerical reconstruction map. Proofs of the theorems and corollaries in this section are given in Section~\ref{app:proofs} of the Supplement.

For weighted mean reconstruction, the reconstruction loss can be separated into approximation by weighted means of the references in $B$ and error in choosing the weights. Take the allowed weight set $\Omega_B$ in \eqref{eq:reconstruction-target} to be nonempty and closed in the simplex and hence compact. The selected weighted mean map $b_B$ in \eqref{eq:barycenter} determines the set $H_B=\{b_B(w):w\in\Omega_B\}$ of possible outputs.
Thus the neural reconstruction map $D_\eta$ in \eqref{eq:decoder} takes values in $H_B$. In a convex Euclidean object space with $\Omega_B=\Delta^{r-1}$,  $H_B$ is simply the convex hull of $B$.

\begin{assumption}\label{ass:decoder-geometry}
The weighted Fr\'echet mean problem in \eqref{eq:barycenter} admits a continuous selection $b_B:\Omega_B\to\M$, which is used for reconstruction.
\end{assumption}
The condition requires the Fr\'echet weighted mean to change continuously as the weights vary. On a compact object space, minimizers exist and uniqueness for every allowed weight implies continuity by the argument of \citet[Lemma~1]{zhou2026e2m}. The condition also holds for the constructions in Section~\ref{sec:geometry-examples} of the Supplement for Euclidean vectors and images, symmetric positive definite (SPD) matrices, graphs, phylogenetic trees and cubical spaces, as well as spherical means within sufficiently concentrated reference groups. For distribution-valued objects, it applies to fixed pairwise Wasserstein geodesics. Section~\ref{app:geometry-examples} of the Supplement provides the corresponding verifications and related geometric results.

Under Assumption~\ref{ass:decoder-geometry}, $H_B$ is compact. We quantify how well it approximates the target distribution by
\begin{equation}\label{eq:family-error}a_B(P)=\left\{\E_P\operatorname{dist}(Y,H_B)^2\right\}^{1/2},\end{equation}
where $\operatorname{dist}(y,A)=\inf_{x\in A}d(y,x)$ denotes the distance from an object $y$ to a set $A\subseteq\M$. %
Unless stated otherwise, the stated  results are conditional  on $L$, $B$ and $\Omega_B$, with independent observations used to fit the flow. %
In Section~\ref{sec:theory-empirical} we discuss sample-based  selection of these quantities.

For an input vector $z$, the smallest reconstruction residual over all allowed weighted means constituting the set $H_B$ is 
\begin{equation}\label{eq:best-reconstruction}\rho_B(z)=\min_{x\in H_B}\norm{\phi_L(x)-z}_\infty.\end{equation}
For a square-integrable scalar function $f$, the root mean square under $Q$ is  $\norm{f}_{L^2(Q)}=\{\E_{Z\sim Q}|f(Z)|^2\}^{1/2}$.  For  a measurable reconstruction map $D:\R^k\to H_B$, write $r_D(z)=\norm{\phi_L\{D(z)\}-z}_\infty$ for the reconstruction residual and define
\begin{equation}\label{eq:projection-error}\eta_{\rm proj}(D)=\norm{r_D-\rho_B}_{L^2(Q)}.\end{equation}
The nonnegative difference $r_D-\rho_B$ measures the additional reconstruction error beyond the best choice of weights. For the neural reconstruction map $D_\eta$ in \eqref{eq:decoder}, $r_{D_\eta}=r_\eta$ from \eqref{eq:decoder-residual}. The following corollary to Theorem~\ref{thm:transfer} focuses on the bound for the weighted mean reconstruction.

\begin{corollary}\label{cor:weighted-mean-transfer}
Under Assumptions~\ref{ass:object-law} and \ref{ass:decoder-geometry}, assume that  the set of landmarks $L$ is finite and nonempty. Let $Q\in\Pcal_2(\R^k)$  and  $D_\eta=b_B\circ w_\eta$ be the reconstruction map in \eqref{eq:decoder}, with a measurable weight map $w_\eta:\R^k\to\Omega_B$. If a measurable numerical approximation $D_\eta^{\mathrm{num}}:\R^k\to\M$ satisfies $\norm{d(D_\eta^{\mathrm{num}},D_\eta)}_{L^2(Q)}\le\beta<\infty$, then
\begin{equation}\label{eq:coverage-projection}\Wp{(D_\eta^{\mathrm{num}})_\#Q,P}{d}\le a_B(P)+2\delta+\eta_{\rm proj}(D_\eta)+2q_2(P,L)+\beta.\end{equation}
\end{corollary}
Here $a_B(P)$ measures how well weighted means approximate target objects, while $q_2(P,L)$ measures distance to the nearest landmark. The remaining terms quantify error in generating landmark representations, choosing reconstruction weights and numerically evaluating the selected mean.
We condition on $L$, $B$, $\Omega_B$, the fitted representation law $Q$, the observed training pairs and the positive scales $s_Z^2$ and $s_Y^2$ in \eqref{eq:decoder-scales} when  drawing the sample used to fit the reconstruction map. 

Let $\mathcal D$ be a class of candidate measurable reconstruction maps $D:\R^k\to H_B$. The neural maps $D_\eta$ in \eqref{eq:decoder}, indexed by their parameters $\eta$, form one such class. Write $\widehat D\in\mathcal D$ for a fitted reconstruction map and $r_{\widehat D}$ for the corresponding residual. In the neural case, the fitted network parameters $\widehat\eta$ result in  $\widehat D=D_{\widehat\eta}$ and $r_{\widehat D}=r_{\widehat\eta}$. Write
\begin{equation}\label{eq:projection-risk}\mathcal J_Q(D)=\E_Qr_D(Z)^2,\qquad \mathcal J_Q^*=\E_Q\rho_B(Z)^2,\qquad \mathcal A_{\mathcal D}=\inf_{D\in\mathcal D}\mathcal J_Q(D)-\mathcal J_Q^*,\end{equation}
where $\mathcal J_Q^*$ is the risk of choosing the best allowed mean at each input and $\mathcal A_{\mathcal D}$ measures the approximation error from restricting reconstruction to $\mathcal D$. Since $r_D\ge\rho_B\ge0$,
\begin{equation}\label{eq:excess-controls-gap}\eta_{\rm proj}(D)^2\le\mathcal J_Q(D)-\mathcal J_Q^*, \end{equation}
providing a bound for the additional residual term on the r.h.s. of  \eqref{eq:coverage-projection}.

The empirical losses $\widehat{\mathcal J}_K(D)$ and $\widehat{\mathcal R}_n(D)$ are defined by substituting $D$ for $D_\eta$ in \eqref{eq:decoder-risk} and \eqref{eq:observed-decoder-risk}. Multiplying the normalized loss in \eqref{eq:normalized-decoder-loss} by $s_Z^2$ implies $\widehat{\mathcal F}_{K,n}(D)=\widehat{\mathcal J}_K(D)+c\widehat{\mathcal R}_n(D)$, where $c=s_Z^2/s_Y^2$. Fitting with only the representation loss is included by taking $c=0$. The corresponding approximation term is
\[\mathcal A_{\mathcal D,c}=\inf_{D\in\mathcal D}\{\mathcal J_Q(D)+c\widehat{\mathcal R}_n(D)\}-\mathcal J_Q^*.\]

For each candidate reconstruction map $D$, write $D^{\mathrm{num}}:\R^k\to\M$ for the numerical implementation, for example with the iteratively computed weighted Fr\'echet mean. Define $\widehat{\mathcal F}^{\mathrm{num}}_{K,n}(D)$ by replacing $D$ with $D^{\mathrm{num}}$ in $\widehat{\mathcal F}_{K,n}(D)$, using the same inputs, observations and scales. For the fitted reconstruction map $\widehat D$, the numerical implementation is denoted by $\widehat D^{\mathrm{num}}$; in the neural case, $\widehat D^{\mathrm{num}}=D_{\widehat\eta}^{\mathrm{num}}$.

To measure the complexity of a class, for a nonempty metric space $(A,\rho)$ and $\varepsilon\ge0$, we employ  the covering number
\[N(A,\varepsilon;\rho)=\inf\left\{m\in\mathbb N:\ \exists a_1,\ldots,a_m\in A,\quad\sup_{a\in A}\min_{1\le j\le m}\rho(a,a_j)\le\varepsilon\right\}.\]
Throughout, covering balls are closed and their centers belong to the set being covered; for the empty set, $N(\varnothing,\varepsilon;\rho)=0$.

\begin{assumption}\label{ass:decoder-fitting}
The class $\mathcal D$ is fixed before independent inputs $Z_1,\ldots,Z_K\sim Q$ are drawn. The corresponding residual class $\mathcal R=\{r_D:D\in\mathcal D\}$ is separable in the uniform distance $d_{\mathcal R}(r,r')=\sup_{z\in\R^k}|r(z)-r'(z)|$. Thus each center in a cover counted by $N(\mathcal R,\varepsilon;d_{\mathcal R})$ is a residual $r_{D_j}$ with $D_j\in\mathcal D$. For tolerances $\epsilon_{\rm comp},\xi_{\rm opt}\ge0$, the numerical criterion $\widehat{\mathcal F}^{\mathrm{num}}_{K,n}$ and a measurably selected fitted reconstruction map $\widehat D\in\mathcal D$ satisfy
\[\sup_{D\in\mathcal D}|\widehat{\mathcal F}^{\mathrm{num}}_{K,n}(D)-\widehat{\mathcal F}_{K,n}(D)|\le\epsilon_{\rm comp},\qquad \widehat{\mathcal F}^{\mathrm{num}}_{K,n}(\widehat D)\le\inf_{D\in\mathcal D}\widehat{\mathcal F}^{\mathrm{num}}_{K,n}(D)+\xi_{\rm opt}.\]
\end{assumption}
Newly  generated inputs are iid from $Q$,  conditional on the fitted flow and selected reference objects. Under the separability condition, the supremum of empirical loss deviations is measurable; see \citet[Sections~1.7 and~2.3.3]{vandervaart1996weak}. On a compact input domain, a jointly continuous residual family indexed by a compact parameter set also has finite covering numbers. These covering numbers control sampling error through standard empirical process bounds \citep[Chapters~11 and~13]{boucheron2013concentration}. The tolerance $\epsilon_{\rm comp}$ provides a bound on the numerical error, %
while $\xi_{\rm opt}$ quantifies the gap to the theoretical infimum over $\mathcal D$.

The bound on sampling error uses the following constants, which are finite by compactness of $H_B$ under Assumption~\ref{ass:decoder-geometry} and the second-moment condition on $Q$:
\begin{equation}\label{eq:decoder-moment-constants}M_B=\sup_{x\in H_B}\norm{\phi_L(x)}_\infty,\qquad m_Q=\{\E_Q\norm Z_\infty^2\}^{1/2},\qquad V_Q=M_B(2m_Q+M_B).\end{equation}
\begin{theorem}\label{thm:projection-oracle}
Under Assumptions~\ref{ass:decoder-geometry} and \ref{ass:decoder-fitting}, for $\varepsilon>0$ with $N(\mathcal R,\varepsilon;d_{\mathcal R})<\infty$ and $\alpha\in(0,1)$, the fitted reconstruction map satisfies, with conditional probability at least $1-\alpha$,
\begin{equation}\label{eq:projection-oracle}\eta_{\rm proj}(\widehat D)^2\le\mathcal A_{\mathcal D,c}+2\Delta_K(\varepsilon,\alpha)+\xi_{\rm opt}+2\epsilon_{\rm comp},\end{equation}
where $\eta_{\rm proj}(\widehat D)$ is the excess residual in \eqref{eq:projection-error} evaluated at the fitted reconstruction map, the constants $M_B,m_Q,V_Q$ are defined in \eqref{eq:decoder-moment-constants} and
\begin{equation}\label{eq:decoder-moment-deviation}\Delta_K(\varepsilon,\alpha)=4(m_Q+M_B)\varepsilon+V_Q\left\{2\sqrt{\frac{2\log\{2N(\mathcal R,\varepsilon;d_{\mathcal R})\}}{K}}+\sqrt{\frac{2}{K\alpha}}\right\}.\end{equation}
For a finite class of reconstruction maps, one may take $\varepsilon=0$ and $N(\mathcal R,0;d_{\mathcal R})\le|\mathcal D|$.
\end{theorem}

The approximation term $\mathcal A_{\mathcal D,c}$ measures how well the class of reconstruction maps can jointly achieve near-optimal matching of landmark representations and reconstruct observed objects. The terms $\xi_{\rm opt}$ and $\epsilon_{\rm comp}$ account for the gap between the attained fitting criterion and its infimum and numerical evaluation of the fitting criterion. The sampling term $\Delta_K(\varepsilon,\alpha)$ arises from using $K$ generated inputs to fit the reconstruction map, conditional on the observations and the  flow. A smaller $\varepsilon$ leads to a finer approximation but at the cost of an increase in the covering number $N(\mathcal R,\varepsilon;d_{\mathcal R})$. The constants $M_B,m_Q,V_Q$ control the scale of loss fluctuations, while $\alpha$ determines the confidence level.

For a fixed finite class of reconstruction maps and confidence level, taking $\varepsilon=0$ yields a sampling error of order $K^{-1/2}$. Increasing the number of generated inputs therefore reduces sampling error, while enlarging the class of reconstruction maps can improve approximation at the cost of greater complexity. Combining the theorem with Corollary~\ref{cor:weighted-mean-transfer} translates these fitting guarantees into a Wasserstein error bound for the generated object law $(\widehat D^{\mathrm{num}})_\#Q$. Section~\ref{sec:fitted-consistency} provides details on sufficient conditions under which the fitted generator is consistent.

\subsection{Learning from a finite sample}\label{sec:theory-empirical}

To separate sampling error from generator error, we compare the generated object law with the empirical law $P_n$. Write $\mu_{L,n}=(\phi_L)_\#P_n$ for the empirical distribution of landmark representations. The following result is a consequence of Theorem~\ref{thm:transfer} and is conditional on the observed sample and fitted generator.

\begin{corollary}\label{cor:metric-empirical}
Under Assumption~\ref{ass:object-law}, consider the observations $Y_1,\ldots,Y_n$ and their empirical law $P_n$. Conditional on the observations, let $Q\in\Pcal_2(\R^k)$,  $L$ be any finite nonempty landmark set and $D:\R^k\to\M$ be a measurable implementation of the  reconstruction map with finite residual $e_Q(D)$,  as defined in \eqref{eq:coordinate-errors}. Suppose $\Wp{Q,\mu_{L,n}}{\norm{\cdot}_\infty}\le\Delta_n$. Then
\begin{equation}\label{eq:metric-empirical}\Wp{D_\#Q,P}{d}\le\Wp{P_n,P}{d}+e_Q(D)+\Delta_n+2q_2(P_n,L).\end{equation}
\end{corollary}

Here  $\Wp{P_n,P}{d}$ quantifies the  sampling error, and the remaining terms describe how accurately the generator reproduces the empirical target. The reconstruction residual includes error from numerically computing the weighted mean, quantified by $\beta$ in Corollary~\ref{cor:weighted-mean-transfer}, because it is evaluated on the generated objects. Conditioning on the observations makes it feasible to select  landmarks and  references  from the same sample used to fit the flow.

For flow matching, the representation error $\Delta_n$ can be bounded in terms of error in estimating the conditional mean velocity, numerical integration error and any endpoint noise. Proposition~\ref{prop:flow-stability} in Section~\ref{app:flow-propagation} of the Supplement gives this bound under Lipschitz and integrability conditions on the fitted velocity field.

\subsection{Asymptotic consistency}\label{sec:fitted-consistency}

Under iid sampling, Assumption~\ref{ass:object-law} ensures that the empirical Wasserstein error tends to zero. Corollary~\ref{cor:metric-empirical} therefore reduces consistency to controlling error in the generated representation law, reconstruction error and the root mean squared distance to the nearest landmark. To control the last quantity, define the optimal root mean squared approximation by at most $k$ objects as
\begin{equation}\label{eq:quantization-error}\varepsilon_k(P)=\inf_{A\subset\M:\,1\le |A|\le k}\left\{\E_P\operatorname{dist}(Y,A)^2\right\}^{1/2}.\end{equation}

Separability and finite second moments imply $\varepsilon_k(P)\to0$ as $k\to\infty$. We choose $1\le k\le n$ empirical medoids by minimizing, over $k$ selected observations, the sum of squared distances from all observations to their nearest selected object. For the resulting landmark set $L$, the root mean squared distance from the observations to their nearest landmark, $q_2(P_n,L)$ from \eqref{eq:landmark-coverage}, satisfies
\begin{equation}\label{eq:medoid-approximation}q_2(P_n,L)\le2\left\{\varepsilon_k(P)+\Wp{P_n,P}{d}\right\}.\end{equation}
\begin{theorem}\label{thm:metric-consistency}
Let $Y_1,Y_2,\ldots$ be iid with law $P$ satisfying Assumption~\ref{ass:object-law} and let $P_n$ be their empirical law. For deterministic $1\le k_n\le n$, select $k_n$ empirical medoids $L_n$ as above. Conditional on the observations and fitting randomness, let $Q_n\in\Pcal_2(\R^{k_n})$ be the generated representation law and $D_n:\R^{k_n}\to\M$ a measurable reconstruction map with finite residual $e_{Q_n}(D_n)$ from \eqref{eq:coordinate-errors}. With  $\delta_n=\Wp{Q_n,\mu_{L_n,n}}{\norm{\cdot}_\infty}$, where $\mu_{L_n,n}$ is the empirical law of the landmark representations, it holds that 
\begin{equation}\label{eq:medoid-generator-bound}\Wp{(D_n)_\#Q_n,P}{d}\le5\Wp{P_n,P}{d}+4\varepsilon_{k_n}(P)+e_{Q_n}(D_n)+\delta_n.\end{equation}
\end{theorem}

Suppose $k_n\to\infty$. If $e_{Q_n}(D_n)+\delta_n\to0$ in probability, then $\Wp{(D_n)_\#Q_n,P}{d}\to0$ in probability. The same implication holds for almost sure convergence. If $\E\{e_{Q_n}(D_n)+\delta_n\}\to0$, then $\E\Wp{(D_n)_\#Q_n,P}{d}\to0$. Uniform subsampling of the observations provides an alternative for landmark selection; Corollary~\ref{cor:metric-empirical} assesses either choice through $q_2(P_n,L)$.

\subsection{Rates of convergence}\label{sec:consistency-rates}

To quantify the sampling and landmark terms in \eqref{eq:medoid-generator-bound}, we control how many small metric balls are needed to approximate bounded parts of the target support $\supp P$ and impose a stronger moment condition. Write $B_d(o,R)=\{y:d(y,o)\le R\}$ and use $N(A,\epsilon;d)$ for the covering number defined in Section~\ref{sec:theory-transfer}, with centers in $A\subseteq\M$.
\begin{assumption}\label{ass:metric-rates}
For a fixed $o\in\M$, there are constants $s>0$, $C_0<\infty$ and $q>4$ such that $N\bigl(\supp P\cap B_d(o,R),\epsilon;d\bigr)\le C_0(R/\epsilon)^s$ for every $R\ge1$ and $0<\epsilon\le R$ and $\E_Pd(Y,o)^q\le M_q<\infty$.
\end{assumption}
The exponent $s$ describes the metric complexity of the target support and may be smaller than the ambient dimension. The covering condition holds with $s=m$ for a law supported on an $m$-dimensional affine subspace of $\R^p$ under Euclidean distance and with $s=1$ for support in a finite metric tree. The moment condition controls the contribution of distant observations to the empirical Wasserstein error.

Recall that $\varepsilon_k(P)$ in \eqref{eq:quantization-error} is the optimal root mean squared approximation error using at most $k$ objects. Under Assumption~\ref{ass:metric-rates}, $\varepsilon_k(P)\le Ck^{-1/s}$ and $\E\Wp{P_n,P}{d}\le Cr_n(s)$, where
\begin{equation}\label{eq:empirical-metric-rate}r_n(s)=\begin{cases}n^{-1/4},&s<4,\\n^{-1/4}\{\log(1+n)\}^{1/2},&s=4,\\n^{-1/s},&s>4.\end{cases}\end{equation}
The two  bounds are established  in Section~\ref{lem_proof:metric-approximation-rates} of the Supplement.  The empirical bound combines the metric partition argument of \citet[Proposition~1]{weed2019sharp} with the shell decomposition and moment estimates of \citet[Sections~2--3]{fournier2015rate}. Recall from Section~\ref{sec:fitted-consistency} that $k$ empirical medoids are selected from the sample of random objects targeting  minimization of  the sum of squared distances from all observations to their nearest selected object. Applying the two above bounds to \eqref{eq:medoid-generator-bound} leads to the following result, with explicit  flow and reconstruction errors.

\begin{theorem}\label{thm:noncompact-rates}
Under Assumptions~\ref{ass:object-law} and \ref{ass:metric-rates}, let $Y_1,\ldots,Y_n$ be iid with law $P$. Select $1\le k\le n$ empirical medoids $L$ from these observations as in Section~\ref{sec:fitted-consistency}. Let $Q$ be the generated representation law on $\R^k$ and $D:\R^k\to\M$ a measurable reconstruction map. Write $\mu_{L,n}=(\phi_L)_\#P_n$ for the empirical representation law and $\delta_n=\Wp{Q,\mu_{L,n}}{\norm{\cdot}_\infty}$ for the corresponding  approximation error. With root mean squared reconstruction residuals $e_Q(D)$ as in \eqref{eq:coordinate-errors} and assuming  $\E\{e_Q(D)+\delta_n\}<\infty$, with expectation over the observations and fitting randomness, it holds that 
\begin{equation}\label{eq:conditional-generator-rate}\E\Wp{D_\#Q,P}{d}\le C\{r_n(s)+k^{-1/s}\}+\E\{e_Q(D)+\delta_n\}.\end{equation}
The constant $C$ may depend on $C_0,s,q,M_q$, but not on $n$ or $k$.
\end{theorem}

The rate of convergence in \eqref{eq:conditional-generator-rate} depends on empirical sampling error $r_n(s)$, landmark approximation error $k^{-1/s}$ and the attained reconstruction and flow errors. For $s=1$ or $s=2$, choosing $k\asymp n^{s/4}$ balances landmark approximation with sampling error, both of order $n^{-1/4}$. The generated law then has expected Wasserstein error $O(n^{-1/4})$ when the expected reconstruction and flow errors also satisfy $\E\{e_Q(D)+\delta_n\}=O(n^{-1/4})$. Greater support complexity is associated with slower rates; for example, $s=8$ with  $r_n(s)=n^{-1/8}$.

\section{Simulation studies}\label{sec:experiments}
Our simulations assess recovery of known probability laws as sample size increases in Euclidean and branching spaces. For each target, we use sample sizes $n=500$, $2000$ and $10{,}000$, with 50 independent replications. Training samples are nested across sample sizes within each replication and comparisons use the same observations for both methods. Replications use independent data and training seeds.

\begin{samepage}
We measure generation accuracy by the empirical Wasserstein metric $W_2$ using the metric $d$ of the object space, as in flow matching evaluations by \citet{tong2024cfm}. For $m$ generated objects $\widehat Y_1,\ldots,\widehat Y_m$ and $q$ test objects $Y_1^{\rm test},\ldots,Y_q^{\rm test}$, this is the 2-Wasserstein distance between their empirical distributions:
\begin{equation}\label{eq:empirical-w2-general}\widehat W_2=\left\{\min_{\gamma\in\Pi_{m,q}}\sum_{i=1}^{m}\sum_{j=1}^{q}\gamma_{ij}d(\widehat Y_i,Y_j^{\rm test})^2\right\}^{1/2},\end{equation}
\end{samepage}
where $\Pi_{m,q}$ is the set of nonnegative $m\times q$ matrices with row sums $1/m$ and column sums $1/q$. For equal sample sizes, optimal transport reduces to a one-to-one assignment. In each simulation replication, we solve this assignment exactly between 5000 generated objects and 5000 independent target observations, with the target sample shared across methods and sample sizes. We report means and sample standard deviations across replications.

The simulations use landmark distances in \eqref{eq:coordinates} and the fitting stages of Algorithm~\ref{alg:distance-fm}, using the fixed normalized reconstruction loss \eqref{eq:normalized-decoder-loss}. For the Euclidean simulations, we implement the reconstruction map by predicting the vector directly within the known domain. This implementation choice uses the linear structure of the observations; the branching space simulations use neural weights and Fr\'echet means. Further details are in section~\ref{app:experiment-implementation} of the Supplement. %

The Euclidean simulations use the conditional flow matching formulation of \citet{lipman2023flow}, denoted FM, fitted directly to the observations. Fr\'echetFlow fits the same objective to landmark representations. The comparisons use the same transformer architecture built from LightningDiT blocks \citep{yao2025reconstruction} and the same number of optimization iterations. Simulation settings are fixed across targets and sample sizes.

\subsection{Euclidean simulations}\label{sec:euclidean-experiments}
We compare generation through landmark distances with direct flow matching on two target laws on $\M=[-1,1]^2$, representing separated modes and curved bands: 
\begin{enumerate}[leftmargin=*,itemsep=3pt]
\item \textit{Eight Gaussians.} The target is an equally weighted mixture of eight Gaussian distributions, each restricted to $\M$, with means $c_j=0.65\{\cos(\pi j/4),\sin(\pi j/4)\}^{\mathsf T}$ for $j=0,\ldots,7$ and covariance $0.065^2I_2$ before restriction.
\item \textit{Two moons.} Let $J\sim\operatorname{Bernoulli}(1/2)$, $\Theta\sim\operatorname{Unif}[0,\pi]$ and $\xi\sim N(0,I_2)$ be independent. Define $V_0(\theta)=(\cos\theta,\sin\theta)^{\mathsf T}$ and $V_1(\theta)=(1-\cos\theta,0.5-\sin\theta)^{\mathsf T}$. The target is the law of $0.6\{V_J(\Theta)-(0.5,0.25)^{\mathsf T}\}+0.04\xi$, conditional on membership in $\M$.
\end{enumerate}

The known domain suggests three corner landmarks, $L=\{(-1,-1),(1,-1),(-1,1)\}$, shared by both targets. The flow generates three-dimensional distance vectors and the reconstruction map directly predicts a point in $[-1,1]^2$. Distances to these landmarks identify every point in the square. The weighted mean interpretation is also available: all points are means of the four corners, thus  $a_B(P)=0$.

\begin{figure}[!htb]
\centering
\includegraphics[width=\textwidth]{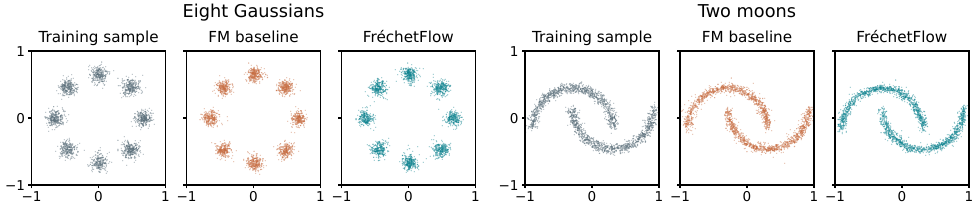}
\caption{Training observations and generated samples at $n=10{,}000$. The left and right groups show eight Gaussians and two moons, respectively; within each group, panels show the training sample, FM baseline \citep{lipman2023flow} and Fr\'echetFlow. Each panel displays 2000 observations or generated draws.}
\label{fig:euclidean-samples}
\end{figure}

Table~\ref{tab:euclidean-w2} shows that mean Wasserstein error decreases with sample size for both methods. Fr\'echetFlow closely matches FM across the two targets, attaining slightly lower mean errors at $n=10{,}000$ in both cases. Figure~\ref{fig:euclidean-samples} shows recovery of the separated modes and curved bands. Learning through landmark distances therefore retains strong accuracy on these Euclidean targets, where direct modeling of the observations is also available.

\begin{table}[htbp]
\centering
\caption{Empirical $W_2$ for  two-dimensional Euclidean targets. Entries are means with sample standard deviations in parentheses.}
\label{tab:euclidean-w2}
\small
\begin{tabular}{@{}lrcc@{}}
\toprule
Target & $n$ & FM baseline & Fr\'echetFlow \\
\midrule
Eight Gaussians & $500$ & $0.0940\;(0.0218)$ & $0.0932\;(0.0213)$ \\
 & $2000$ & $0.0666\;(0.0151)$ & $0.0679\;(0.0141)$ \\
 & $10{,}000$ & $0.0559\;(0.0110)$ & $0.0533\;(0.0121)$ \\
\addlinespace
Two moons & $500$ & $0.0656\;(0.0180)$ & $0.0660\;(0.0191)$ \\
 & $2000$ & $0.0463\;(0.0108)$ & $0.0460\;(0.0110)$ \\
 & $10{,}000$ & $0.0380\;(0.0111)$ & $0.0362\;(0.0098)$ \\
\bottomrule
\end{tabular}
\end{table}

\subsection{Graphs, trees and a cubical space}\label{sec:graph-tree-experiments}
Here we assess non-Euclidean scenarios, specifically whether Fr\'echetFlow recovers the probabilities of different branches, tree topologies, or cells together with continuous variation within them. In each setting, weighted means of the chosen reference objects cover the full domain.

\begin{figure}[!htbp]
\centering
\includegraphics[width=\textwidth]{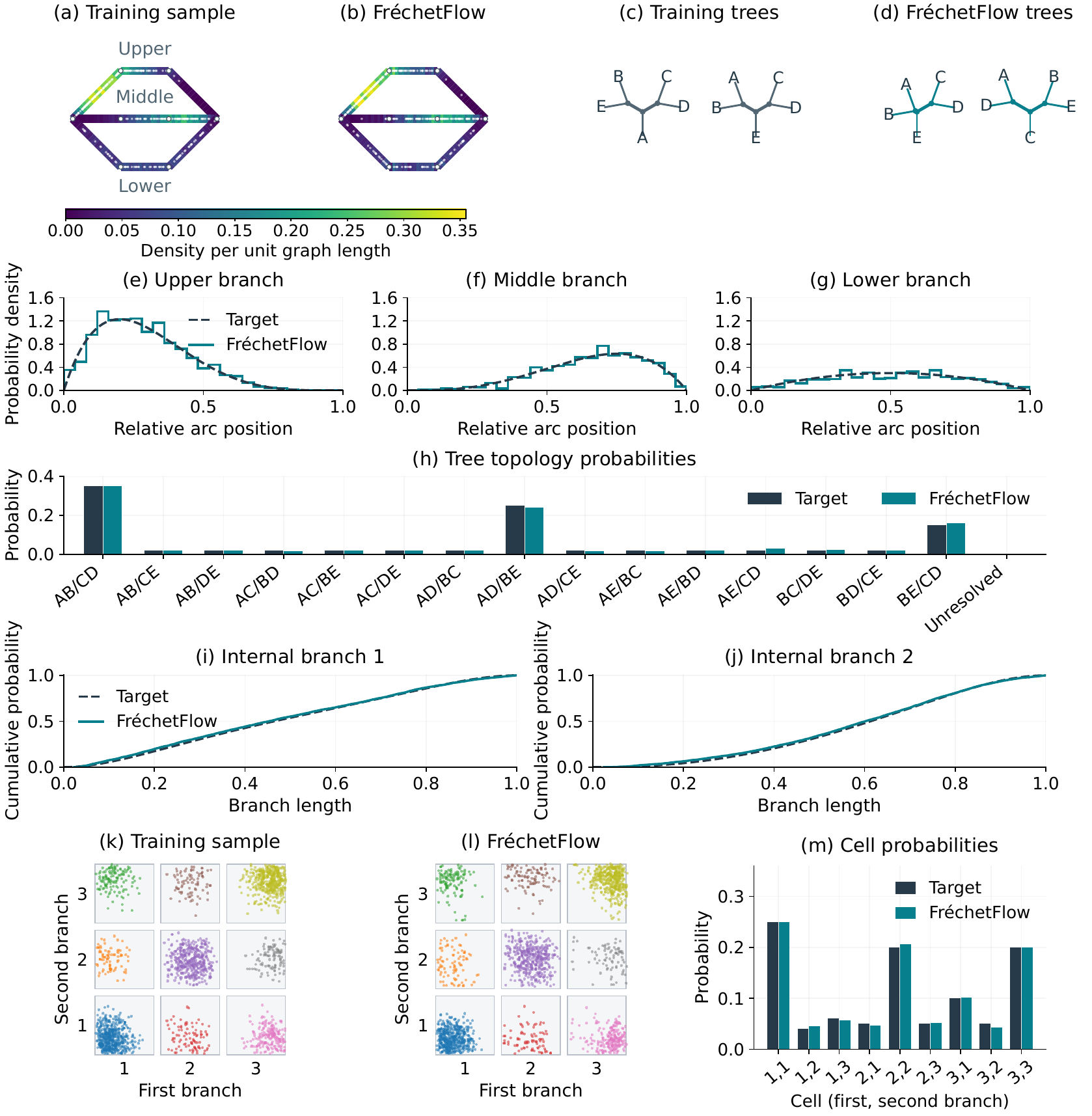}
\caption{Graph, whole-tree and cubical simulations at $n=10{,}000$. Panels (a)--(b) display graph densities for the training and generated samples on a common scale; (c)--(d) show two observed and two generated trees. Panels (e)--(g) compare generated branch subdensities with the generating law, (h) compares topology probabilities and (i)--(j) compare marginal internal branch length distributions. Panels (k)--(l) display 2000 training and generated cubical objects in within-cell coordinates; (m) compares cell probabilities. Probability and distribution summaries use all 5000 generated objects from the displayed fits. The dashed curves and dark bars denote the generating laws.}
\label{fig:graph-tree-cubical-samples}\label{fig:graph-cubical-samples}\label{fig:metric-graph-samples}\label{fig:phylogenetic-tree-samples}\label{fig:cat0-cubical-samples}
\end{figure}

\paragraph{Locations on a metric graph.}
The graph has eight vertices and nine edges arranged as three branches joining two endpoints, as shown in Figure~\ref{fig:graph-tree-cubical-samples}(a)--(b). An observation is a location on the graph, with shortest path distance as the metric. We select the upper, middle or lower branch with probabilities $0.5$, $0.3$ and $0.2$, respectively, and draw the fractional position along it from $\operatorname{Beta}(2,5)$, $\operatorname{Beta}(4,2)$ or $\operatorname{Beta}(2,2)$. The graph geometry is specified in Section~\ref{app:metric-graph-implementation} of the Supplement.

\paragraph{Phylogenetic trees.}
An observation is an unrooted binary tree on five labeled taxa A--E at its leaves, with terminal branch lengths fixed at one and two internal lengths in $[0,1]$. There are 15 topologies, or branching patterns, and separation is measured by the Billera--Holmes--Vogtmann (BHV) metric \citep{billera2001geometry}. Within a topology, this is Euclidean distance between the two internal lengths; paths between topologies pass through trees with at least one zero internal length. The generating law assigns probabilities $0.35$, $0.25$ and $0.15$ to AB/CD, AD/BE and BE/CD, respectively, and distributes the remaining mass equally among the others. For AB/CD, leaves A and B share an internal vertex, leaves C and D share another, and E attaches to the vertex between them. Conditional on topology, the internal lengths are independent beta variables with topology-specific parameters. We provide additional details in Section~\ref{app:phylogenetic-implementation} of the Supplement. %

\paragraph{Joint positions on two branching tracks.}\label{sec:cubical-experiment}
We consider two robots, each confined to its own Y-shaped track with three unit branches labeled 1--3. One observation records both positions: for example, $((1,0.3),(3,0.8))$ places the robots $0.3$ and $0.8$ units along branches 1 and 3, respectively.

For fixed branch labels $(a,b)$, the distances $(u,v)$ from the junctions range over $[0,1]^2$. These nine squares meet along edges where a robot reaches its junction, forming a CAT(0) cubical space. The metric is the root sum of squared shortest-path distances on the two tracks. We draw branch pairs with probabilities
\[\{P(a,b)\}_{a,b=1}^3=\begin{pmatrix}0.25&0.04&0.06\\0.05&0.20&0.05\\0.10&0.05&0.20\end{pmatrix}.\]
Conditional on $(a,b)$, each robot's distance from its junction is drawn independently, using $\operatorname{Beta}(2,5)$, $\operatorname{Beta}(3,3)$ or $\operatorname{Beta}(5,2)$ for branches 1, 2 and 3, respectively. These laws concentrate positions near the junction, midpoint and outer endpoint, respectively; the branch choices are dependent. More details on geometry and generating law can be found in Example~\ref{sec:cubical-object-examples} and Section~\ref{app:cubical-implementation} of the Supplement. 

\paragraph{Results.}
Mean Wasserstein error decreases with sample size in all three spaces under fixed fitting settings (Table~\ref{tab:graph-tree-w2}). Figure~\ref{fig:graph-tree-cubical-samples} demonstrates  recovery of their structural features at $n=10{,}000$. The graph samples reproduce the different concentrations along the branches, and the generated trees recover the dominant topology modes and marginal internal branch length distributions. The cubical samples capture unequal cell probabilities and variation within cells, including dependence between the component branch choices. These results illustrate simultaneous recovery of structural membership and variation within branches, topologies and cells.

\begin{table}[htbp]
\centering
\caption{Empirical $W_2$ for Fr\'echetFlow on a metric graph, in phylogenetic tree space and in a CAT(0) cubical space. Entries are means with sample standard deviations in parentheses. The ground metrics are graph shortest path distance, Billera--Holmes--Vogtmann distance and the intrinsic product metric, respectively.}
\label{tab:graph-tree-w2}
\small
\begin{tabular}{@{}rccc@{}}
\toprule
$n$ & Metric graph & Phylogenetic trees & Cubical space \\
\midrule
$500$ & $0.1577\;(0.0514)$ & $0.1621\;(0.0270)$ & $0.1256\;(0.0247)$ \\
$2000$ & $0.1008\;(0.0413)$ & $0.1150\;(0.0178)$ & $0.0891\;(0.0148)$ \\
$10{,}000$ & $0.0876\;(0.0330)$ & $0.0994\;(0.0120)$ & $0.0726\;(0.0129)$ \\
\bottomrule
\end{tabular}
\end{table}

\section{Real data applications}\label{sec:data-applications}
For spherical locations, cell neighborhoods and brain networks, we compare generated and held-out observations by the empirical Wasserstein distance in \eqref{eq:empirical-w2-general}, using the metric of the object space. We  use ten repeated partitions, with identical training, validation and test subsets used by the two methods in each repetition, and report mean Wasserstein errors with sample standard deviations across partitions. 

\subsection{Earth data on the sphere}\label{sec:sphere-experiments}
The earthquake, fire, flood and volcano datasets \citep{chen2024riemannian} describe geographic distributions of events or sites on the Earth's surface. We compare Fr\'echetFlow with the Riemannian flow matching (RFM) method from that study. Each observation is a latitude--longitude pair mapped to the unit sphere, with great-circle distance measured in radians.
For each partition, we assign approximately 80\%, 10\% and 10\% of observations to training, validation and testing, with equally sized validation and test sets. Evaluation uses exact discrete transport between 5000 generated points and all held-out test points.

\subsection{MERFISH cell neighborhoods}\label{sec:merfish-experiment}
The MERFISH data \citep{zhang2021merfish} are used to quantify variation in the cellular environments of somatostatin (Sst) neurons in the mouse motor cortex. We compare Fr\'echetFlow with the point cloud Wasserstein flow matching (WFM) implementation of \citet{haviv2025wasserstein}, using the supplied spatial transcriptomic atlas of 276,556 cells and 254 measured genes. One observation is the empirical distribution of the 16 neighboring cells closest in physical location to an Sst neuron within its tissue section. 
Each neighboring cell is represented by a vector in $\R^{16}$ containing the first 16 principal component scores of normalized log gene expression. For two neighborhoods $Y=16^{-1}\sum_{j=1}^{16}\delta_{x_j}$ and $Y'=16^{-1}\sum_{j=1}^{16}\delta_{x'_j}$, the object metric is
\begin{equation}\label{eq:merfish-object-metric}d(Y,Y')^2=\Wp{Y,Y'}{\norm{\cdot}_2}^2=\frac{1}{16}\min_{\pi\in\mathfrak S_{16}}\sum_{j=1}^{16}\norm{x_j-x'_{\pi(j)}}_2^2.\end{equation}
Thus each generated object describes a distribution of expression states within a cellular neighborhood.

The 64 available tissue sections are divided into 51 training, six validation and seven test sections, keeping neighborhoods that share cells together. Both methods share normalization, principal components and scaling fitted on training sections. Training and test samples contain 3243--3475 and 344--554 neighborhoods, respectively. Evaluation compares all test neighborhoods with an equally sized generated sample. We first use $d$ in \eqref{eq:merfish-object-metric} to compare expression distributions within two neighborhoods, then use these distances in \eqref{eq:empirical-w2-general} to compare the distributions of whole neighborhoods. Both levels use exact assignments.

\subsection{Brain connectivity matrices}\label{sec:brain-experiment}
Here we aim at the distribution of functional connectivity networks across subjects. We use the Autism Brain Imaging Data Exchange (ABIDE) input available through DiffeoCFM 
\citep{collas2025diffeocfm}, with resting-state functional magnetic resonance imaging measurements from 846 subjects on 39 common brain regions. Each subject's connectivity  is represented by a full rank correlation matrix. %
Write $C=AA^{\mathsf T}$ with lower triangular Cholesky factor $A$ and denote by  $\Psi(C)$ the strictly lower triangular entries of $\operatorname{diag}(A)^{-1}A$, where $\operatorname{diag}(A)$ is the diagonal matrix formed from the diagonal of $A$. The metric $d(C,C')=\norm{\Psi(C)-\Psi(C')}_2$ is the normalized Cholesky pullback distance used in DiffeoCFM. The map $\Psi$ provides an invertible vector representation of the correlation matrices and the  %
weighted Fr\'echet mean is obtained by taking the weighted average of the reference vectors $\Psi(b_j)$ and mapping back through $\Psi^{-1}$, producing a valid correlation matrix. %

Each random partition contains 684 training subjects, 77 validation subjects and 85 test subjects. DiffeoCFM models the normalized Cholesky coordinates directly. %
We compute empirical $W_2$ \eqref{eq:empirical-w2-general}, comparing 85 generated matrices with  85 held-out matrices.

\subsection{Results for the scientific datasets}
\begin{table}[htbp]
\centering
\caption{Empirical $W_2$ on the scientific datasets. Entries are means with sample standard deviations in parentheses over ten paired partitions; smaller values indicate greater accuracy. The ground metrics are great-circle distance in radians for the Earth datasets, neighborhood Wasserstein distance for MERFISH and normalized Cholesky distance for ABIDE. RFM and WFM denote Riemannian and Wasserstein flow matching, respectively.}
\label{tab:real-data-w2}\label{tab:sphere-w2}\label{tab:merfish-w2}\label{tab:brain-w2}
\small
\begin{tabular}{@{}llcc@{}}
\toprule
Dataset & Comparison method & Comparison $W_2$ & Fr\'echetFlow $W_2$ \\
\midrule
Earthquakes & RFM & $0.1379\;(0.0230)$ & $0.1280\;(0.0188)$ \\
Fires & RFM & $0.1055\;(0.0249)$ & $0.1277\;(0.0332)$ \\
Floods & RFM & $0.1424\;(0.0254)$ & $0.1426\;(0.0282)$ \\
Volcanoes & RFM & $0.3188\;(0.0600)$ & $0.3104\;(0.0644)$ \\
\midrule
MERFISH & WFM & $0.6407\;(0.0091)$ & $0.5797\;(0.0063)$ \\
ABIDE & DiffeoCFM & $5.2946\;(0.0396)$ & $5.1756\;(0.0660)$ \\
\bottomrule
\end{tabular}
\end{table}

\begin{figure}[!htbp]
\centering
\includegraphics[width=\textwidth]{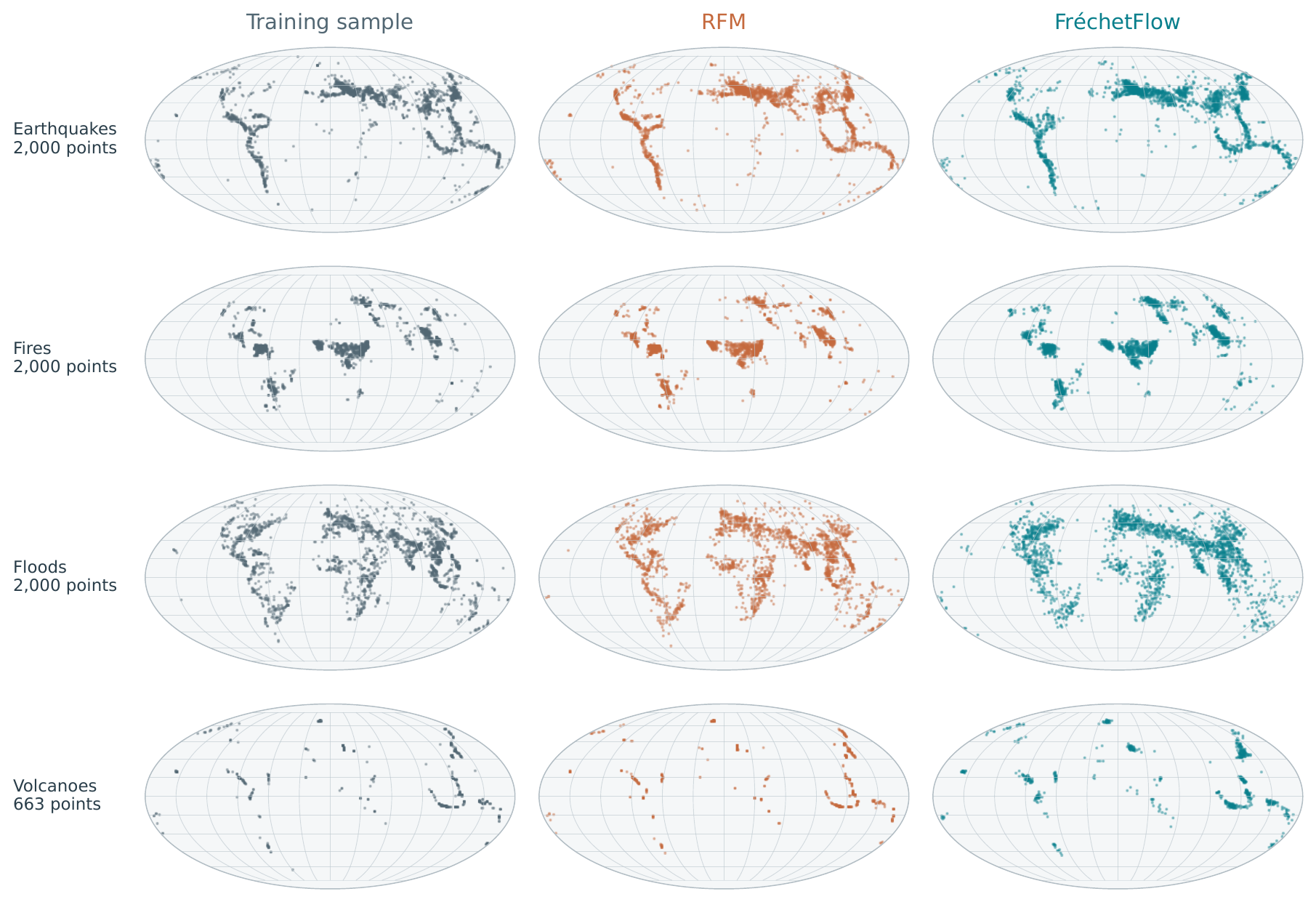}
\caption{Earth locations and generated samples in a Mollweide projection. Rows show earthquakes, fires, floods and volcanoes; columns show training observations, Riemannian FM and Fr\'echetFlow. Panels display 2000 points, or 663 for volcanoes to match the training count. Table~\ref{tab:real-data-w2} reports evaluation against held-out observations over all ten partitions.}
\label{fig:sphere-samples}
\end{figure}

Table~\ref{tab:real-data-w2} compares generation accuracy under the metric of each object space. On the Earth datasets, Fr\'echetFlow has lower mean error on earthquakes and volcanoes, nearly equal error on floods and higher error than the  more specialized Riemannian Flow Matching on fires. Figure~\ref{fig:sphere-samples} shows that both methods reproduce the broad geographic concentrations, including the narrow bands in the volcano and earthquake data and the more dispersed flood locations.

\begin{figure}[!htbp]
\centering
\includegraphics[width=\textwidth]{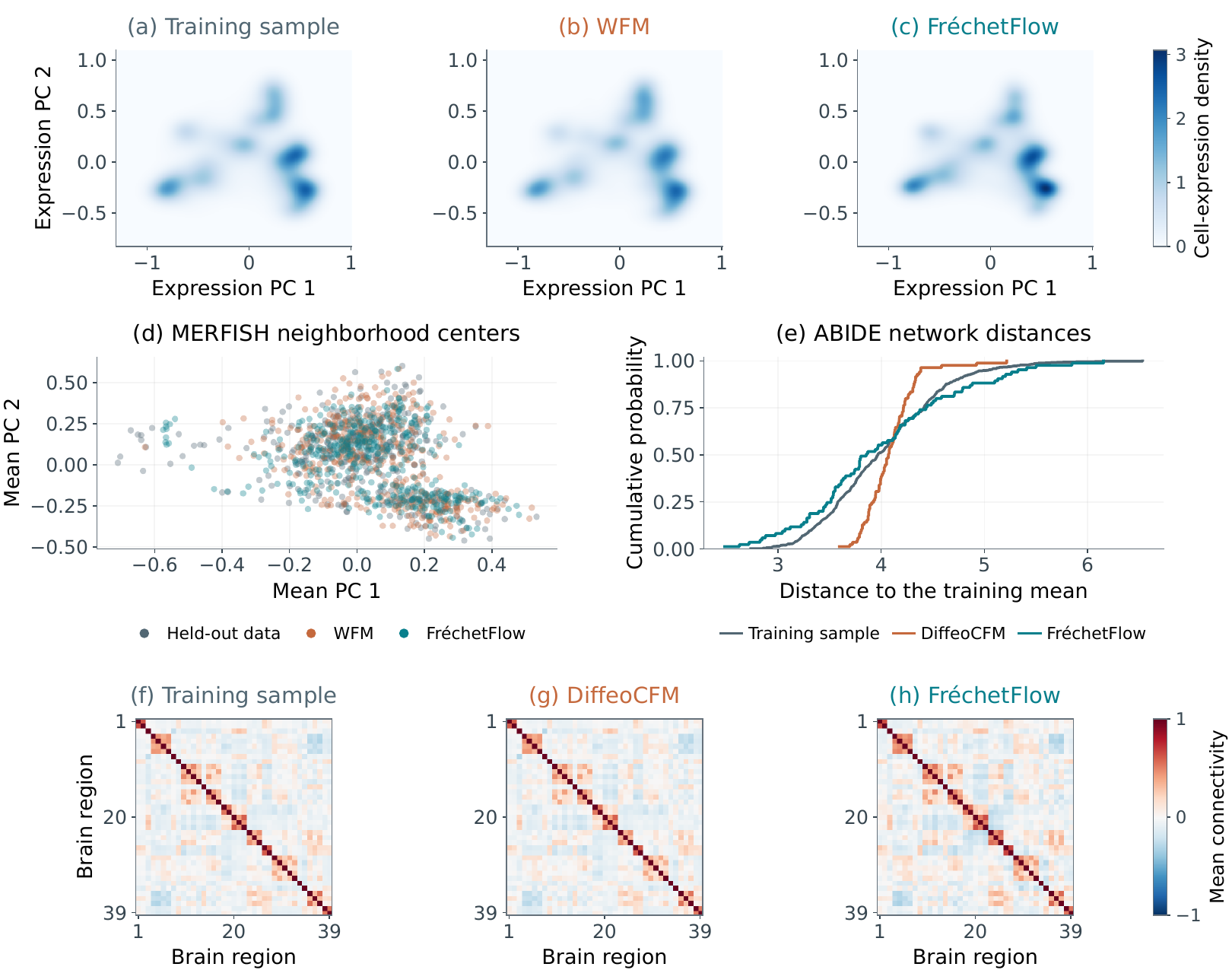}
\caption{MERFISH cell neighborhoods and ABIDE connectivity, each using a shared partition across methods. (a)--(c) Cell expression densities in the first two principal components, with equal weight per neighborhood and common axes, color scale and Gaussian smoothing bandwidth chosen from the training data. (d) Means of the 16 cell expression vectors in held-out and generated neighborhoods, plotted in the same principal components. (e) Empirical cumulative distributions of distances from training and generated networks to the training Fr\'echet mean under normalized Cholesky distance. (f)--(h) Entrywise mean correlation matrices with common region ordering and color scale. WFM denotes Wasserstein flow matching.}
\label{fig:merfish-brain-samples}\label{fig:merfish-neighborhood-samples}\label{fig:brain-connectivity-samples}
\end{figure}

\begin{samepage}
For MERFISH, Fr\'echetFlow reduces mean empirical $W_2$ by 9.5\% compared to the more specialized Wasserstein Flow Matching,   with lower error in all ten paired partitions. This comparison assesses gene expression variation among cells within neighborhoods and variation among whole neighborhoods. Figure~\ref{fig:merfish-brain-samples}(a)--(c) shows that both methods reproduce the main concentrations of cell expression values. Panel (d) compares the means of the 16 cell expression vectors within each neighborhood and shows agreement with the broad distribution of held-out neighborhood centers.
\end{samepage}

\begin{samepage}
Fr\'echetFlow also attains lower mean Wasserstein error than DiffeoCFM on the ABIDE connectivity matrices. In Figure~\ref{fig:merfish-brain-samples}(e), distances to the training Fr\'echet mean provide a view of variation across networks: Fr\'echetFlow captures the broad range of distances in the training sample. Panels (f)--(h) show that both methods reproduce the broad patterns in average correlations between brain regions.
\end{samepage}

\section{Discussion}\label{sec:discussion}
The proposed Fr\'echetFlow learns distributions of metric space-valued observations through landmark distances and object reconstruction, irrespective of the specific geometry of the metric space. The associated Wasserstein bounds reflect  the effects of sampling, landmark selection and fitting. Under mild regularity conditions, they  yield consistency and convergence rates. Our  experiments demonstrate accurate generation in Euclidean, curved and branching spaces, including specific settings of scientific interest  such as variation in tree topology and cubical cell membership.

In future research, the effects of metrics that are informed by scientific knowledge could further distinguish biological variation from measurement variation. The error bounds also suggest studying landmark and reference selection jointly: landmarks determine the information in the representation, while the reconstructions guided by reference sets  determine which objects can be generated. Relating these choices to the complexity of the flow and reconstruction map might be utilized to  guide model sizes as the sample grows. The proposed Fr\'echetFlow is a very general approach that also applies to Euclidean vectors, probability distributions and manifold-valued data, with competitive performance in the reported comparisons. It could be further refined for random objects in metric spaces with specific characteristics. %

Conditional generation is a further direction of interest for future research.  Letting the flow and reconstruction weights depend on covariates will allow us to generate conditional distributions of random objects, conditioning on subject characteristics, which would be of special interest for brain networks, as would be generating   cellular neighborhoods conditional on tissue location.  %

\section*{Acknowledgements}
H.G.M. was partially supported by NSF grant DMS-2310450. The authors declare no conflict of interest.

\FloatBarrier
\bibliographystyle{plainnat}
\bibliography{references}

\clearpage
\pdfbookmark[0]{Supplementary material}{supplement}
\hypersetup{linkcolor=black,citecolor=black,urlcolor=black}
\begin{center}
\large\bfseries SUPPLEMENTARY MATERIAL
\end{center}

\setcounter{section}{0}
\renewcommand{\thesection}{S.\arabic{section}}
\renewcommand{\theHsection}{supplement.\arabic{section}}
\counterwithout{theorem}{section}
\setcounter{theorem}{0}
\renewcommand{\thetheorem}{S\arabic{theorem}}
\renewcommand{\theHtheorem}{supplement.\arabic{theorem}}
\setcounter{equation}{0}
\renewcommand{\theequation}{S.\arabic{equation}}
\renewcommand{\theHequation}{supplement.\arabic{equation}}
\setcounter{table}{0}
\renewcommand{\thetable}{S\arabic{table}}
\renewcommand{\theHtable}{supplement.\arabic{table}}
\setcounter{figure}{0}
\renewcommand{\thefigure}{S\arabic{figure}}
\renewcommand{\theHfigure}{supplement.\arabic{figure}}
\allowdisplaybreaks
\section{Examples of metric space-valued data}\label{sec:geometry-examples}
Fr\'echetFlow uses a common framework for flow optimization and minimization of the reconstruction loss \eqref{eq:normalized-decoder-loss} across the examples below, where each object space is furnished with space-specific distance and weighted Fr\'echet means. Landmarks determine how well the landmark representations distinguish objects and mean references determine the available outputs; $a_B(P)$ in \eqref{eq:family-error} measures approximation by those means. Section~\ref{sec:theory-transfer} of the main paper connects the actual reconstruction residual to error in the generated object distribution through distance to the nearest landmark and Section~\ref{app:geometry-examples} collects the geometric constructions and verifies their conditions.

\begin{example}[Euclidean vectors and images]\label{sec:vector-matrix-examples}
Under Euclidean distance, weighted Fr\'echet means are ordinary weighted averages. For a concrete example, consider observations in the unit square $\M=[0,1]^2$ and choose three corners as landmarks $L$ and all four corners as mean references $B$. Distances to $L$ identify every point, whereas weighted averages of $B$ fill the square. The landmark representation has a Lipschitz inverse.

Images provide a higher-dimensional Euclidean example central to generative modeling \citep{lipman2023flow}. An image $I$ with $p$ intensity values across its pixels and channels, each scaled to $[0,1]$, is an element of $[0,1]^p$. A multiscale distance compares both individual pixels and broader spatial patterns. Let $A_s$ average nonoverlapping $s\times s$ pixel blocks within each channel, with $A_1$ leaving the image unchanged and let $p_s$ be the number of entries in $A_sI$. For positive scale constants $a_s^2$, set $D_0^2=\frac13\sum_{s\in\{1,2,4\}}p_s/a_s^2$ and define
\begin{equation}\label{eq:multiscale-image-metric}d(I,I')^2=\frac{1}{3D_0^2}\sum_{s\in\{1,2,4\}}\frac{\norm{A_sI-A_sI'}_2^2}{a_s^2}.\end{equation}
The full-resolution term retains fine detail, while the two coarser terms compare local averages. The constants $a_s^2$ balance these contributions and $D_0$ normalizes the diameter of the image cube to one. This is Euclidean distance after an injective linear transformation, so the corresponding weighted Fr\'echet means are pixelwise weighted averages and remain valid images.
\end{example}

\begin{example}[Symmetric positive-definite matrices]\label{sec:spd-examples}
Symmetric positive-definite (SPD) matrices describe full-rank covariance structures and diffusion tensors \citep{dryden2009covariance}. Different metrics express different notions of similarity and lead to different ways of averaging reference matrices. Under the Frobenius distance, $d_F(A,A')=\norm{A-A'}_F$, the weighted Fr\'echet mean is the ordinary matrix average $\sum_jw_jb_j$, which remains positive definite. Log-Euclidean distance, $d_{\mathrm{LE}}(A,A')=\norm{\log A-\log A'}_F$, instead averages the matrix logarithms and maps the result back through the exponential, giving $\exp\{\sum_jw_j\log b_j\}$ \citep{arsigny2006logeuclidean}. The Log-Cholesky metric follows a similar principle, averaging the strictly lower triangular entries and log-diagonal entries of the Cholesky factors before mapping back to an SPD matrix \citep{lin2019cholesky}.

The affine-invariant metric, $d_{\mathrm{AI}}(A,A')=\norm{\log(A^{-1/2}A'A^{-1/2})}_F$, is useful when separation should remain unchanged by a common invertible linear transformation of the variables \citep{pennec2006tensor}. Under this metric, the SPD matrices form a Hadamard manifold with nonpositive curvature, so weighted means are unique and obtained by intrinsic optimization. %

In the brain-connectivity application, each observation is an SPD matrix with unit diagonal, that is, a full-rank correlation matrix describing associations among brain regions. We use the normalized-Cholesky metric of \citet{collas2025diffeocfm}, whose bijective map $\Psi$ leads to Euclidean coordinates with $d(C,C')=\norm{\Psi(C)-\Psi(C')}_2$. Averaging in these coordinates and mapping back yields the weighted mean $\Psi^{-1}\{\sum_jw_j\Psi(b_j)\}$, a valid correlation matrix for every simplex weight vector.
\end{example}

\begin{example}[Locations on the sphere]\label{sec:spherical-objects}
Earth events such as earthquakes and volcanic activity are naturally recorded as locations on the sphere \citep{chen2024riemannian}. An observation is a point $y\in\mathbb S^2$, with separation measured by great-circle distance $d(x,y)=\arccos\langle x,y\rangle$. Distances to the three coordinate directions $e_1,e_2,e_3$ are $\arccos y_1$, $\arccos y_2$ and $\arccos y_3$, so these three landmarks identify every location. Weighted means are unique when the locations lie in a geodesic ball of radius $0<\rho<\pi/4$. The vertices of a sufficiently fine geodesic triangulation form such local groups, whose means fill the triangles and together cover the sphere.
\end{example}

\begin{example}[Locations on trees and metric graphs]\label{sec:network-location-examples}
Trees and metric graphs provide natural domains for observations located on a fixed network, such as traffic accidents on roads or dendritic spines on neuronal branches \citep{baddeley2021networks}. An observation is a single location, specified by a branch and position along the branch, where shortest-path distance $d$ quantifies separation along the available routes. On a finite metric tree $\mathcal T$, choosing all leaves as the landmarks $L$ provides a representation that preserves the tree metric exactly: $\norm{\phi_L(x)-\phi_L(y)}_\infty=d(x,y)$, $x,y\in \mathcal T$ \citep[Remark~11]{ilmavirta2025lipschitz}. Every point of the tree is also a weighted mean of at most two leaves \citep[Remark~6.4]{sturm2003barycenters}.

Cycles introduce additional routes in a metric graph. In the graph used in Section~\ref{sec:graph-tree-experiments}, each edge is the unique shortest path between its endpoints. Weighted means of the endpoints traverse the edge, so that weighted means of vertices cover the entire graph. %
\end{example}

\begin{example}[Phylogenetic trees]\label{sec:branching-object-examples}
In phylogenetics, the observation is an entire tree whose topology and branch lengths describe evolutionary relationships among labelled taxa. Collections of trees inferred from genetic sequences can represent variation across genes or uncertainty about these relationships. The Billera--Holmes--Vogtmann (BHV) geometry represents each topology by a region of branch lengths and joins these regions along their shared unresolved trees \citep{billera2001geometry}. The five-taxon domain used in Section~\ref{sec:graph-tree-experiments} fixes terminal lengths at one and allows internal lengths in $[0,1]$, so each fully resolved topology corresponds to a square of branch lengths. Distances to the star tree and single-branch trees identify every split length, while weighted means of each square's corners fill that square.
\end{example}

\begin{example}[Joint positions on two branching tracks]\label{sec:cubical-object-examples}
Cubical configuration spaces arise in robotic motion planning, where a point records the configuration of the whole system and a square represents two compatible motions that can occur simultaneously \citep{ardila2012robots}. Consider two robots, each confined to its own fixed Y-shaped track $\mathcal T_3$ with three unit branches meeting at a junction. One observation is their joint position $y=(y_1,y_2)\in\mathcal T_3\times\mathcal T_3$, where $y_1=(a,u)$ and $y_2=(b,v)$ specify the branch labels $a,b\in\{1,2,3\}$ and distances $u,v\in[0,1]$ from the junctions. On either track, the distance between $(a,u)$ and $(a',u')$ is $|u-u'|$ if $a=a'$ and $u+u'$ otherwise, since movement between branches passes through the junction. The distance between two joint positions is $d(y,y')=\{d_{\mathcal T_3}(y_1,y'_1)^2+d_{\mathcal T_3}(y_2,y'_2)^2\}^{1/2}$. For a fixed branch pair, the two distances from the junctions vary over $[0,1]^2$. The nine branch pairs therefore give nine square cells, joined along shared edges where a robot reaches its junction. Three cells meet at an interior point of a shared edge, where no manifold chart exists. Distances to the sixteen vertices distinguish all configurations and weighted means of a cell's corners fill that cell. The CAT(0) geometry ensures unique weighted means even where cells meet.
\end{example}

\begin{example}[Distributions represented by point clouds]\label{sec:distribution-object-examples}
An observation can itself be a probability distribution. In the MERFISH application, a tissue neighborhood is represented by the empirical distribution of its cell-expression vectors. Wasserstein distance compares two neighborhoods by optimally matching their constituent cells. Generating whole neighborhoods therefore models variation both between neighborhoods and among cells within each neighborhood. For equally weighted point clouds of a common size, weighted Fr\'echet means minimize a weighted sum of squared Wasserstein distances over clouds of that size. With two reference clouds, this construction reduces to interpolation of optimally matched points along a Wasserstein geodesic.

Images also admit a distribution-valued interpretation: normalizing nonnegative intensities to unit mass yields a distribution over pixel locations. Wasserstein distance measures spatial transport of intensity, complementing the pixelwise Euclidean geometry in Example~\ref{sec:vector-matrix-examples}.
\end{example}

\section{Proofs}\label{app:proofs}

\subsection{Proof of Theorem~\ref{thm:transfer}}\label{app:landmark-coverage}
Recall the law $\mu_L=(\phi_L)_\#P$ of the landmark representation under the target law $P$, the representation errors $\delta$ and $e_Q(D)$ from \eqref{eq:coordinate-errors}, and the root mean squared distance $q_2(P,L)$ to the nearest landmark from \eqref{eq:landmark-coverage}.

By Assumption~\ref{ass:object-law}, there exists $y_0\in\M$ such that $\E_P d(Y,y_0)^2<\infty$. Since $L$ is finite and nonempty, the triangle inequality implies
\[
\norm{\phi_L(y)}_\infty=\max_{1\le j\le k}d(y,\ell_j)\le d(y,y_0)+\max_{1\le j\le k}d(y_0,\ell_j),
\]
and we applied Minkowski's inequality to show that $\mu_L\in\Pcal_2(\R^k)$:
\[
\left\{\E_P\norm{\phi_L(Y)}_\infty^2\right\}^{1/2}\le\left\{\E_P d(Y,y_0)^2\right\}^{1/2}+\max_{1\le j\le k}d(y_0,\ell_j)<\infty.
\]

We then take the independent random coupling $Z\sim Q$ and $Z^\star\sim\mu_L$. By the definition of $\delta$, we derive the following result using triangle inequality and Minkowski's inequality:
\[
\delta\le\left\{\E\norm{Z-Z^\star}_\infty^2\right\}^{1/2}\le\left\{\E_Q\norm Z_\infty^2\right\}^{1/2}+\left\{\E_P\norm{\phi_L(Y)}_\infty^2\right\}^{1/2}<\infty.
\]
Furthermore, $q_2(P,L)\le\{\E_P d(Y,\ell_1)^2\}^{1/2}<\infty$.

For $x,y\in\M$, choose a landmark $\ell\in L$ nearest to $y$, so that $d(y,\ell)=\min_jd(y,\ell_j)$. The triangle inequality implies
\begin{equation}\label{eq:pointwis_bound}
\begin{aligned}
d(x,y)&\le d(x,\ell)+d(y,\ell)\\
&=d(x,\ell)-d(y,\ell)+2d(y,\ell)\\
&\le |d(x,\ell)-d(y,\ell)|+2d(y,\ell)\\
&\le\norm{\phi_L(x)-\phi_L(y)}_\infty+2\min_jd(y,\ell_j).
\end{aligned}
\end{equation}

Fix $\varepsilon>0$ and let $(Z,Z^\star)\sim\pi$, where $\pi$ is a coupling of $Q$ and $\mu_L$ chosen so that
\[
\left\{\E\norm{Z-Z^\star}_\infty^2\right\}^{1/2} = \left\{\int_{\R^k\times\R^k}\norm{z-z^\star}_\infty^2\,\pi(dz,dz^\star)\right\}^{1/2}\le\delta+\varepsilon.
\]
The landmark map $\phi_L$ is continuous so it's Borel measurable. Under Assumption~\ref{ass:object-law}, $\M$ is Polish, so the disintegration theorem provides a regular conditional distribution $K(z^\star,\cdot)$ of $Y$ given $\phi_L(Y)=z^\star$ \citep[Chapter~1]{villani2009optimal}. For every Borel set $A\subseteq\M$,
\[
P(A)=\int_{\R^k}K(z^\star,A)\,\mu_L(dz^\star),
\]
and $K(z^\star,\{y:\phi_L(y)=z^\star\})=1$ for $\mu_L$-almost every $z^\star$. We define a joint law of $(Z,Z^\star,Y)$ by
\[
\Pi(dz,dz^\star,dy)=\pi(dz,dz^\star)K(z^\star,dy).
\]
The marginal law of $Y$ is $P$ since
\[
\Pi(Y\in A)=\int_{\R^k\times\R^k}K(z^\star,A)\,\pi(dz,dz^\star)=\int_{\R^k}K(z^\star,A)\,\mu_L(dz^\star)=P(A)
\]
for every Borel set $A\subseteq\M$. Under this joint law, $Y\sim P$ and $\phi_L(Y)=Z^\star$ almost surely, so
\[
\left\{\E\norm{Z-Z^\star}_\infty^2\right\}^{1/2}=\left\{\E\norm{Z-\phi_L(Y)}_\infty^2\right\}^{1/2}\le\delta+\varepsilon.
\]
Applying the bound in equation \eqref{eq:pointwis_bound} to $D(Z)$ and $Y$, where $D:\R^k\to\M$ and inserting $Z$ between their landmark representations, we have
\[
\begin{aligned}
d(D(Z),Y)&\le\norm{\phi_L(D(Z))-\phi_L(Y)}_\infty+2\min_jd(Y,\ell_j)\\
&\le\norm{\phi_L(D(Z))-Z}_\infty+\norm{Z-\phi_L(Y)}_\infty+2\min_jd(Y,\ell_j).
\end{aligned}
\]
We again use the Minkowski's inequality,
\[
\begin{aligned}
\left\{\E d(D(Z),Y)^2\right\}^{1/2}
&\le\left\{\E\norm{\phi_L(D(Z))-Z}_\infty^2\right\}^{1/2}\\
&\quad+\left\{\E\norm{Z-\phi_L(Y)}_\infty^2\right\}^{1/2}
+2\left\{\E\min_jd(Y,\ell_j)^2\right\}^{1/2}\\
&\le e_Q(D)+\delta+\varepsilon+2q_2(P,L),
\end{aligned}
\]
where $Z\sim Q$ and $Y\sim P$, respectively. Since the right-hand side is finite and under Assumption~\ref{ass:object-law}, another application of Minkowski's inequality leads to 
\[
\left\{\E d(D(Z),y_0)^2\right\}^{1/2}\le\left\{\E d(D(Z),Y)^2\right\}^{1/2}+\left\{\E_P d(Y,y_0)^2\right\}^{1/2}<\infty,
\]
which shows $D_\#Q\in\Pcal_2(\M)$. Finally, the joint law of $(D(Z),Y)$ is a coupling of $D_\#Q$ and $P$, so
\[
\Wp{D_\#Q,P}{d}\le\left\{\E d(D(Z),Y)^2\right\}^{1/2}\le e_Q(D)+\delta+\varepsilon+2q_2(P,L).
\]
Letting $\varepsilon\downarrow0,$ we can have equation \eqref{eq:distance-transfer}.\hfill$\square$

\subsection{Proof of Corollary~\ref{cor:weighted-mean-transfer}}\label{app:weighted-mean-transfer}

Recall the mean-family approximation error $a_B(P)$ from \eqref{eq:family-error}, the smallest reconstruction residual $\rho_B$ from \eqref{eq:best-reconstruction} and the reconstruction map's excess residual $\eta_{\rm proj}(D_\eta)$ from \eqref{eq:projection-error}.

Assumption~\ref{ass:decoder-geometry} makes $H_B$ compact, so $\phi_L(H_B)$ is bounded. For each $y\in\M$, we can always find a nearest mean $x_y\in H_B$, with $d(y,x_y)=\operatorname{dist}(y,H_B)$ by compactness of $H_B.$ Evaluating the reconstruction objective at $x_y$ and using the 1-Lipschitz property of $\phi_L$ give
\[\rho_B(z)\le\norm{\phi_L(x_y)-z}_\infty\le\operatorname{dist}(y,H_B)+\norm{\phi_L(y)-z}_\infty.\]
For $\varepsilon>0$, use the coupling of $Z\sim Q$ and $Y\sim P$ constructed in the proof of Theorem~\ref{thm:transfer}, whose root mean squared discrepancy between landmark representations is at most $\delta+\varepsilon$. Minkowski's inequality shows that $\norm{\rho_B}_{L^2(Q)}\le a_B(P)+\delta+\varepsilon.$ Letting $\varepsilon\downarrow0$ and using $r_\eta=\rho_B+(r_\eta-\rho_B),$ we have the following bound for the root mean squared reconstruction residual $e_Q(D_\eta)$ from \eqref{eq:coordinate-errors},
\begin{align*}
    e_Q(D_\eta)=\norm{r_\eta}_{L^2(Q)}\le\norm{\rho_B}_{L^2(Q)}+\eta_{\rm proj}(D_\eta)\le a_B(P)+\delta+\eta_{\rm proj}(D_\eta).
\end{align*}
Substituting this inequality into Theorem~\ref{thm:transfer} leads to the bound for the exact reconstruction map. Coupling the numerical and exact outputs at the same $Z\sim Q$ yields
\[\Wp{(D_\eta^{\mathrm{num}})_\#Q,(D_\eta)_\#Q}{d}\le\left\{\E_Qd\{D_\eta^{\mathrm{num}}(Z),D_\eta(Z)\}^2\right\}^{1/2}\le\beta.\]
The Wasserstein triangle inequality then proves \eqref{eq:coverage-projection}.\hfill$\square$

\subsection{Proof of Theorem~\ref{thm:projection-oracle}}\label{sec:projection-fitting}
We use the constants $M_B$, $m_Q$ and $V_Q$ defined in \eqref{eq:decoder-moment-constants}. Here $M_B$ bounds the sup norm of the landmark representations in $H_B$, while $m_Q$ measures the root mean squared sup norm of the generated vectors $Z\sim Q$, and $V_Q=M_B(2m_Q+M_B)$. Taking expectations with respect to $Z\sim Q$ on both sides of $(r_D(Z)-\rho_B(Z))^2\le r_D(Z)^2-\rho_B(Z)^2$ yields \eqref{eq:excess-controls-gap}. Under Assumption~\ref{ass:decoder-fitting}, the inputs $Z_1,\ldots,Z_K$ are independent with law $Q$. For an integrable function $f$, write
\[
\mathbb Q_Kf=\frac1K\sum_{i=1}^Kf(Z_i),\qquad Qf=\E_{Z\sim Q}f(Z).
\]
Recall that $r_D(z)=\norm{\phi_L(D(z))-z}_\infty$ and that $D(z)\in H_B$. By the definition of $M_B$, we have $\norm{\phi_L(D(z))}_\infty\le M_B$, and therefore
\[
r_D(z)\le\norm z_\infty+M_B.
\]
To obtain the bound under the finite second moment assumption on $Q$, we define
\[
f_D(z)=r_D(z)^2-\norm z_\infty^2,
\]
and $\mathbb Q_Kf_D=\widehat{\mathcal J}_K(D)-\frac1K\sum_{i=1}^K\norm{Z_i}_\infty^2.$ Similarly, for any $D,D'\in\mathcal D$, where $\mathcal D$ is the class of candidate measurable
reconstruction maps,
\[
Qf_D-Qf_{D'}=\mathcal J_Q(D)-\mathcal J_Q(D').
\]
We next obtain a common bound for the absolute values of the functions $f_D$. The reverse triangle inequality implies 
\[
\left|r_D(z)-\norm z_\infty\right|
=
\left|\norm{\phi_L(D(z))-z}_\infty-\norm z_\infty\right|
\le\norm{\phi_L(D(z))}_\infty
\le M_B.
\]
Factoring the difference of squares and using $r_D(z)\le\norm z_\infty+M_B$, we obtain
\[
\begin{aligned}
|f_D(z)|=\left|r_D(z)^2-\norm z_\infty^2\right|=\left|r_D(z)-\norm z_\infty\right|\left\{r_D(z)+\norm z_\infty\right\}\le M_B(2\norm z_\infty+M_B).
\end{aligned}
\]
Define $F(z)=M_B(2\norm z_\infty+M_B).$ The function $F$ bounds the class $\{f_D:D\in\mathcal D\}$, where $|f_D(z)|\le F(z)$ for every $D$ and $z$. By Minkowski's inequality,
\[
\begin{aligned}
\norm F_{L^2(Q)}
&\le 2M_B\left\{\E_Q\norm Z_\infty^2\right\}^{1/2}+M_B^2\\
&=M_B(2m_Q+M_B)=V_Q.
\end{aligned}
\]
By the definition of $d_{\mathcal R}(r_D,r_{D'})$ in Assumption \ref{ass:decoder-fitting}, if $d_{\mathcal R}(r_D,r_{D'})\le\varepsilon$, $|r_D(z)-r_{D'}(z)|\le\varepsilon$ for every $z$. The common subtracted term cancels, giving
\[
\begin{aligned}
|f_D(z)-f_{D'}(z)|
&=|r_D(z)^2-r_{D'}(z)^2|\\
&=|r_D(z)-r_{D'}(z)|\{r_D(z)+r_{D'}(z)\}\\
&\le 2\varepsilon(\norm z_\infty+M_B).
\end{aligned}
\]
We now bound the largest discrepancy between empirical and population averages over the whole class:
\[
S_K=\sup_{D\in\mathcal D}|(\mathbb Q_K-Q)f_D|.
\]
Assumption~\ref{ass:decoder-fitting} provides a countable uniformly dense subclass of residuals. The supremum $S_K$ can be taken over this subclass and is therefore measurable, since for any $D,D'\in\mathcal D$,
\[
|(\mathbb Q_K-Q)(f_D-f_{D'})|
\le 2d_{\mathcal R}(r_D,r_{D'})(\mathbb Q_K+Q)(\norm z_\infty+M_B).
\]
We choose an $\varepsilon$-cover of $\mathcal R$ with $N=N(\mathcal R,\varepsilon;d_{\mathcal R})<\infty$ centers $r_{D_1},\ldots,r_{D_N}\in\mathcal R$, and write $f_j=f_{D_j}$. For each $D$, we choose a center $r_{D_j}$ with $d_{\mathcal R}(r_D,r_{D_j})\le\varepsilon$. Then
\[
\begin{aligned}
|(\mathbb Q_K-Q)f_D|
&\le |(\mathbb Q_K-Q)f_j|+|(\mathbb Q_K-Q)(f_D-f_j)|\\
&\le |(\mathbb Q_K-Q)f_j|+\mathbb Q_K|f_D-f_j|+Q|f_D-f_j|\\
&\le |(\mathbb Q_K-Q)f_j|+2\varepsilon(\mathbb Q_K+Q)(\norm z_\infty+M_B).
\end{aligned}
\]
Taking the supremum over $D$ leads to 
\[
S_K\le\max_{1\le j\le N}|(\mathbb Q_K-Q)f_j|+2\varepsilon(\mathbb Q_K+Q)(\norm z_\infty+M_B).
\]
Since $\E\mathbb Q_Kh=Qh$ for an integrable function $h$, and $\E_Q\norm Z_\infty\le m_Q$,
\[
\E S_K\le\E\max_{1\le j\le N}|(\mathbb Q_K-Q)f_j|+4\varepsilon(m_Q+M_B).
\]

To bound the expected maximum over the finite set of centers, we introduce an independent sample $Z_1',\ldots,Z_K'$ following law $Q$, and let $\mathbb Q_K'f=K^{-1}\sum_{i=1}^Kf(Z_i')$. Conditional on $Z_1,\ldots,Z_K$,
\[
(\mathbb Q_K-Q)f_j=\E\bigl[(\mathbb Q_K-\mathbb Q_K')f_j\mid Z_1,\ldots,Z_K\bigr].
\]
Since the function $(a_1,\ldots,a_N)\mapsto\max_j|a_j|$ is convex, we apply the conditional Jensen's inequality
\[
\E\max_{1\le j\le N}|(\mathbb Q_K-Q)f_j|
\le
\E\max_{1\le j\le N}\left|\frac1K\sum_{i=1}^K\{f_j(Z_i)-f_j(Z_i')\}\right|.
\]

Let $\epsilon_1,\ldots,\epsilon_K$ be independent Rademacher variables, independent of both $\{Z_i\}_{i=1}^{K}$ and $\{Z'_i\}_{i=1}^{K}$, with $\Pr(\epsilon_i=1)=\Pr(\epsilon_i=-1)=1/2$. For each $i$, if we swap $Z_i$ and $Z_i'$, it changes the sign of the vector $\{f_j(Z_i)-f_j(Z_i')\}_{j=1}^N$ without changing its distribution. Consequently, inserting the independent signs does not change the expected maximum:
\[
\begin{aligned}
&\E\max_{1\le j\le N}\left|\frac1K\sum_{i=1}^K\{f_j(Z_i)-f_j(Z_i')\}\right|=
\E\max_{1\le j\le N}\left|\frac1K\sum_{i=1}^K\epsilon_i\{f_j(Z_i)-f_j(Z_i')\}\right|.
\end{aligned}
\]
Applying the triangle inequality and then using the identical distributions of the two samples ($\{Z_i\}_{i=1}^{K}$ and $\{Z'_i\}_{i=1}^{K}$) yields
\[
\begin{aligned}
\E\max_{1\le j\le N}|(\mathbb Q_K-Q)f_j|
&\le
\E\max_{1\le j\le N}\left|\frac1K\sum_{i=1}^K\epsilon_i f_j(Z_i)\right|
+
\E\max_{1\le j\le N}\left|\frac1K\sum_{i=1}^K\epsilon_i f_j(Z_i')\right|\\
&=
2\E\max_{1\le j\le N}\left|\frac1K\sum_{i=1}^K\epsilon_i f_j(Z_i)\right|.
\end{aligned}
\]

To bound the signed sums, we condition on $Z_1,\ldots,Z_K$ and write
\[
T_j=\frac1K\sum_{i=1}^K\epsilon_i f_j(Z_i),\qquad v_K^2=\frac1{K^2}\sum_{i=1}^KF(Z_i)^2.
\]
Let $\E_\epsilon$ denote expectation over the signs. As the signs are independent and the inequality $\cosh(u)\le\exp(u^2/2)$ holds, for every $\lambda>0$, we have
\[
\begin{aligned}
\E_\epsilon e^{\lambda T_j}
=\prod_{i=1}^K\cosh\left(\frac{\lambda f_j(Z_i)}K\right)\le\exp\left\{\frac{\lambda^2}{2K^2}\sum_{i=1}^Kf_j(Z_i)^2\right\}\le\exp\left(\frac{\lambda^2v_K^2}{2}\right).
\end{aligned}
\]
The same bound holds for $-T_j$. Hence
\[
\begin{aligned}
\exp\left\{\lambda\E_\epsilon\max_{1\le j\le N}|T_j|\right\}
&\le\E_\epsilon\exp\left\{\lambda\max_{1\le j\le N}|T_j|\right\}\\
&\le\sum_{j=1}^N\left(\E_\epsilon e^{\lambda T_j}+\E_\epsilon e^{-\lambda T_j}\right)\\
&\le2N\exp\left(\frac{\lambda^2v_K^2}{2}\right).
\end{aligned}
\]
Taking logarithms on both sides and dividing by $\lambda$ we arrive at 
\[
\E_\epsilon\max_{1\le j\le N}|T_j|
\le\frac{\log(2N)}{\lambda}+\frac{\lambda v_K^2}{2}.
\]
For $v_K>0$, choosing $\lambda=\sqrt{2\log(2N)}/v_K$ yields
\[
\E_\epsilon\max_{1\le j\le N}|T_j|\le v_K\sqrt{2\log(2N)}.
\]
Finally,
\[
\E v_K\le\{\E v_K^2\}^{1/2}
=\left\{\frac{\E_QF(Z)^2}{K}\right\}^{1/2}
\le\frac{V_Q}{\sqrt K}.
\]
Combining this bound with symmetrization leads to 
\[
\E\max_{1\le j\le N}|(\mathbb Q_K-Q)f_j|
\le2V_Q\sqrt{\frac{2\log(2N)}K}.
\]
Therefore,
\[
\E S_K\le4(m_Q+M_B)\varepsilon+2V_Q\sqrt{\frac{2\log(2N)}K}.
\]

We next control the fluctuations of $S_K$ around its expectation. Since $|f_D|\le F$,
\[
S_K\le\mathbb Q_KF+QF.
\]
The right-hand side is square-integrable because $F\in L^2(Q)$, so $\operatorname{Var}(S_K)$ is finite. Replace $Z_i$ by an independent copy $Z_i'$ and denote the resulting supremum by $S_K^{(i)}$. For every $D$, this replacement changes $(\mathbb Q_K-Q)f_D$ by $\{f_D(Z_i)-f_D(Z_i')\}/K.$ The absolute difference between two supremums is bounded by the supremum of the absolute differences. It follows that
\[
|S_K-S_K^{(i)}|
\le\frac1K\sup_{D\in\mathcal D}|f_D(Z_i)-f_D(Z_i')|
\le\frac{F(Z_i)+F(Z_i')}{K}.
\]
Using $(a+b)^2\le2a^2+2b^2$ and $\E_QF(Z)^2\le V_Q^2$, we obtain
\[
\E(S_K-S_K^{(i)})^2
\le\frac{4V_Q^2}{K^2}.
\]
The Efron--Stein inequality \citep[Proposition~1]{boucheron2005moment} then implies 
\[
\operatorname{Var}(S_K)
\le\frac12\sum_{i=1}^K\E(S_K-S_K^{(i)})^2
\le\frac{2V_Q^2}{K}.
\]
For $V_Q>0$, Chebyshev's inequality implies
\[
\Pr\left\{S_K>\E S_K+V_Q\sqrt{\frac{2}{K\alpha}}\right\}\le\alpha.
\]
With conditional probability at least $1-\alpha$,
\[
\begin{aligned}
S_K
&\le4(m_Q+M_B)\varepsilon
+2V_Q\sqrt{\frac{2\log(2N)}K}
+V_Q\sqrt{\frac{2}{K\alpha}}=\Delta_K(\varepsilon,\alpha),
\end{aligned}
\]
which leads to  the definition of $\Delta_K(\varepsilon,\alpha)$ in \eqref{eq:decoder-moment-deviation}.

Here, we compare the fitted map $\widehat D$ with an arbitrary $D\in\mathcal D$. The numerical approximation condition in Assumption~\ref{ass:decoder-fitting} implies 
\[
\widehat{\mathcal F}_{K,n}(\widehat D)
\le\widehat{\mathcal F}^{\mathrm{num}}_{K,n}(\widehat D)+\epsilon_{\rm comp},
\qquad
\widehat{\mathcal F}^{\mathrm{num}}_{K,n}(D)
\le\widehat{\mathcal F}_{K,n}(D)+\epsilon_{\rm comp}.
\]
The optimization condition leads to 
\[
\widehat{\mathcal F}^{\mathrm{num}}_{K,n}(\widehat D)
\le\widehat{\mathcal F}^{\mathrm{num}}_{K,n}(D)+\xi_{\rm opt}.
\]
Combining these three inequalities yields
\[
\widehat{\mathcal F}_{K,n}(\widehat D)-\widehat{\mathcal F}_{K,n}(D)
\le\xi_{\rm opt}+2\epsilon_{\rm comp}.
\]

Since the common term $\norm z_\infty^2$ cancels when two losses are subtracted, we have
\[
\mathcal J_Q(\widehat D)-\mathcal J_Q(D)=Q(f_{\widehat D}-f_D).
\]
Moreover,
\[
\begin{aligned}
Q(f_{\widehat D}-f_D)
&=\mathbb Q_K(f_{\widehat D}-f_D)
+(Q-\mathbb Q_K)f_{\widehat D}
-(Q-\mathbb Q_K)f_D\\
&\le\mathbb Q_K(f_{\widehat D}-f_D)+2S_K.
\end{aligned}
\]

Adding the observed-object loss difference and using the definition of $\widehat{\mathcal F}_{K,n}$, we obtain
\[
\begin{aligned}
&\mathcal J_Q(\widehat D)+c\widehat{\mathcal R}_n(\widehat D)
-\mathcal J_Q(D)-c\widehat{\mathcal R}_n(D)\\
&\qquad\le
\mathbb Q_K(f_{\widehat D}-f_D)
+c\{\widehat{\mathcal R}_n(\widehat D)-\widehat{\mathcal R}_n(D)\}
+2S_K\\
&\qquad=
\widehat{\mathcal F}_{K,n}(\widehat D)-\widehat{\mathcal F}_{K,n}(D)+2S_K\\
&\qquad\le
\xi_{\rm opt}+2\epsilon_{\rm comp}+2\Delta_K(\varepsilon,\alpha).
\end{aligned}
\]
This holds for every $D\in\mathcal D$, so taking the infimum over $D$ yields 
\[
\mathcal J_Q(\widehat D)+c\widehat{\mathcal R}_n(\widehat D)
\le
\inf_{D\in\mathcal D}\{\mathcal J_Q(D)+c\widehat{\mathcal R}_n(D)\}
+\xi_{\rm opt}+2\epsilon_{\rm comp}+2\Delta_K(\varepsilon,\alpha).
\]
Since $c\widehat{\mathcal R}_n(\widehat D)\ge0$, we may discard this term from the left-hand side. Subtracting $\mathcal J_Q^*$ and using the definition of $\mathcal A_{\mathcal D,c}$ then yields
\[
\mathcal J_Q(\widehat D)-\mathcal J_Q^*
\le
\mathcal A_{\mathcal D,c}
+\xi_{\rm opt}+2\epsilon_{\rm comp}+2\Delta_K(\varepsilon,\alpha).
\]
Finally, \eqref{eq:excess-controls-gap} implies 
\[
\eta_{\rm proj}(\widehat D)^2
\le
\mathcal A_{\mathcal D,c}
+2\Delta_K(\varepsilon,\alpha)
+\xi_{\rm opt}+2\epsilon_{\rm comp},
\]
which proves \eqref{eq:projection-oracle}. \hfill$\square$

\subsection{Proof of Corollary~\ref{cor:metric-empirical}}\label{app:raw-empirical}

We first introduce the definitions used in this proof. The root mean squared distance to the nearest landmark, $q_2(P_n,L)$, is defined by \eqref{eq:landmark-coverage} with $P_n$ in place of $P$. The reconstruction error $e_Q(D)$ in \eqref{eq:coordinate-errors} uses the residual of the implemented output, including its numerical approximation.

Condition on the observations $Y_1,\ldots,Y_n$, the landmarks $L=(\ell_1,\ldots,\ell_k)$ and the fitted generator specified by the representation law $Q$ and reconstruction map $D$, so that $P_n$, $L$, $Q$ and $D$ are fixed. For any fixed $y_0\in\M$, the empirical law satisfies
\[
\int_\M d(y,y_0)^2\,P_n(dy)=\frac1n\sum_{i=1}^n d(Y_i,y_0)^2<\infty.
\]
Thus $P_n\in\Pcal_2(\M)$. Its landmark representation law is
\[
\mu_{L,n}=(\phi_L)_\#P_n=\frac1n\sum_{i=1}^n\delta_{\phi_L(Y_i)}.
\]
Since this law is supported on finitely many vectors in $\R^k$, it also has finite second moments. Write
\[
\delta_n=\Wp{Q,\mu_{L,n}}{\norm{\cdot}_\infty}.
\]
We take an independent coupling of $Z\sim Q$ and $Z^\star\sim\mu_{L,n}$ and apply Minkowski's inequality,
\[
\begin{aligned}
\delta_n
&\le\left\{\E\norm{Z-Z^\star}_\infty^2\right\}^{1/2}\le\left\{\E_Q\norm Z_\infty^2\right\}^{1/2}
+\left\{\frac1n\sum_{i=1}^n\norm{\phi_L(Y_i)}_\infty^2\right\}^{1/2}
<\infty.
\end{aligned}
\]
The root mean squared distance to the nearest landmark is finite as well:
\[q_2(P_n,L)=\left\{\frac1n\sum_{i=1}^n\min_{1\le j\le k}d(Y_i,\ell_j)^2\right\}^{1/2}<\infty.\]
We then apply Theorem~\ref{thm:transfer} with $P_n$ as the target law
\[
\Wp{D_\#Q,P_n}{d}\le e_Q(D)+\delta_n+2q_2(P_n,L).
\]
Since we have $\delta_n\le\Delta_n$,
\[
\Wp{D_\#Q,P_n}{d}\le e_Q(D)+\Delta_n+2q_2(P_n,L).
\]
Since $P\in\Pcal_2(\M)$, the Wasserstein triangle inequality now yields
\[
\begin{aligned}
\Wp{D_\#Q,P}{d}
&\le\Wp{D_\#Q,P_n}{d}+\Wp{P_n,P}{d}\\
&\le\Wp{P_n,P}{d}+e_Q(D)+\Delta_n+2q_2(P_n,L).
\end{aligned}
\]
This proves \eqref{eq:metric-empirical}.

\hfill$\square$

\subsection{Proof of Theorem \ref{thm:metric-consistency}} \label{app:metric-consistency}

Condition on the observations and fitting randomness. Since $Q_n\in\Pcal_2(\R^{k_n})$ and the empirical representation law has finite support, $\delta_n$ is finite. Corollary~\ref{cor:metric-empirical} therefore applies with $Q=Q_n$, $D=D_n$, $L=L_n$ and $\Delta_n=\delta_n$. Substituting \eqref{eq:medoid-approximation} into \eqref{eq:metric-empirical} implies \eqref{eq:medoid-generator-bound}.

The following lemma provides the approximation and sampling bounds used in the proof.
\begin{lemma}\label{lem:medoid-approximation}
Under Assumption~\ref{ass:object-law}, $\varepsilon_k(P)\to0$ in \eqref{eq:quantization-error}. For any finite nonempty $A$ and $P,P'\in\Pcal_2(\M)$,
\[|q_2(P,A)-q_2(P',A)|\le\Wp{P,P'}{d}.\]
If $L$ consists of $k\le n$ empirical medoids of an empirical law $P_n$, then the bound \eqref{eq:medoid-approximation} holds.
For iid observations from $P$, $\Wp{P_n,P}{d}\to0$ almost surely and in mean.
\end{lemma}

For the consistency conclusions following Theorem~\ref{thm:metric-consistency}, Lemma~\ref{lem:medoid-approximation} shows that $\Wp{P_n,P}{d}\to0$ almost surely and in mean, while $\varepsilon_{k_n}(P)\to0$ for deterministic $k_n\to\infty$. The bound then implies  convergence in probability or almost surely whenever the fitting errors vanish in the corresponding mode. If $\E\{e_{Q_n}(D_n)+\delta_n\}\to0$, taking expectations in \eqref{eq:medoid-generator-bound} also yields convergence in mean.\qed

\subsection{Proof of Theorem \ref{thm:noncompact-rates}} \label{app:metric-rate-proofs}
Throughout the proof, $C$ denotes a positive constant that may change from line to line. The condition $\E\{e_Q(D)+\delta_n\}<\infty$ implies that $e_Q(D)$ and $\delta_n$ are finite almost surely. Since $\mu_{L,n}$ has finite support, it has a finite second moment. We apply Minkowski's inequality to couplings of $Q$ and $\mu_{L,n}$,
\[
\left\{\E_Q\norm Z_\infty^2\right\}^{1/2}\le\delta_n+\left\{\E_{\mu_{L,n}}\norm Z_\infty^2\right\}^{1/2}<\infty.
\]
Thus $Q\in\Pcal_2(\R^k)$ almost surely, and the finite-sample bound \eqref{eq:medoid-generator-bound} applies:
\[
\Wp{D_\#Q,P}{d}\le5\Wp{P_n,P}{d}+4\varepsilon_k(P)+e_Q(D)+\delta_n,
\]
where $\varepsilon_k(P)$ is defined in \eqref{eq:quantization-error}. The following lemma bounds the sampling and landmark approximation errors.
\begin{lemma}\label{lem:metric-approximation-rates}
Suppose Assumptions~\ref{ass:object-law} and~\ref{ass:metric-rates} hold. Let $Y_1,\ldots,Y_n$ be iid with law $P$ and let $P_n=n^{-1}\sum_{i=1}^n\delta_{Y_i}$. There exists a constant $C$, depending only on $C_0,s,q,M_q$, such that, for every $k,n\ge1$,
\[
\varepsilon_k(P)\le Ck^{-1/s},
\qquad
\E\Wp{P_n,P}{d}\le Cr_n(s),
\]
where
\[
r_n(s)=
\begin{cases}
n^{-1/4},&s<4,\\
n^{-1/4}\{\log(1+n)\}^{1/2},&s=4,\\
n^{-1/s},&s>4.
\end{cases}
\]
\end{lemma}
Taking expectations and applying Lemma~\ref{lem:metric-approximation-rates} leads to 
\[
\E\Wp{D_\#Q,P}{d}\le C\{r_n(s)+k^{-1/s}\}+\E\{e_Q(D)+\delta_n\}.
\]\hfill$\square$

\subsection{Proof of Lemma \ref{lem:medoid-approximation}}\label{lem_proof:medoid-approximation}
We first show that the quantization error $\varepsilon_k(P)$ defined in \eqref{eq:quantization-error} tends to zero. Fix $o\in\M$ such that $\E_Pd(Y,o)^2<\infty$. Since $\M$ is separable, it contains a countable dense subset $\{z_1,z_2,\ldots\}$. Set $x_1=o$ and $x_{j+1}=z_j$ for $j\ge1$, and define $B_k=\{x_1,\ldots,x_k\}.$ Moreover, $B_k\subseteq B_{k+1}$, so
\[
\operatorname{dist}(y,B_{k+1})\le\operatorname{dist}(y,B_k)
\]
for every $y\in\M$.

To identify the limit, fix $y\in\M$ and $\epsilon>0$, we can always find $z_j$ satisfying $d(y,z_j)<\epsilon$. For every $k\ge j+1$, this point belongs to $B_k$, and therefore
\[
\operatorname{dist}(y,B_k)\le d(y,z_j)<\epsilon.
\]
Since $\epsilon>0$ is arbitrary, $\operatorname{dist}(y,B_k)\to0$ for every $y\in\M$. Also, $o\in B_k$ for every $k$, so
\[
0\le\operatorname{dist}(y,B_k)^2\le d(y,o)^2.
\]
The upper bound is integrable under $P$. Then by dominated convergence
\[
\E_P\operatorname{dist}(Y,B_k)^2\longrightarrow0.
\]
By the definition of $\varepsilon_k(P)$ as the infimum over all nonempty sets with at most $k$ points,
\[
0\le\varepsilon_k(P)\le q_2(P,B_k)
=\left\{\E_P\operatorname{dist}(Y,B_k)^2\right\}^{1/2}\longrightarrow0.
\]

We next prove the bound for $q_2(P,A)$, where $A\subset\M$ is any finite nonempty set. For $y,y'\in\M$, choose $a'\in A$ nearest to $y'$. Then
\[
\operatorname{dist}(y,A)=\min_{a\in A}d(y,a)\le d(y,a')\le d(y,y')+d(y',a')
=d(y,y')+\operatorname{dist}(y',A).
\]
Interchanging $y$ and $y'$ leads to 
\[
|\operatorname{dist}(y,A)-\operatorname{dist}(y',A)|\le d(y,y').
\]

Let $(Y,Y')$ have any coupling of $P$ and $P'$. Applying the following pointwise bound and Minkowski's inequality yields
\[
\begin{aligned}
q_2(P,A)
&=\left\{\E\operatorname{dist}(Y,A)^2\right\}^{1/2}\\
&\le\left\{\E\bigl(\operatorname{dist}(Y',A)+d(Y,Y')\bigr)^2\right\}^{1/2}\\
&\le\left\{\E\operatorname{dist}(Y',A)^2\right\}^{1/2}
+\left\{\E d(Y,Y')^2\right\}^{1/2}\\
&=q_2(P',A)+\left\{\E d(Y,Y')^2\right\}^{1/2}.
\end{aligned}
\]
The same argument with $P$ and $P'$ interchanged leads to 
\[
|q_2(P,A)-q_2(P',A)|\le\left\{\E d(Y,Y')^2\right\}^{1/2}.
\]
This holds for every coupling of $P$ and $P'$. Taking the infimum over these couplings proves
\[
|q_2(P,A)-q_2(P',A)|\le\Wp{P,P'}{d}.
\]

We now establish the empirical medoid bound. Fix a realized sample $Y_1,\ldots,Y_n$ and an integer $1\le k\le n$. Let $A=\{a_1,\ldots,a_r\}$ be any nonempty set of centers in $\M$, with $r\le k$. Assign each observation to a nearest center, choosing the smallest center index in a tie. In this case, for each observation, it belongs to exactly one group. Denote the resulting index sets by $I_1,\ldots,I_r$. Thus, for $i\in I_j$, $d(Y_i,a_j)=\operatorname{dist}(Y_i,A).$

For every nonempty set $I_j$, let $m_j=|I_j|$. We choose an index $i_j\in I_j$ minimizing $d(Y_i,a_j)^2$ within that set, and set $b_j=Y_{i_j}$. Then $d(b_j,a_j)^2\le\frac1{m_j}\sum_{i\in I_j}d(Y_i,a_j)^2.$ For each $i\in I_j$, the triangle inequality and $(u+v)^2\le2u^2+2v^2$ imply
\[
d(Y_i,b_j)^2\le\{d(Y_i,a_j)+d(a_j,b_j)\}^2
\le2d(Y_i,a_j)^2+2d(a_j,b_j)^2.
\]
Summing over the set and using the choice of $b_j$ leads to 
\[
\begin{aligned}
\sum_{i\in I_j}d(Y_i,b_j)^2
&\le2\sum_{i\in I_j}d(Y_i,a_j)^2+2m_jd(a_j,b_j)^2\le4\sum_{i\in I_j}d(Y_i,a_j)^2.
\end{aligned}
\]

Let $B$ consist of the selected observations $b_j$ from the nonempty sets. It contains at most $r\le k$ points. For an observation assigned to set $I_j$, its distance to $B$ is no greater than its distance to $b_j$. Consequently,
\[
\begin{aligned}
q_2(P_n,B)^2
&=\frac1n\sum_{i=1}^n\operatorname{dist}(Y_i,B)^2\\
&\le\frac1n\sum_{j:I_j\ne\varnothing}\sum_{i\in I_j}d(Y_i,b_j)^2\\
&\le\frac4n\sum_{j:I_j\ne\varnothing}\sum_{i\in I_j}d(Y_i,a_j)^2=4q_2(P_n,A)^2.
\end{aligned}
\]
If fewer than $k$ observation indices have been selected, we can simply add further indices until there are $k$. Since $L$ is an empirical medoid selection, it minimizes the squared approximation error over all selections of $k$ observation indices. Its error is therefore no larger than that of the selection just constructed. Taking square roots one finds
\[
q_2(P_n,L)\le2q_2(P_n,A).
\]
Applying the bound with $P_n$ and $P$ yields
\[
q_2(P_n,L)\le2\left\{q_2(P,A)+\Wp{P_n,P}{d}\right\}.
\]
The left hand side does not depend on the arbitrary set $A$. Taking the infimum over all nonempty $A\subset\M$ with $|A|\le k$ therefore leads to 
\[
q_2(P_n,L)\le2\left\{\varepsilon_k(P)+\Wp{P_n,P}{d}\right\},
\]
which proves \eqref{eq:medoid-approximation}.

Finally, suppose that $Y_1,Y_2,\ldots$ are iid with law $P$. We prove that $\Wp{P_n,P}{d}\to0$ almost surely and in mean. Fix $\epsilon>0$. The convergence $q_2(P,B_k)\to0$ established at the beginning of the proof allows us to choose a finite nonempty set $A=B_k=\{a_1,\ldots,a_r\}$ such that $q_2(P,A)<\epsilon$. Define $a(y)$ to be a nearest point to $y$ in $A$. The coupling $(Y,a(Y))$ leads to 
\[
\Wp{P,a_\#P}{d}\le\left\{\E_Pd(Y,a(Y))^2\right\}^{1/2}=q_2(P,A).
\]
Similarly, coupling each observation with its selected representative,
\[
\Wp{P_n,a_\#P_n}{d}\le\left\{\frac1n\sum_{i=1}^nd(Y_i,a(Y_i))^2\right\}^{1/2}=q_2(P_n,A).
\]
The Wasserstein triangle inequality consequently yields
\[
\Wp{P_n,P}{d}\le q_2(P_n,A)+\Wp{a_\#P_n,a_\#P}{d}+q_2(P,A).
\]

We first examine the empirical approximation term. Since $\operatorname{dist}(Y,A)^2$ is integrable, the strong law of large numbers implies
\[
q_2(P_n,A)^2=\frac1n\sum_{i=1}^n\operatorname{dist}(Y_i,A)^2
\longrightarrow\E_P\operatorname{dist}(Y,A)^2=q_2(P,A)^2
\]
almost surely. Taking square roots shows that $q_2(P_n,A)\to q_2(P,A)$ almost surely. For $j=1,\ldots,r$, write
\[
p_j=P\{a(Y)=a_j\},\qquad
\widehat p_{n,j}=\frac1n\sum_{i=1}^n\mathbf1\{a(Y_i)=a_j\}.
\]
These are the masses of $a_\#P$ and $a_\#P_n$ at $a_j$. The strong law implies $\widehat p_{n,j}\to p_j$ almost surely for every $j$. Since $A$ is finite, these convergences hold simultaneously almost surely.

To construct a transport plan, keep the common mass $\min(\widehat p_{n,j},p_j)$ at each $a_j$. The total remaining mass is
\[
1-\sum_{j=1}^r\min(\widehat p_{n,j},p_j)
=\frac12\sum_{j=1}^r|\widehat p_{n,j}-p_j|.
\]
Since both the starting and destination points belong to $A$, the distance over which any remaining mass is transported is at most $\diam(A)$. So the squared transport cost is bounded by
\begin{align}\label{eq:bound_lem1}
\Wp{a_\#P_n,a_\#P}{d}^{\,2}
\le\frac{\diam(A)^2}{2}\sum_{j=1}^r|\widehat p_{n,j}-p_j|
\longrightarrow0
\end{align}
almost surely. Combining this convergence with the triangle inequality leads to 
\[
\limsup_{n\to\infty}\Wp{P_n,P}{d}\le2q_2(P,A)<2\epsilon
\]
almost surely.

We choose a deterministic sequence $\epsilon_m\downarrow0$ and, for each $m$, a finite set $A_m$ satisfying $q_2(P,A_m)<\epsilon_m$. For each $m$, the previous strong law argument leads to an event $E_m$ of probability one corresponding to $\limsup_{n\to\infty}\Wp{P_n,P}{d}\le2\epsilon_m$. The countable intersection $E=\bigcap_{m=1}^{\infty}E_m$ also has probability one. On $E$, this bound holds for every $m$, and letting $m\to\infty$ proves $\Wp{P_n,P}{d}\to0$ almost surely.

Since both $a_\#P_n$ and $a_\#P$ are supported on $A$,
\[
0\le\Wp{a_\#P_n,a_\#P}{d}\le\diam(A).
\]
Using the almost sure convergence established in \eqref{eq:bound_lem1} and the bound $\Wp{a_\#P_n,a_\#P}{d}\le\diam(A)$, we apply the dominated convergence theorem to obtain
\[
\E\Wp{a_\#P_n,a_\#P}{d}\longrightarrow0.
\]
Furthermore, we use Jensen's inequality for the square root
\[
\begin{aligned}
\E q_2(P_n,A)
&=\E\left\{\frac1n\sum_{i=1}^n\operatorname{dist}(Y_i,A)^2\right\}^{1/2}\le\left\{\frac1n\sum_{i=1}^n\E\operatorname{dist}(Y_i,A)^2\right\}^{1/2}=q_2(P,A).
\end{aligned}
\]
Taking expectations and then taking the limit superior on both sides, we have
\[
\limsup_{n\to\infty}\E\Wp{P_n,P}{d}\le2q_2(P,A)<2\epsilon.
\]
Since $\epsilon>0$ is arbitrary, $\E\Wp{P_n,P}{d}\to0$. This completes the proof.\hfill$\square$

\subsection{Proof of Lemma \ref{lem:metric-approximation-rates}}\label{lem_proof:metric-approximation-rates}
Throughout the proof, $C$ denotes a positive constant that may change from line to line. We first partition $\supp P$ into the disjoint Borel sets
\[
A_0=\supp P\cap B_d(o,1),\qquad A_j=\supp P\cap\{y:2^{j-1}<d(y,o)\le2^j\},\quad j\ge1.
\]
Since $(\M,d)$ is Polish, $\supp P$ has probability one. We write $p_j=P(A_j),$ apply Markov's inequality and the moment assumption
\[
p_j\le P\{d(Y,o)>2^{j-1}\}\le M_q2^{-(j-1)q},\quad j\ge1.
\]
Together with $p_0\le1$, this yields
\[
p_j\le C(1+M_q)2^{-jq},\quad j\ge0.
\]
For $0<\epsilon\le2^j$, cover $\supp P\cap B_d(o,2^j)$ by balls of radius $\epsilon$. Discard balls that do not meet $A_j$ and replace each remaining center by a point in its ball's intersection with $A_j$. The triangle inequality shows that the new balls of radius $2\epsilon$ cover $A_j$. Hence, by Assumption~\ref{ass:metric-rates},
\[
N(A_j,2\epsilon;d)\le N\bigl(\supp P\cap B_d(o,2^j),\epsilon;d\bigr)\le C_0(2^j/\epsilon)^s.
\]

For each nonempty $A_j$ and integer $m\ge C_0$, choosing $\epsilon=2^j(C_0/m)^{1/s}\le2^j$ yields  at most $m$ centers in $A_j$ with covering radius $2C_0^{1/s}2^jm^{-1/s}$. For $1\le m<C_0$, one point in $A_j$ suffices after enlarging the constant, since $\diam(A_j)\le2^{j+1}$. Thus, for every integer $m\ge1$, there is a set $B_{j,m}\subset A_j$ with $|B_{j,m}|\le m$ and $\operatorname{dist}(y,B_{j,m})\le C2^jm^{-1/s}$ for all $y\in A_j$.

Choose $a$ satisfying $0<a<\frac{s(q-2)}2,$ and set $c=(1-2^{-a})/2$. For sufficiently large $k$, allocate $m_j=\lfloor ck2^{-aj}\rfloor$ centers to set $A_j$ whenever $m_j\ge1$. The total allocation satisfies
\[
\sum_{j\ge0}m_j\le ck\sum_{j\ge0}2^{-aj}=\frac{k}{2}.
\]

Let $J_k=\max\{j\ge0:ck2^{-aj}\ge1\}$. The definition implies $ck2^{-aJ_k}\ge1$ and $ck2^{-a(J_k+1)}<1$, and therefore $(ck)^{1/a}/2<2^{J_k}\le(ck)^{1/a}$. Also, $\lfloor x\rfloor\ge x/2$ for $x\ge1$, so $m_j\ge ck2^{-aj}/2$ for $0\le j\le J_k$.

For each nonempty allocated set, choose $B_{j,m_j}$ and set
\[
B_k=\{o\}\cup\bigcup_{\substack{0\le j\le J_k\\A_j\ne\varnothing}}B_{j,m_j}.
\]
Since $|B_{j,m_j}|\le m_j$ and $\sum_{j=0}^{J_k}m_j\le k/2$, the union together with $o$ contains at most $1+k/2\le k$ points for $k\ge2$. For $j\le J_k$,
\[
\begin{aligned}
\int_{A_j}\operatorname{dist}(y,B_k)^2\,P(dy)\le Cp_j2^{2j}m_j^{-2/s}\le Ck^{-2/s}p_j2^{2j+2aj/s}\le Ck^{-2/s}2^{j(2+2a/s-q)}.
\end{aligned}
\]
Since $2+2a/s-q<0$, summing over the allocated sets leads to 
\[
\sum_{j=0}^{J_k}\int_{A_j}\operatorname{dist}(y,B_k)^2\,P(dy)
\le Ck^{-2/s}\sum_{j=0}^{J_k}2^{j(2+2a/s-q)}
\le Ck^{-2/s}.
\]

On the remaining sets, since their union is contained in $\{y:d(y,o)>2^{J_k}\}$,
\[
\begin{aligned}
\sum_{j>J_k}\int_{A_j}\operatorname{dist}(y,B_k)^2\,P(dy)
&\le\E\left[d(Y,o)^2\mathbf1\{d(Y,o)>2^{J_k}\}\right]\le M_q2^{-J_k(q-2)}\le Ck^{-(q-2)/a}.
\end{aligned}
\]
The second inequality uses $d(Y,o)^2\le d(Y,o)^q2^{-J_k(q-2)}$ on the event $\{d(Y,o)>2^{J_k}\}$. Since $(q-2)/a>2/s$, this tail contribution is bounded by $Ck^{-2/s}$. Consequently we have,
\[
\varepsilon_k(P)^2\le\E_P\operatorname{dist}(Y,B_k)^2\le Ck^{-2/s}.
\]
Taking square roots proves the bound for sufficiently large $k$. For $1\le k<k_0$, the set $\{o\}$ has $\varepsilon_k(P)\le\{\E d(Y,o)^2\}^{1/2}\le M_q^{1/q}$. Since $M_q^{1/q}\le M_q^{1/q}k_0^{1/s}k^{-1/s}$ for these values of $k$, increasing $C$ to at least $M_q^{1/q}k_0^{1/s}$ establishes the bound for all $k\ge1$.

For the empirical Wasserstein bound, we combine the metric partition argument of \citet[Proposition~1]{weed2019sharp} with the shell decomposition and moment estimates of \citet[Sections~2--3]{fournier2015rate}. Fix an integer $H\ge1$. For every nonempty $A_j$ and each $h=1,\ldots,H$, the covering bound supplies a cover with centers in $A_j$, radius $r_{j,h}=2^j4^{-h}$ and at most $C2^{2sh}$ centers.

We assign each point of $A_j$ to a nearest center at level $H$. For $h=1,\ldots,H-1$, we assign each level-$(h+1)$ center to the nearest level-$h$ center. These groups form a Borel partition of $A_j$. Denote this partition, after empty sets are discarded, by $\mathcal E_{j,h}$, and set $\mathcal E_{j,0}=\{A_j\}$. Each partition set at level $h+1$ is contained in a unique parent set at level $h$.

Since each nonempty partition set corresponds to a center, $|\mathcal E_{j,h}|\le C2^{2sh}$. For a point belonging to a partition set indexed by a level-$h$ center, the distance to that center is at most $\sum_{\ell=h}^{H}r_{j,\ell}\le\frac43r_{j,h}$. Any two points in the same partition set are therefore at distance at most $\frac83r_{j,h}$ from each other. Consequently,
\[
\max_{E\in\mathcal E_{j,h}}\diam(E)\le C2^j4^{-h}.
\]
The single set at level zero is $A_j$, which satisfies $\diam(A_j)\le2^{j+1}$.

For any two probability measures $\mu,\nu$ supported on $A_j$, these partitions imply
\begin{align}\label{eq:lem2_W2}
\Wp{\mu,\nu}{d}^{\,2}\le C2^{2j}\left[2^{-4H}+\sum_{h=1}^{H}2^{-4h}\sum_{E\in\mathcal E_{j,h}}|\mu(E)-\nu(E)|\right].
\end{align}
To verify this inequality, we first match mass $\min\{\mu(E),\nu(E)\}$ inside each finest-level partition set $E\in\mathcal E_{j,H}$. Every such transport has both endpoints in $E$, so its squared distance is at most $\diam(E)^2$. Since the total mass matched is at most one, the total cost is bounded by $C2^{2j}2^{-4H}$.

We then proceed from finer to coarser partitions. After matching within a partition set $E$, its remaining signed mass is $\mu(E)-\nu(E)$: a positive value represents surplus mass from $\mu$, while a negative value represents mass still required by $\nu$. Within each parent set at level $h-1$, we match surplus and deficit masses from its child sets at level $h$. Summing over all parent sets, the total mass matched at this stage is therefore at most
\[
\frac12\sum_{E\in\mathcal E_{j,h}}|\mu(E)-\nu(E)|.
\]
Every such transport stays inside a level-$(h-1)$ parent set, whose squared diameter is at most $C2^{2j}4^{-2(h-1)}$. Since $4^{-2(h-1)}=16 \times 2^{-4h}$, absorbing the fixed factor into $C$ bounds the cost at this stage by
\[
C2^{2j}2^{-4h}\sum_{E\in\mathcal E_{j,h}}|\mu(E)-\nu(E)|.
\]
At level zero, all remaining mass is matched within $A_j$, because $\mu(A_j)=\nu(A_j)=1$. Adding the costs over all levels leads to a transport plan with the stated bound, and we prove the inequality for $\Wp{\mu,\nu}{d}^{\,2}$ in \eqref{eq:lem2_W2}.

Write $\widehat p_j=P_n(A_j)$ and $b_j=\min\{p_j,\widehat p_j\}$. When $p_j>0$ and $\widehat p_j>0$, define the conditional probability measures given $A_j$ by
\[
P_j(E)=P(E\mid A_j)=\frac{P(E\cap A_j)}{p_j},\qquad
P_{n,j}(E)=P_n(E\mid A_j)=\frac{P_n(E\cap A_j)}{\widehat p_j}.
\]
Within $A_j$, match mass $b_j$ using the preceding transport construction for $P_j$ and $P_{n,j}$. If either $p_j$ or $\widehat p_j$ is zero, there is no common mass to match within $A_j$. When both masses are positive,
\[
b_j\sum_{E\in\mathcal E_{j,h}}\left|\frac{P_n(E)}{\widehat p_j}-\frac{P(E)}{p_j}\right|
\le\sum_{E\in\mathcal E_{j,h}}|P_n(E)-P(E)|+|\widehat p_j-p_j|.
\]
For example, if $\widehat p_j\le p_j$, the left-hand side is
\[
\sum_{E\in\mathcal E_{j,h}}\left|P_n(E)-\frac{\widehat p_j}{p_j}P(E)\right|,
\]
which is bounded by
\[
\begin{aligned}
&\sum_{E\in\mathcal E_{j,h}}|P_n(E)-P(E)|
+\left(1-\frac{\widehat p_j}{p_j}\right)\sum_{E\in\mathcal E_{j,h}}P(E)\\
&\qquad=\sum_{E\in\mathcal E_{j,h}}|P_n(E)-P(E)|+p_j-\widehat p_j.
\end{aligned}
\]
The other case follows by symmetry.

After matching the common mass within each set $A_j$, the remaining mass in $A_j$ is $|\widehat p_j-p_j|$, belonging to the measure that assigns more mass to $A_j$. The two remaining measures have the same total mass and can be coupled, for example by their normalized product when that mass is positive. Since
\[
d(x,y)^2\le2d(x,o)^2+2d(y,o)^2,
\]
the cost of this remaining transport is at most $C\sum_{j\ge0}2^{2j}|\widehat p_j-p_j|.$

Combining this cost with the transport bounds within the sets $A_j$ and using $\sum_{h=1}^{H}2^{-4h}\le C$, we obtain
\[
\begin{aligned}
\Wp{P_n,P}{d}^{\,2}
&\le C\sum_{j\ge0}2^{2j}\Bigg[(p_j+\widehat p_j)2^{-4H}+|\widehat p_j-p_j|\\
&\qquad\qquad\qquad+\sum_{h=1}^{H}2^{-4h}\sum_{E\in\mathcal E_{j,h}}|P_n(E)-P(E)|\Bigg].
\end{aligned}
\]
The countable sum of the couplings within the sets $A_j$ and the coupling of the remaining measures has marginals $P_n$ and $P$. Finite second moments ensure that the cost of transporting the remaining mass is finite almost surely.

The partitions constructed from the covers are chosen independently of the observations. For each partition set $E$, the empirical mass $P_n(E)$ has variance $P(E)\{1-P(E)\}/n$. Hence
\[
\E|P_n(E)-P(E)|\le\left\{\frac{P(E)\{1-P(E)\}}{n}\right\}^{1/2}\le\sqrt{\frac{P(E)}{n}}.
\]
By Cauchy--Schwarz and the bound on the number of partition sets,
\[
\begin{aligned}
\sum_{E\in\mathcal E_{j,h}}\E|P_n(E)-P(E)|
&\le\frac1{\sqrt n}\sum_{E\in\mathcal E_{j,h}}\sqrt{P(E)}\le\sqrt{\frac{|\mathcal E_{j,h}|}{n}\sum_{E\in\mathcal E_{j,h}}P(E)}\le C2^{sh}\sqrt{\frac{p_j}{n}}.
\end{aligned}
\]
Likewise,
\[
\E\widehat p_j=p_j,\qquad \E|\widehat p_j-p_j|\le\sqrt{\frac{p_j}{n}}.
\]

Taking expectations in the transport bound and applying Tonelli's theorem yields
\[
\E\Wp{P_n,P}{d}^{\,2}\le C\sum_{j\ge0}2^{2j}\left[p_j2^{-4H}+\sqrt{\frac{p_j}{n}}\left\{1+\sum_{h=1}^{H}2^{(s-4)h}\right\}\right].
\]
The bound on $p_j=P(A_j)$ implies 
\[
\sum_{j\ge0}2^{2j}p_j\le C\sum_{j\ge0}2^{-j(q-2)}<\infty
\]
and
\[
\sum_{j\ge0}2^{2j}\sqrt{p_j}\le C\sum_{j\ge0}2^{-j(q/2-2)}<\infty.
\]
The second series converges because $q>4$. Consequently, for every integer $H\ge1$,
\[
\E\Wp{P_n,P}{d}^{\,2}\le C\left[2^{-4H}+n^{-1/2}\left\{1+\sum_{h=1}^{H}2^{(s-4)h}\right\}\right].
\]

The resulting rate is determined by the growth of the geometric sum $\sum_{h=1}^{H}2^{(s-4)h}$. For $s<4$, we let $H\to\infty$ to eliminate the approximation term $2^{-4H}$. For $s\ge4$ and $n\ge2$, we choose $H=\lceil(\log_2 n)/(2s)\rceil$. At $s=4$, the sum equals $H$ with a logarithmic factor, while $2^{-4H}\le n^{-1/2}$. For $s>4$, the sum is bounded by $C2^{(s-4)H}$ and $n^{-1/2}2^{(s-4)H}\le Cn^{-2/s}$. Consequently,
\[
\E\Wp{P_n,P}{d}^{\,2}\le C
\begin{cases}
n^{-1/2}, & s<4,\\
n^{-1/2}\log(1+n), & s=4,\\
n^{-2/s}, & s>4.
\end{cases}
\]

In all three cases, Jensen's inequality leads to 
\[
\E\Wp{P_n,P}{d}\le\left\{\E\Wp{P_n,P}{d}^{\,2}\right\}^{1/2}\le Cr_n(s).
\]\hfill$\square$

\section{Verification of weighted-mean continuity}\label{app:geometry-examples}

This section verifies Assumption~\ref{ass:decoder-geometry} for the constructions in Section~\ref{sec:geometry-examples} by establishing existence and continuity of the weighted mean map on the corresponding allowed weight sets $\Omega_B$.
\subsection{Pairwise means in geodesic spaces}\label{sec:pairwise-means}

\begin{proposition}\label{prop:pairwise-means}
Let $(\M,d)$ be a geodesic space and $B=(b_1,\ldots,b_r)$ a finite reference tuple. For each pair in a nonempty collection $E_B\subseteq\{(i,j):i<j\}$, fix a constant-speed geodesic $\gamma_{ij}:[0,1]\to\M$ from $b_i$ to $b_j$. With $e_j$ denoting the $j$th standard basis vector in $\R^r$, define
\[
\Omega_B=\bigcup_{(i,j)\in E_B}\{(1-t)e_i+te_j:0\le t\le1\},\qquad b_B\{(1-t)e_i+te_j\}=\gamma_{ij}(t).
\]
Then $b_B$ is a continuous selection of weighted Fr\'echet means, so Assumption~\ref{ass:decoder-geometry} holds.
\end{proposition}

\begin{proof}
For any $x\in\M$ and $t\in[0,1]$,
\[
(1-t)d(x,b_i)^2+td(x,b_j)^2\ge t(1-t)\{d(x,b_i)+d(x,b_j)\}^2\ge t(1-t)d(b_i,b_j)^2.
\]
Equality holds at $x=\gamma_{ij}(t)$, since its distances to $b_i$ and $b_j$ are $t\,d(b_i,b_j)$ and $(1-t)d(b_i,b_j)$. Thus the selected point is a weighted mean. The selections are continuous on each simplex edge and agree at shared vertices. Since these edges form a finite closed cover of $\Omega_B$, the pasting lemma proves continuity.
\end{proof}

\subsection{Hadamard spaces}\label{sec:hadamard-example}

Suppose $\M$ is a nonempty closed geodesically convex subset of a Hadamard space. For every simplex weight vector, the weighted Fr\'echet mean exists uniquely. The barycenter contraction inequality \citep[Theorem~6.3]{sturm2003barycenters} implies 
\begin{equation}\label{eq:prototype-contraction}
d\{b_B(w),b_B(w')\}\le W_1\!\left(\sum_jw_j\delta_{b_j},\sum_jw'_j\delta_{b_j};d\right)\le\frac{R_B}{2}\norm{w-w'}_1,
\end{equation}
where $R_B=\max_{i,j}d(b_i,b_j)$. The second inequality follows by keeping common mass fixed and transporting the remaining mass over distance at most $R_B$. Hence $b_B$ is continuous on the simplex and on any allowed subset $\Omega_B$, verifying Assumption~\ref{ass:decoder-geometry}.

\subsection{Images with pixelwise and multiscale distances}\label{app:image-verification}

Let $\M=[0,1]^p$ and $d(I,I')=\norm{A(I-I')}_2$, where $A$ is an injective linear map. This includes root mean squared pixel distance and the multiscale distance in \eqref{eq:multiscale-image-metric}. For $\bar b_w=\sum_jw_jb_j$,
\[
\sum_jw_j\norm{A(x-b_j)}_2^2=\norm{A(x-\bar b_w)}_2^2+\sum_jw_j\norm{A(\bar b_w-b_j)}_2^2.
\]
Since $\bar b_w\in\M$ and $A$ is injective, the unique weighted mean is $b_B(w)=\bar b_w$. Its continuity verifies Assumption~\ref{ass:decoder-geometry}.

\subsection{Symmetric positive-definite matrices}\label{app:spd-verification}

Let $\mathcal S_m^+$ denote the positive-definite symmetric matrices of size $m\times m$. Under Frobenius distance, the unique weighted mean is $b_B(w)=\sum_jw_jb_j$. It remains positive definite because $\lambda_{\min}\{b_B(w)\}\ge\min_j\lambda_{\min}(b_j)>0$, and it depends continuously on $w$.

For Log-Euclidean distance, $\Psi_{\mathrm{LE}}(A)=\log A$ is a bijective isometry onto the symmetric matrices with Frobenius norm \citep{arsigny2006logeuclidean}. For Log-Cholesky distance, write $A=R_AR_A^{\mathsf T}$ with $R_A$ lower triangular and positive on the diagonal. The map $\Psi_{\mathrm{LC}}$ collecting its strictly lower triangular entries and logarithmic diagonal entries is a bijective isometry onto Euclidean space \citep{lin2019cholesky}. For either coordinate map $\Psi$, the unique weighted mean is
\[
b_B(w)=\Psi^{-1}\left\{\sum_jw_j\Psi(b_j)\right\}.
\]
Continuity of $\Psi^{-1}$ implies  continuity of $b_B$. Under the affine-invariant metric, $\mathcal S_m^+$ is a Hadamard manifold \citep{pennec2006tensor}, so existence, uniqueness and continuity follow from Section~\ref{sec:hadamard-example}. Thus Assumption~\ref{ass:decoder-geometry} holds for each of these metrics.

\subsection{Brain connectivity matrices}\label{app:matrix-verification}

For full rank correlation matrices, use the normalized Cholesky coordinate isometry $\Psi$ from Example~\ref{sec:spd-examples}. Under the corresponding metric, the unique weighted mean is
\[
b_B(w)=\Psi^{-1}\left\{\sum_jw_j\Psi(b_j)\right\}.
\]
The coordinate average depends continuously on $w$, and $\Psi^{-1}$ is continuous. Hence Assumption~\ref{ass:decoder-geometry} holds.

\subsection{The sphere with great-circle distance}\label{sec:sphere-geometry}

Let $\M=\mathbb S^2$ with geodesic distance $d(x,y)=\arccos\langle x,y\rangle$, and let $\Omega_B$ consist of weight vectors supported within one of the local reference groups in Section~\ref{app:sphere-implementation}. Each group is contained in a geodesic ball of radius less than $\pi/4$ about its central vertex. For each $w\in\Omega_B$, the objective
\[
F_w(x)=\sum_{j=1}^r w_jd(x,b_j)^2
\]
is continuous on the compact sphere and therefore attains its minimum. The radius condition implies uniqueness of this global minimizer for every allowed weight vector \citep[Theorem~2.6]{afsari2011gradient}. Denote it by $b_B(w)$. We show that $b_B$ is continuous on $\Omega_B$.

Let $w_n\to w$ in $\Omega_B$ and write $x_n=b_B(w_n)$. Consider any convergent subsequence $x_{n_\ell}\to x_*$. Such subsequences exist by compactness of $\mathbb S^2$. Since $x_{n_\ell}$ is a global minimizer, for every $y\in\mathbb S^2$,
\[
F_{w_{n_\ell}}(x_{n_\ell})\le F_{w_{n_\ell}}(y).
\]
Joint continuity of $(w,x)\mapsto F_w(x)$ allows us to pass to the limit and obtain
\[
F_w(x_*)\le F_w(y),\qquad y\in\mathbb S^2.
\]
Thus $x_*$ minimizes $F_w$, and uniqueness implies $x_*=b_B(w)$. If $x_n$ did not converge to $b_B(w)$, some subsequence would remain at distance at least $\epsilon>0$ from $b_B(w)$. Compactness would give a contradiction. Therefore $b_B(w_n)\to b_B(w)$, proving continuity and verifying Assumption~\ref{ass:decoder-geometry}.

\subsection{Finite metric trees}\label{sec:finite-tree-geometry}

A finite metric tree with positive edge lengths is a Hadamard space. Its weighted Fr\'echet means therefore exist uniquely and depend continuously on the weights by Section~\ref{sec:hadamard-example}, verifying Assumption~\ref{ass:decoder-geometry}.

\subsection{Locations on a metric graph}\label{app:metric-graph-verification}

For the graph in Section~\ref{sec:graph-tree-experiments}, let $B$ contain its vertices and restrict $\Omega_B$ to the simplex edges corresponding to graph edges. Provided each graph edge is a shortest path between its endpoints, Proposition~\ref{prop:pairwise-means} supplies a continuous selection of weighted means along these edges. Thus Assumption~\ref{ass:decoder-geometry} holds for this construction.

\subsection{Phylogenetic trees}\label{app:phylogenetic-verification}

Consider the five-taxon tree domain in Example~\ref{sec:branching-object-examples}, with fixed terminal lengths and two internal lengths in $[0,1]$. Use the corner references of each topology's branch length square and allow weights supported on the four corners of one square. Each square is a geodesically convex Euclidean subset of BHV space, so the weighted mean of its corners is their Euclidean weighted average and remains in the specified domain. Since BHV space is Hadamard, its weighted mean map is unique and continuous. Restricting that map to the allowed weights verifies Assumption~\ref{ass:decoder-geometry}.

\subsection{A CAT(0) cubical space}\label{app:cubical-verification}

Let $\M=\mathcal T_3\times\mathcal T_3$ have the product metric from Example~\ref{sec:cubical-object-examples}, where $\mathcal T_3$ is the three-branch metric tree. This product is a Hadamard space, so its weighted means exist uniquely and depend continuously on the weights. Thus Assumption~\ref{ass:decoder-geometry} holds.

\subsection{Distributions represented by point clouds}\label{app:distribution-verification}

The MERFISH objects are empirical distributions of sixteen equally weighted points in $\R^{16}$, with the Wasserstein metric in \eqref{eq:merfish-object-metric}. For finite references $B$, let $C$ be the convex hull of all reference atoms. The weighted Fr\'echet objective can consequently be minimized over the compact set $C^{16}$, where it is continuous. Thus a weighted mean exists for every simplex weight vector.

Assume additionally that the global weighted Fr\'echet mean is unique as a probability distribution for every $w\in\Omega_B$, and denote it by $b_B(w)$. Let $\mathcal K_C=\{16^{-1}\sum_{i=1}^{16}\delta_{x_i}:x_i\in C\}$. This set is compact in the Wasserstein metric because it is the continuous image of the compact set $C^{16}$. Indeed, matching atoms with the same index leads to 
\[
W_2^2\left(\frac1{16}\sum_{i=1}^{16}\delta_{x_i},\frac1{16}\sum_{i=1}^{16}\delta_{y_i}\right)\le\frac1{16}\sum_{i=1}^{16}\|x_i-y_i\|_2^2.
\]
The existence argument above and uniqueness imply that $b_B(w)\in\mathcal K_C$ for every allowed weight vector.

To prove continuity, let $w_n\to w$ in $\Omega_B$ and write $\nu_n=b_B(w_n)$. Compactness of $\mathcal K_C$ implies there is  a convergent subsequence $\nu_{n_\ell}\to\nu_*$. For every $\nu\in\mathcal K_C$, the minimizing property implies 
\[
\sum_{j=1}^r w_{n_\ell,j}W_2^2(\nu_{n_\ell},b_j)\le\sum_{j=1}^r w_{n_\ell,j}W_2^2(\nu,b_j).
\]
Passing to the limit shows that $\nu_*$ minimizes the objective over $\mathcal K_C$. Since this compact set contains a global minimizer, $\nu_*$ is also a global minimizer, and uniqueness implies $\nu_*=b_B(w)$. The same argument applies to every convergent subsequence. Compactness therefore implies $\nu_n\to b_B(w)$, proving continuity and verifying Assumption~\ref{ass:decoder-geometry}.

\section{Empirical flow and propagation of fitting error}\label{app:flow}

We establish the ideal empirical flow and control the discrepancy between its target distribution and the generated landmark vectors. All statements are conditional on the observations, selected landmarks and fitted parameters; the landmarks may be selected from the same observations used to fit the flow.

\subsection{The ideal empirical flow}\label{app:empirical-flow}

Write $Z_i^\star=\phi_L(Y_i)$ and $\mu_{L,n}=n^{-1}\sum_{i=1}^n\delta_{Z_i^\star}$. Let $Z^\star\sim\mu_{L,n}$ and let $G,H\sim N(0,I_k)$ be mutually independent and independent of $Z^\star$. For $\sigma\ge0$, define
\[
\epsilon=\frac{\sigma}{\sqrt{k}},\qquad V=Z^\star+\epsilon H,\qquad X_t=(1-t)G+tV,\qquad U=V-G,\qquad p_{n,t}=\law(X_t).
\]
For $t<1$, set
\[
s_t^2=(1-t)^2+t^2\epsilon^2,\qquad a_t=\frac{t\epsilon^2-(1-t)}{s_t^2},\qquad b_t=\frac{1-t}{s_t^2}.
\]
Gaussian conditioning leads to  $u_{n,t}(x)=\E(U\mid X_t=x)=a_tx+b_tM_{n,t}(x)$ where
\begin{align}\label{eq:empirical-reference}
&\omega_i(t,x)=\frac{\exp\{-\norm{x-tZ_i^\star}_2^2/(2s_t^2)\}}{\sum_{j=1}^n\exp\{-\norm{x-tZ_j^\star}_2^2/(2s_t^2)\}},\qquad M_{n,t}(x)=\sum_{i=1}^n\omega_i(t,x)Z_i^\star.
\end{align}

For a fitted velocity $\widehat v=v_{\widehat\theta}$, define its approximation error relative to the ideal empirical velocity by
\begin{equation}\label{eq:empirical-velocity-error}\mathcal E_n(\widehat v)=\int_0^1\E_{p_{n,t}}\norm{\widehat v_t(X)-u_{n,t}(X)}_2^2\,dt.\end{equation}
This error integrates uniformly over time. An independent time draw with density positive on $(0,1)$ preserves the conditional regression target $u_{n,t}$ but changes the time weighting of the fitting loss.

\begin{proposition}\label{prop:reference-flow}
The ODE with velocity $u_n$ and initial value $G$ has a unique solution $X_t^{u_n}$ for $t<1$, with $\law(X_t^{u_n})=p_{n,t}$. It has an almost sure and $L^2$ endpoint limit $X_1^{u_n}$ with law $\law(V)$. In particular, when $\sigma=0$, this endpoint law is $\mu_{L,n}$. Moreover, for an independent $T\sim\operatorname{Unif}[0,1]$, the risk $\mathcal R_n(v)=\E\norm{v_T(X_T)-U}_2^2$ satisfies
\[
\mathcal R_n(v)-\mathcal R_n(u_n)=\int_0^1\E_{p_{n,t}}\norm{v_t(X)-u_{n,t}(X)}_2^2\,dt=\mathcal E_n(v)
\]
whenever $\mathcal R_n(v)<\infty$.
\end{proposition}

\noindent\textit{Proof.} Conditional on $Z^\star$, the pair $(U,X_t)$ is jointly Gaussian, and by regression for Gaussians,   $\E(U\mid X_t=x,Z^\star)=a_tx+b_tZ^\star$. By Bayes' formula one obtains the weights in \eqref{eq:empirical-reference}. Since the empirical target has finite support and $M_{n,t}$ is bounded,  by differentiation 
\[
\nabla_xM_{n,t}(x)=\frac{t}{s_t^2}\operatorname{Cov}(Z^\star\mid X_t=x).
\]
For every $T_0<1$, the covariance is uniformly bounded and $s_t^2\ge(1-T_0)^2$ on $[0,T_0]$. Thus $u_n$ is smooth, globally Lipschitz in space and of linear growth on this interval, so its ODE has a unique solution.

For every smooth compactly supported $f$, differentiation along the interpolant yields
\[
\frac{d}{dt}\E f(X_t)=\E\{\nabla f(X_t)^{\mathsf T}U\}=\E\{\nabla f(X_t)^{\mathsf T}u_{n,t}(X_t)\}.
\]
To identify its law with the ODE law, let $\Phi_{t,T_0}$ denote the ODE flow from time $t$ to $T_0$ and set $f_t(x)=f(\Phi_{t,T_0}(x))$. These functions satisfy $\partial_t f_t+\nabla f_t^{\mathsf T}u_{n,t}=0$ and have support in a common compact set. Then by the  preceding calculation  $\frac{d}{dt}\E f_t(X_t)=0$, hence $\E f(X_{T_0})=\E f(\Phi_{0,T_0}(G))$. This proves $\law(X_t^{u_n})=p_{n,t}$ for every $t<1$.

Put $M_U=(\E\norm U_2^2)^{1/2}<\infty$. Conditional Jensen's inequality and the marginal identity imply $\E\norm{u_{n,t}(X_t^{u_n})}_2^2\le M_U^2$. Consequently, for $0\le s<t<1$,
\[
\left\{\E\norm{X_t^{u_n}-X_s^{u_n}}_2^2\right\}^{1/2}\le\int_s^t\left\{\E\norm{u_{n,r}(X_r^{u_n})}_2^2\right\}^{1/2}\,dr\le(t-s)M_U.
\]
Thus the trajectories have an $L^2$ endpoint limit. Also, $\E\int_0^1\norm{u_{n,t}(X_t^{u_n})}_2\,dt\le M_U$, so they have finite total path length almost surely and converge almost surely to the same limit. Since $X_t-V=(1-t)(G-V)$, the interpolant converges to $V$ in $L^2$, identifying $\law(X_1^{u_n})=\law(V)$. Finally, $\E\{U-u_{n,T}(X_T)\mid T,X_T\}=0$, so expansion of the squared loss proves the excess-risk identity. \hfill$\square$

\subsection{Propagation of fitting error}\label{app:flow-propagation}

Let $Z_t^{\widehat v}$ solve the fitted ODE in \eqref{eq:fm-generation} from $G\sim N(0,I_k)$, and let $\widetilde Z_1$ be its numerical endpoint. The following conditions control the accumulation of velocity error along the fitted trajectories.

\begin{assumption}\label{ass:learned-flow}
The fitted field $\widehat v$ is globally Lipschitz in space with a nonnegative bound $\Lambda(t)$ satisfying $\int_0^1\Lambda(t)\,dt<\infty$ and $\int_0^1\norm{\widehat v_t(0)}_2\,dt<\infty$. For an integrable function $\lambda:[0,1]\to\R$,
\[\langle\widehat v_t(x)-\widehat v_t(y),x-y\rangle\le\lambda(t)\norm{x-y}_2^2\qquad(x,y\in\R^k)\]
for almost every $t$. The numerical and exact endpoints, coupled by the same Gaussian initial value, satisfy $\{\E\norm{\widetilde Z_1-Z_1^{\widehat v}}_2^2\}^{1/2}\le\eta_{\mathrm{num}}<\infty$.
\end{assumption}

The Lipschitz and growth conditions ensure that the fitted trajectories exist uniquely. They hold, for example, for finite feedforward velocity networks with fixed weights and Lipschitz activations, with time included as an input. Together with the numerical error bound, these conditions give $Q\in\Pcal_2(\R^k)$ for a Gaussian source, including after inverse affine scaling and projection. The signed function $\lambda$ controls how quickly trajectories separate; negative values allow contraction to reduce earlier errors. Such one-sided bounds are standard in ODE stability analysis \citep[Section~I.10]{hairer1993solving}. The term $\eta_{\mathrm{num}}$ bounds the root mean squared numerical integration error. The resulting error propagation factor is

\begin{equation}\label{eq:flow-propagation-factor}K_\lambda=\left[\int_0^1\exp\left\{2\int_t^1\lambda(s)\,ds\right\}dt\right]^{1/2}.\end{equation}

Let $\mathcal Z$ be the nonempty closed convex constraint set in \eqref{eq:clipped-code}, containing every $Z_i^\star$, and set $Q=\law\{\Pi_{\mathcal Z}(\widetilde Z_1)\}$.

\begin{proposition}\label{prop:flow-stability}
Suppose Assumption~\ref{ass:learned-flow} holds and $\mathcal E_n(\widehat v)<\infty$. Then $Q\in\Pcal_2(\R^k)$ and
\begin{equation}\label{eq:empirical-delta}
\begin{aligned}
\Wp{Q,\mu_{L,n}}{\norm{\cdot}_\infty}&\le\Wp{Q,\mu_{L,n}}{\norm{\cdot}_2}\le\Delta_n,\\
\Delta_n&=\eta_{\mathrm{num}}+K_\lambda\sqrt{\mathcal E_n(\widehat v)}+\sigma,
\end{aligned}
\end{equation}
where $K_\lambda$ is defined in \eqref{eq:flow-propagation-factor}.
\end{proposition}

\noindent\textit{Proof.} The Lipschitz and growth conditions in Assumption~\ref{ass:learned-flow} give
\[
\sup_{0\le t\le1}\norm{Z_t^{\widehat v}}_2\le\exp\left\{\int_0^1\Lambda(s)\,ds\right\}\left\{\norm G_2+\int_0^1\norm{\widehat v_s(0)}_2\,ds\right\}.
\]
The right-hand side has finite second moment. The fitted trajectory is therefore continuous in $L^2$, including at time one. The numerical error bound implies $\widetilde Z_1\in L^2$, and nonexpansiveness of projection, together with $\norm{\Pi_{\mathcal Z}(z)}_2\le\norm z_2+\norm{\Pi_{\mathcal Z}(0)}_2$, implies $Q\in\Pcal_2(\R^k)$.

Start the fitted and ideal empirical flows at the same $G$, and write $D_t=Z_t^{\widehat v}-X_t^{u_n}$. The one-sided bound in Assumption~\ref{ass:learned-flow} yields, almost everywhere for $t<1$,
\[
\frac12\frac{d}{dt}\norm{D_t}_2^2\le\lambda(t)\norm{D_t}_2^2+\norm{D_t}_2\norm{\widehat v_t(X_t^{u_n})-u_{n,t}(X_t^{u_n})}_2.
\]
By Gronwall's inequality for $\norm{D_t}_2$, followed by Minkowski's inequality, 
\[
\left\{\E\norm{D_T}_2^2\right\}^{1/2}\le\int_0^T\exp\left\{\int_t^T\lambda(s)\,ds\right\}\left\{\E_{p_{n,t}}\norm{\widehat v_t(X)-u_{n,t}(X)}_2^2\right\}^{1/2}\,dt,\qquad T<1.
\]
Both trajectories have $L^2$ endpoints. Since $\lambda$ is integrable and $\mathcal E_n(\widehat v)<\infty$, dominated convergence permits $T\uparrow1$. Cauchy--Schwarz implies 
\[
\left\{\E\norm{Z_1^{\widehat v}-X_1^{u_n}}_2^2\right\}^{1/2}\le K_\lambda\sqrt{\mathcal E_n(\widehat v)}.
\]
The ideal endpoint has law $\law(V)$, and coupling $V=Z^\star+\sigma H/\sqrt{k}$ with $Z^\star$ leads to  $\Wp{\law(V),\mu_{L,n}}{\norm{\cdot}_2}\le\sigma$. Combining this with the numerical error bound and the Wasserstein triangle inequality yields
\[
\Wp{\law(\widetilde Z_1),\mu_{L,n}}{\norm{\cdot}_2}\le\eta_{\mathrm{num}}+K_\lambda\sqrt{\mathcal E_n(\widehat v)}+\sigma.
\]
Projection onto $\mathcal Z$ is nonexpansive and fixes every empirical target vector, so the same bound holds for $Q$. Finally, $\norm{\cdot}_\infty\le\norm{\cdot}_2$ proves the stated maximum norm bound. By Cauchy--Schwarz, the Lipschitz condition implies  the one-sided inequality with $\lambda=\Lambda$. 
\hfill$\square$

For the affine scaling $S$ used in Algorithm~\ref{alg:distance-fm} of the main paper, apply the argument above to the target $S_\#\mu_{L,n}$ and evaluate the velocity, numerical and endpoint-noise errors in the scaled space. The inverse map $S^{-1}$ multiplies the resulting upper bound by its Euclidean Lipschitz constant $\Lip(S^{-1})$, and projection in the original distance units cannot increase it. Accordingly, in Corollary~\ref{cor:metric-empirical} one may take $\Delta_n=\Lip(S^{-1})\{\eta_{\mathrm{num}}+K_\lambda\sqrt{\mathcal E_n(\widehat v)}+\sigma\}$ with all quantities computed for the scaled flow.

\begin{corollary}\label{cor:metric-empirical-flow}
Under Assumptions~\ref{ass:object-law} and~\ref{ass:learned-flow}, consider observations $Y_1,\ldots,Y_n$ with empirical law $P_n$ and a finite nonempty landmark set $L$. Fit the flow to $\mu_{L,n}$ using source $N(0,I_k)$ and endpoint noise level $\sigma\ge0$, and let $Q$ be its projected numerical endpoint law in \eqref{eq:clipped-code}. If $\mathcal E_n(\widehat v)<\infty$ and the implemented reconstruction map $D:\R^k\to\M$ is measurable with finite residual $e_Q(D)$ in \eqref{eq:coordinate-errors}, then
\begin{equation}\label{eq:metric-empirical-flow}\begin{aligned}\Wp{D_\#Q,P}{d}&\le\Wp{P_n,P}{d}+e_Q(D)+2q_2(P_n,L)\\&\quad+\eta_{\mathrm{num}}+K_\lambda\sqrt{\mathcal E_n(\widehat v)}+\sigma.\end{aligned}\end{equation}
\end{corollary}
\noindent\textit{Proof.} Proposition~\ref{prop:flow-stability} implies $Q\in\Pcal_2(\R^k)$ and the bound \eqref{eq:empirical-delta}. Substitution into Corollary~\ref{cor:metric-empirical} proves the result. \hfill$\square$

\section{Implementation details}\label{app:experiment-implementation}

For equally sized generated and test samples, the empirical Wasserstein distance in \eqref{eq:empirical-w2-general} reduces to
\begin{equation}\label{eq:empirical-w2-experiment}\widehat W_2=\left\{\frac{1}{m}\min_{\pi\in\mathfrak S_m}\sum_{i=1}^m d\bigl(\widehat Y_i,Y^{\rm test}_{\pi(i)}\bigr)^2\right\}^{1/2},\end{equation}
where $\mathfrak S_m$ is the set of permutations of $\{1,\ldots,m\}$. The simulations solve this assignment exactly with $m=5000$.

Fr\'echetFlow and the Euclidean FM baseline use Meta's \texttt{flow\_matching} package with the straight paths setting in \eqref{eq:fm-training-path}. For the velocity networks, we use LightningDiT blocks \citep{yao2025reconstruction}, with one token per vector entry, learned positional embeddings, RMS normalization and SwiGLU layers. Velocity networks for simulations have three width-64 blocks and four attention heads, and for applications, velocity networks have six width-256 blocks and eight heads. All simulation flows use 20,000 optimization iterations, and generation uses 64 Heun steps; these settings are fixed across targets and sample sizes. Inputs are standardized entrywise using the training data. The fitted flow is selected by minimizing validation velocity mean squared error.

\subsection{Euclidean simulations}\label{app:euclidean-implementation}

The targets and sampling designs are specified in Section~\ref{sec:euclidean-experiments}. Both methods use the common simulation flow settings, separately standardizing object vectors and landmark representations. For points in the square $[-1,1]^2$, Fr\'echetFlow uses the three landmarks $L=\{(-1,-1),(1,-1),(-1,1)\}$, representing each point by its three Euclidean distances to these landmarks. Since the square has diameter $\sqrt8$, we clip each generated distance to $[0,\sqrt8]$. The resulting vector $z$ is reconstructed as
\[
D_\eta(z)=-1+2\operatorname{sigmoid}\!\left(f_\eta(z/\sqrt8)\right),
\]
where $f_\eta:\mathbb R^3\to\mathbb R^2$ has three width-64 hidden layers with SiLU activations and $D_\eta$ is the reconstruction map in \eqref{eq:decoder}. For the direct FM baseline, which generates point coordinates rather than landmark distances, we clip each coordinate of the final ODE output to $[-1,1]$.

\subsection{Locations on a metric graph}\label{app:metric-graph-implementation}

The graph consists of three paths joining the common endpoints $a=(-1.5,0)$ and $b=(1.5,0)$. For each $h\in\{-1,0,1\}$, define the intermediate vertices $u_h=(-0.5,h)$ and $v_h=(0.5,h)$, with edges connecting $a$ to $u_h$, $u_h$ to $v_h$, and $v_h$ to $b$. Thus, the graph has eight vertices and nine edges. The three paths meet only at $a$ and $b$; there are no edges connecting intermediate vertices at different levels. Each edge is assigned its Euclidean segment length: $\ell(\{a,u_h\})=\ell(\{v_h,b\})=\sqrt{1+h^2}$ and $\ell(\{u_h,v_h\})=1$. Consequently, the middle path has total length $3$, whereas the upper and lower paths each have total length $1+2\sqrt{2}$. An object is a point anywhere on the graph, including the interior of an edge. The distance $d_G(x,y)$ is the length of the shortest path connecting $x$ and $y$ along the graph. All eight vertices serve as landmarks. Writing $L=(q_1,\ldots,q_8)$ for a fixed ordering of these vertices, the landmark representation of an object $x$ is $\phi_L(x)=\bigl(d_G(x,q_1),\ldots,d_G(x,q_8)\bigr)^{\mathsf T}\in\mathbb{R}^8$.

The eight vertices also serve as mean references. The reconstruction network supplies two softmax weights for the endpoints of each edge, giving a point along that edge. We return the candidate with the smallest maximum-norm reconstruction residual.

\subsection{Phylogenetic trees}\label{app:phylogenetic-implementation}

An observation is an unrooted binary tree on five labelled taxa A--E, with three branches meeting at each internal vertex. Its five terminal branches connect to the leaves and have length one, while its two internal lengths vary in $[0,1]$. Each topology (branching pattern) has two leaf pairs sharing internal vertices and one remaining leaf: five choices for that leaf and three pairings of the other four give $15$ topologies. For example, AB/CD pairs A with B and C with D, with E at the central vertex. The BHV metric \citep{billera2001geometry} compares internal lengths by Euclidean distance within a topology; shortest paths between topologies pass through trees with zero internal branch lengths. We index the topology labels lexicographically by $j=0,\ldots,14$. Topologies AB/CD, AD/BE and BE/CD ($j=0,7,14$) have probabilities $0.35$, $0.25$ and $0.15$, respectively; each remaining topology has probability $0.25/12$. Conditional on the topology, the two internal lengths are drawn independently from the following distributions, with the first length corresponding to the alphabetically first pair in the label:
\[
\begin{array}{c cc}
\text{Topology index }j & \text{First branch length} & \text{Second branch length} \\
0,3,6,9,12 & \operatorname{Beta}(2,5) & \operatorname{Beta}(5,2) \\
1,4,7,10,13 & \operatorname{Beta}(5,2) & \operatorname{Beta}(3,3) \\
2,5,8,11,14 & \operatorname{Beta}(2,2) & \operatorname{Beta}(2,2)
\end{array}
\]

\begin{figure}[htbp]
\centering
\begingroup
\definecolor{treeblue}{RGB}{34,93,158}
\definecolor{treeorange}{RGB}{190,93,25}

\newcommand{\treepanel}[6]{%
  \begin{scope}[shift={(#1,#2)}]
    \filldraw[fill=black!2,draw=black!12,rounded corners=5pt]
      (-2.4,-1.65) rectangle (2.4,1.65);
    \node[font=\small\bfseries] at (0,1.33)
      {#3\quad $(a,b)=(#4,#5)$};
    \node[font=\scriptsize,text=black!65] at (0,0.98) {#6};
    \begin{scope}[yshift=-0.1cm]
      \coordinate (L) at ({-1.05*#4},0);
      \coordinate (M) at (0,0);
      \coordinate (R) at ({1.05*#5},0);
      \draw[leafbranch] (L) -- ++(-0.48,0.58) node[above left] {$A$};
      \draw[leafbranch] (L) -- ++(-0.48,-0.58) node[below left] {$B$};
      \draw[leafbranch] (R) -- ++(0.48,0.58) node[above right] {$C$};
      \draw[leafbranch] (R) -- ++(0.48,-0.58) node[below right] {$D$};
      \draw[leafbranch] (M) -- (0,-0.95) node[below] {$E$};
      \ifdim #4pt>0pt
        \draw[internalbranch] (L) -- node[above=1pt,text=treeblue] {$a$} (M);
      \fi
      \ifdim #5pt>0pt
        \draw[internalbranch] (M) -- node[above=1pt,text=treeblue] {$b$} (R);
      \fi
      \foreach \p in {L,M,R} \fill[black!75] (\p) circle (1.8pt);
    \end{scope}
  \end{scope}%
}

\resizebox{\linewidth}{!}{%
\begin{tikzpicture}[
    x=1cm,y=1cm,
    every node/.style={font=\small},
    leafbranch/.style={line width=0.8pt,black!65},
    internalbranch/.style={line width=1.7pt,treeblue},
    corner/.style={circle,fill=treeblue,inner sep=2.6pt},
    example/.style={circle,fill=treeorange,inner sep=2.6pt}
]
\begin{scope}[shift={(1.0,7.8)},x=3.2cm,y=3.2cm]
  \fill[treeblue!5] (0,0) rectangle (1,1);
  \draw[black!25] (0,0) rectangle (1,1);
  \draw[->,black!70] (0,0) -- (1.20,0) node[right] {$a$};
  \draw[->,black!70] (0,0) -- (0,1.20) node[above] {$b$};
  \node[corner,label={[text=treeblue]below left:T1}] at (0,0) {};
  \node[corner,label={[text=treeblue]below right:T2}] at (1,0) {};
  \node[corner,label={[text=treeblue]above left:T3}] at (0,1) {};
  \node[corner,label={[text=treeblue]above right:T4}] at (1,1) {};
  \node[example,label={[text=treeorange]above right:T5}] at (0.35,0.70) {};
  \node[example,label={[text=treeorange]above left:T6}] at (0.80,0.30) {};
  \node[below=3pt] at (0.5,0) {$0\leq a\leq1$};
  \node[rotate=90,above=3pt] at (0,0.5) {$0\leq b\leq1$};
\end{scope}

\node[anchor=north west,align=left,text width=8.2cm]
  at (6.0,11.6) {%
    Each labeled point corresponds to the tree shown in the matching panel below.};
\draw[internalbranch] (6.0,10.4) -- (6.6,10.4);
\node[anchor=west] at (6.8,10.4) {Blue: internal branches $a$ and $b$.};
\draw[leafbranch] (6.0,9.8) -- (6.6,9.8);
\node[anchor=west] at (6.8,9.8) {Gray: fixed branches leading to the leaves.};
\node[anchor=north west,align=left,text width=8.2cm]
  at (6.0,9.15) {%
    A zero internal length merges its two endpoints.\par\medskip
    T1--T4 are the four corner references.\par
    T5 and T6 are examples inside the square.};

\treepanel{2.5}{5.35}{T1}{0}{0}{Star tree}
\treepanel{7.5}{5.35}{T2}{1}{0}{AB internal branch only}
\treepanel{12.5}{5.35}{T3}{0}{1}{CD internal branch only}
\treepanel{2.5}{1.75}{T4}{1}{1}{Two unit internal branches}
\treepanel{7.5}{1.75}{T5}{0.35}{0.70}{Shorter $a$; longer $b$}
\treepanel{12.5}{1.75}{T6}{0.80}{0.30}{Longer $a$; shorter $b$}
\end{tikzpicture}
}%
\endgroup

\caption{Six trees corresponding to points in the branch-length square
for topology AB/CD and its collapsed boundary trees. The left internal
branch has length $a$ and the right has length $b$. Blue branch lengths
are drawn proportionally, while all terminal branch lengths remain
fixed across panels. T1--T4 are the four reference trees; T5 and T6
illustrate possible reconstructed trees.}
\label{fig:tree-reconstruction}
\end{figure}
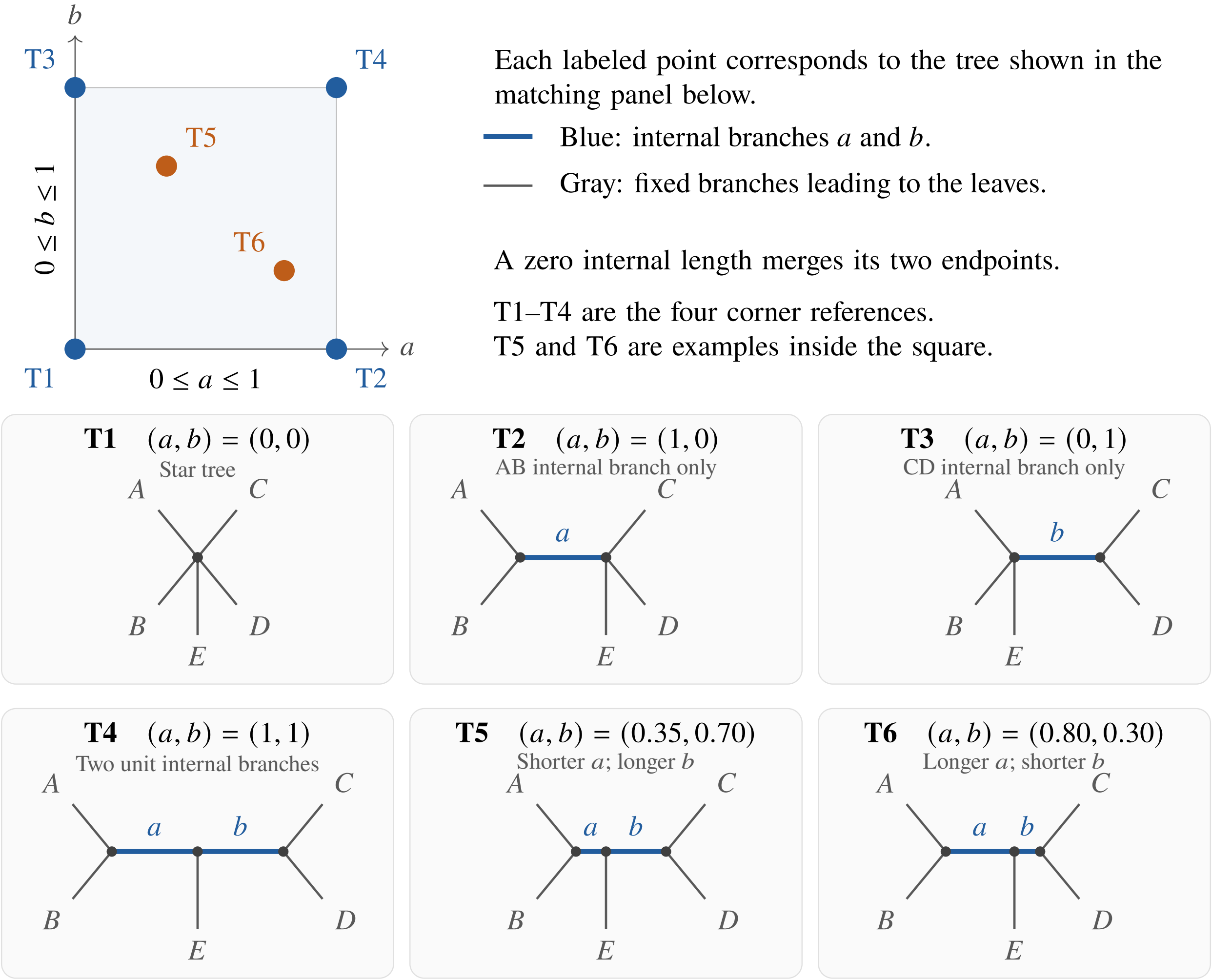

Figure~\ref{fig:tree-reconstruction} illustrates the branch-length square for topology AB/CD and its boundary trees.

The four corners of the square represent four simple trees. At $(0,0)$, both internal branches have collapsed, giving the star tree. At $(1,0)$ or $(0,1)$, only one internal branch remains, with length one. At $(1,1)$, both internal branches have length one. We collect these corner trees from all 15 branching patterns. These trees serve as both landmarks for measuring distances and references for reconstruction. Distances between trees are measured using the BHV metric.

The flow generates a vector of 26 distances, one to each landmark tree. We first clip these distances to $[0,2\sqrt2]$. Using the distances to the star tree and the single branch references, we estimate the two internal branch lengths for each of the 15 topologies. We compare the landmark distances of these initial trees with the generated vector and retain the three topologies with the smallest maximum absolute errors.

We then reconstruct one tree within each retained topology. The reconstruction network predicts scores for the reference trees, and a softmax over the four corner scores of each topology yields  four weights. The weighted average of the corner coordinates determines the reconstructed branch lengths. This yields three candidate trees, one for each retained topology. For each candidate, we compute its distances to all 26 landmarks and return the candidate with the smallest sup norm discrepancy from the generated distance vector as the final predicted tree.
\subsection{\texorpdfstring{Joint positions on two branching tracks}{Joint positions on two branching tracks}}\label{app:cubical-implementation}

We simulate joint positions of two robots on the Y-shaped tracks in Example~\ref{sec:cubical-object-examples}, using the intrinsic product metric defined there. Let $a,b\in\{1,2,3\}$ denote the branches occupied by the first and second robots, respectively, and let $u,v\in[0,1]$ be their distances from the junctions along the selected unit branches. We draw $(a,b)$ jointly using the following probability matrix, whose rows index the first robot's branch and columns index the second robot's branch:
\[\{P(a,b)\}_{a,b=1}^3=\begin{pmatrix}0.25&0.04&0.06\\0.05&0.20&0.05\\0.10&0.05&0.20\end{pmatrix}.\]
Conditional on $(a,b)$, we draw $u$ and $v$ independently from $\operatorname{Beta}(\alpha_a,\beta_a)$ and $\operatorname{Beta}(\alpha_b,\beta_b)$, respectively, where $(\alpha_1,\beta_1)=(2,5)$, $(\alpha_2,\beta_2)=(3,3)$ and $(\alpha_3,\beta_3)=(5,2)$. These beta laws concentrate positions toward the junction on branch 1, around the midpoint on branch 2 and toward the outer endpoint on branch 3. The joint branch probabilities make the two robot positions dependent.

\begin{figure}[t]
\centering
\begin{tikzpicture}[
    font=\small,
    vertex/.style={circle,fill=black,inner sep=1.6pt},
    sample/.style={circle,fill=red!80!black,inner sep=2.2pt},
    >=Latex
]

\begin{scope}[shift={(0,0)}]
    \node at (0,1.7) {$T_1$};
    \coordinate (o) at (0,0);
    \coordinate (a1) at (-0.9,1);
    \coordinate (a2) at (0.9,1);
    \coordinate (a3) at (0,-1.2);

    \draw[thick] (a1)--(o)--(a2);
    \draw[thick] (o)--(a3);
    \draw[blue!70!black,very thick] (o)--(a2);

    \node[vertex,label=left:$o_1$] at (o) {};
    \node[vertex,label=above:$a_1$] at (a1) {};
    \node[vertex,label=above:$a_2$] at (a2) {};
    \node[vertex,label=below:$a_3$] at (a3) {};
    \node[sample,label=right:$p$] at (0.54,0.6) {};
\end{scope}

\node at (1.65,0) {\Large $\times$};

\begin{scope}[shift={(3.3,0)}]
    \node at (0,1.7) {$T_2$};
    \coordinate (o) at (0,0);
    \coordinate (b1) at (-0.9,1);
    \coordinate (b2) at (0.9,1);
    \coordinate (b3) at (0,-1.2);

    \draw[thick] (b1)--(o)--(b2);
    \draw[thick] (o)--(b3);
    \draw[blue!70!black,very thick] (o)--(b3);

    \node[vertex,label=left:$o_2$] at (o) {};
    \node[vertex,label=above:$b_1$] at (b1) {};
    \node[vertex,label=above:$b_2$] at (b2) {};
    \node[vertex,label=below:$b_3$] at (b3) {};
    \node[sample,label=right:$q$] at (0,-0.48) {};
\end{scope}

\draw[->,thick] (4.6,0) -- (5.4,0);

\begin{scope}[shift={(5.9,-1.9)}]
    \node at (2,4.35) {$T_1\times T_2$: nine square cells};

    \foreach \i in {1,2,3} {
        \foreach \j in {1,2,3} {
            \begin{scope}[
                shift={({1.45*(\i-1)},{1.45*(3-\j)})}
            ]
                \ifnum\i=2
                    \ifnum\j=3
                        \fill[blue!15] (0,0) rectangle (1.1,1.1);
                    \fi
                \fi
                \draw[thick] (0,0) rectangle (1.1,1.1);
                \foreach \x in {0,1.1} {
                    \foreach \y in {0,1.1} {
                        \node[vertex] at (\x,\y) {};
                    }
                }
                \node at (0.55,0.82) {$C_{\i\j}$};
            \end{scope}
        }
    }

    \node[sample] at (2.11,0.44) {};
    \node[font=\scriptsize,anchor=west]
        at (2.20,.35) {$x=(p,q)$};
\end{scope}

\node[align=center] at (4.4,-3.15) {
    $C_{ij}=[o_1,a_i]\times[o_2,b_j]$,
    \qquad
    $L=B=V=\{o_1,a_1,a_2,a_3\}\times\{o_2,b_1,b_2,b_3\}$\\[5pt]
    Each point $x$ is represented by
    $\phi_L(x)=\bigl(d(x,v)\bigr)_{v\in V}\in\mathbb R^{16}$.
};

\end{tikzpicture}
\caption{
Geometry of the cubical dataset.
Each point is a pair of positions on two three-branch metric trees.
The nine branch pairs form nine square cells, and the
$4\times4=16$ vertex pairs serve as landmarks and reference points.
The red points illustrate a sample in $C_{23}$.
}
\label{fig:cubical-dataset}
\end{figure}
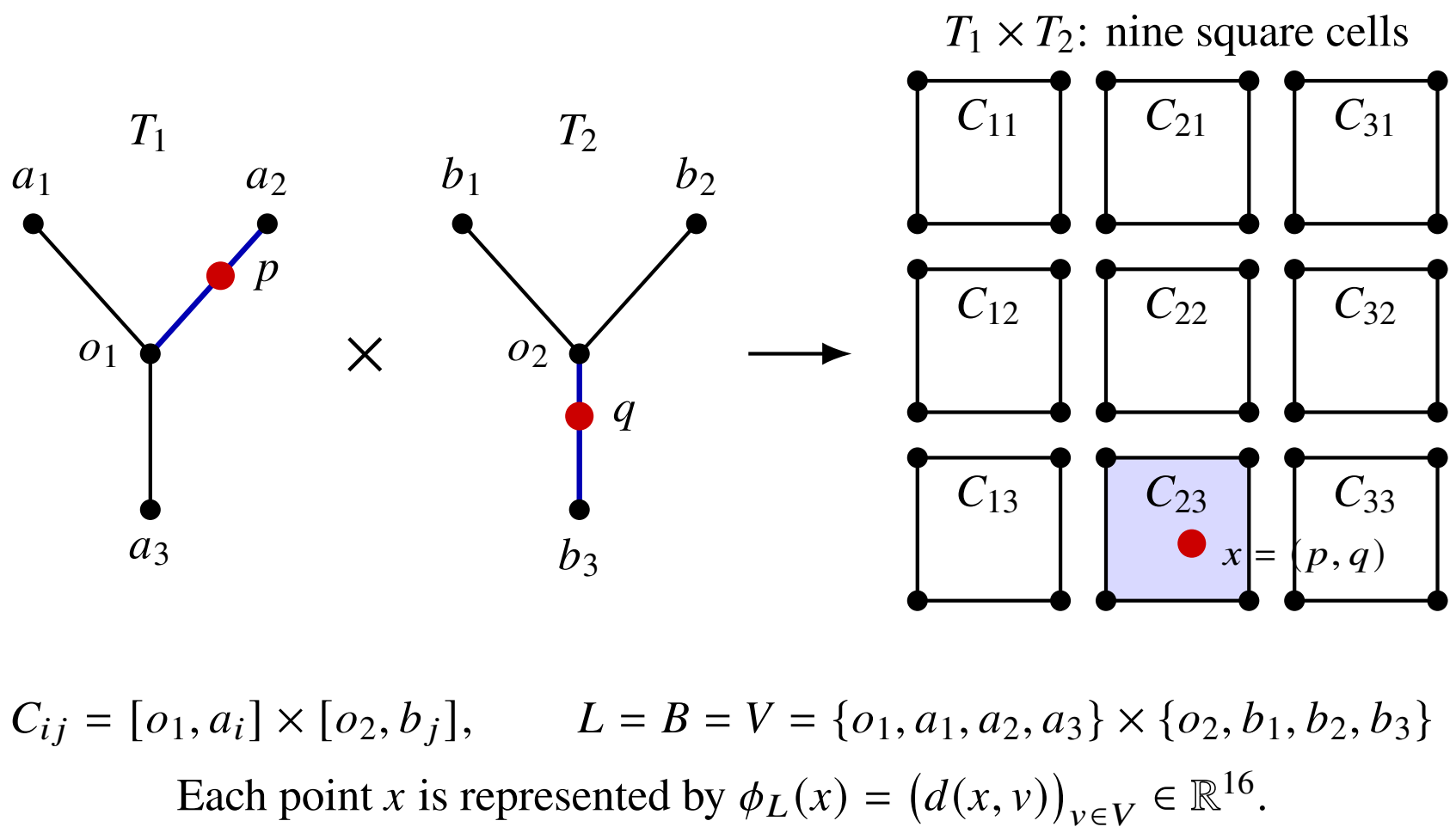
The space in Section~\ref{sec:graph-tree-experiments}, illustrated in Figure~\ref{fig:cubical-dataset}, is the product of two Y-shaped metric trees. Each point is a pair of positions, one on each tree. The nine possible branch pairs form nine square cells, joined along shared edges and vertices. The sixteen pairs of tree vertices serve as both distance landmarks and reference points for weighted means. Each data point is represented by its distances to these sixteen landmarks.

To reconstruct a point, we first project the generated distance vector onto $[0,2\sqrt{2}]^{16}$. A preliminary search using grid points and squared-distance equations selects three candidate cells by their smallest attained reconstruction residuals. The network outputs sixteen vertex scores. Within each retained cell, we normalize its four corner scores by softmax and use the resulting weights to average the corners, which defines a candidate point. We return the candidate whose landmark distance vector has the smallest sup norm discrepancy.

\subsection{Earth data on the sphere}\label{app:sphere-implementation}

We represent each point on the sphere by its great-circle distances to $e_1,e_2,e_3$ and clip generated distances to $[0,\pi]$. We then use 42 reference points from a once-subdivided icosahedron. Each reference point and its mesh neighbors form a group. Given a distance vector $z$, we choose the three groups whose centers have the smallest maximum distance error. Within each group, the network's scores are normalized by softmax and an intrinsic weighted mean is computed iteratively. We return the candidate whose landmark distance vector has the smallest sup norm discrepancy.

\subsection{MERFISH cell neighborhoods}\label{app:merfish-implementation}

Each observation consists of the gene expression vectors of 16 cells near an Sst neuron, as described in Section~\ref{sec:merfish-experiment}. For each of ten data splits, we divide the 64 tissue sections into 51 training, six validation, and seven test sections. 

Starting from a randomly chosen training neighborhood, farthest-point selection successively adds the neighborhood farthest from its nearest selected reference under the Wasserstein metric \eqref{eq:merfish-object-metric}. The 512 selected neighborhoods form $B$, and the first 64 form $L$. Generated distances are projected onto $[0,\infty)^{64}$. The reconstruction network predicts softmax weights over all 512 references. An approximate weighted Wasserstein mean is computed by alternating optimal assignments of the current cloud's atoms to each reference cloud with weighted averages of the matched atoms.

\subsection{Brain connectivity matrices}\label{app:brain-decoding}

ABIDE time series use the 39 regions' MSDL atlas from the DiffeoCFM release \citep{collas2025diffeocfm}. Oracle approximating shrinkage covariance estimates are converted to correlations by diagonal normalization. We reserve 10\% for testing and 10\% of the remainder for validation, using the rest for training, yielding 684 training, 77 validation and 85 test subjects per partition. Both methods use the normalized Cholesky metric in Section~\ref{sec:brain-experiment}.

Write $x_i=\Psi(C_i)\in\R^{741}$ and let $\bar x$ be the training mean. Set $p=741$, $n_{\rm fit}=684$ and
\[
a^2=\frac1{n_{\rm fit}}\sum_i\norm{x_i-\bar x}_2^2,\qquad R_0=1.25\max_i\norm{x_i-\bar x}_1.
\]
Let $v_1,\ldots,v_{p+1}$ be the unit vertices of a centered regular simplex with fixed Helmert orientation, satisfying $\sum_jv_j=0$ and $\sum_jv_jv_j^{\mathsf T}=(p+1)I_p/p$. The landmarks and references are
\[
L=\{\Psi^{-1}(\bar x+av_j):1\le j\le p+1\},\qquad B=\{\Psi^{-1}(\bar x\pm R_0e_j):1\le j\le p\}.
\]
The reconstruction network predicts softmax weights over all reference matrices. Their normalized-Cholesky coordinates are averaged with these weights and mapped back through $\Psi^{-1}$ to obtain a correlation matrix.

\FloatBarrier
\end{document}